\documentclass[pra,reprint,a4paper,showpacs]{revtex4-1}
\usepackage{color}
\usepackage[pdftex]{graphicx}

\usepackage[cmex10]{amsmath}
\begin{document}

\title{A Review on {\it Ab Initio} Approaches for Multielectron Dynamics}

\author{Kenichi L. Ishikawa}
\email[Electronic mail:]{ishiken@n.t.u-tokyo.ac.jp}
\author{Takeshi Sato}
\email[Electronic mail:]{sato@atto.t.u-tokyo.ac.jp}
\affiliation{
Department of Nuclear Engineering and Management, School of Engineering,
The University of Tokyo, 7-3-1 Hongo, Bunkyo-ku, Tokyo 113-8656, Japan
}
\affiliation{
Photon Science Center, School of Engineering, 
The University of Tokyo, 7-3-1 Hongo, Bunkyo-ku, Tokyo 113-8656, Japan
}

\date{\today}

\begin{abstract}
In parallel with the evolution of femtosecond and attosecond laser as well as free-electron laser technology, a variety of theoretical methods have been developed to describe the behavior of atoms, molecules, clusters, and solids under the action of those laser pulses. Here we review major {\it ab initio} wave-function-based numerical approaches to simulate multielectron dynamics in atoms and molecules driven by intense long-wavelength and/or ultrashort short-wavelength laser pulses. Direct solution of the time-dependent Schr\"odinger equation (TDSE), though its applicability is limited to He, ${\rm H}_2$, and Li, can provide an exact description and has been greatly contributing to the understanding of dynamical electron-electron correlation. Multiconfiguration self-consistent-field (MCSCF) approach offers a flexible framework from which a variety of methods can be derived to treat both atoms and molecules, with possibility to systematically control the accuracy. The equations of motion of configuration interaction coefficients and molecular orbitals for general MCSCF ansatz have recently been derived. Time-dependent extension of the $R$-matrix theory, originally developed for electron-atom collision, can realistically and accurately describe laser-driven complex multielectron atoms.  
\end{abstract}


%

\maketitle

\section{Introduction}
%
%
%
%
Atoms and molecules, subject to visible (VIS)-to-midinfrared (MIR) laser pulses with an intensity typically higher than $10^{14}\,{\rm W/cm}^2$, exhibit highly nonlinear phenomena including above-threshold ionization (ATI), tunneling ionization, high-harmonic generation (HHG), and nonsequential double ionization (NSDI) \cite{Protopapas1997RPP,Brabec2000RMP}. HHG is, especially, more and more widely used as an ultrashort (down to attoseconds) coherent light source in the extreme-ultraviolet (XUV) and soft x-ray (SX) spectral ranges \cite{Popmintchev2012Nature,Chang2011,AttosecondPhysics}. In addition, another type of ultrashort, intense, coherent XUV and x-ray sources, i.e. free-electron lasers have emerged. These developments have triggered novel research activities such as ultrafast molecular probing, attosecond science, and XUV nonlinear optics \cite{Itatani2004Nature,Haessler2010NatPhys,Salieres2012RPP,Agostini2004RPP,Krausz2009RMP,Gallmann2012ARPC,Sekikawa2004Nature,Nabekawa2005PRL}.

Time-dependent simulations of the electronic dynamics in atoms and molecules still remain a challenge. For high-field phenomena, direct solution of the time-dependent Schr\"odinger equation (TDSE) within the single-active electron (SAE) approximation is widely used, in which only the outermost electron is explicitly treated, with the influence of the others expressed by a frozen effective potential. Naturally, however, this approximation fails to treat multielectron and multichannel effects \cite{Haessler2010NatPhys,Gordon2006PRL,Rohringer2009PRA,Smirnova2009Nature,Smirnova2009PNAS,Akagi2009Science,Boguslavskiy2012Science}, which are attracting increasing interest. Thus, various many-electron methods have recently been under active development.

In this article, we give an overview of {\it ab initio (first-principles) wave-function-based} approaches to simulate multielectron dynamics in atoms and molecules driven by intense long-wavelength (VIS to MIR) and/or ultrashort short-wavelength (XUV to SX) laser pulses. 
Let us consider, within the Born-Oppenheimer (BO) (or fixed-nuclei) and dipole approximations, that an $N$-electron atomic or molecular system is driven by an external laser electric field ${\bf E}(t)$. The dynamics of the electronic system is governed by the time-dependent Schr\"odinger equation,
\begin{equation}
\label{eq:TDSE}
i\frac{\partial\Psi (t)}{\partial t} = \hat{H}(t)\Psi (t)
\end{equation}
where the time-dependent Hamiltonian 
\begin{equation}
\hat{H}(t)=\hat{H}_1(t)+\hat{H}_2,
\end{equation}
is decomposed into the one-electron part,
\begin{equation}
\label{eq:H1}
\hat{H}_1(t) = \sum_i \hat{h}({\bf r}_i,t) 
\end{equation}
and the two-electron part,
\begin{equation}
\hat{H}_2 = \sum_{i=1}^N \sum_{j < i} \frac{1}{|{\bf r}_i - {\bf r}_j|},
\end{equation}
for the interelectronic Coulomb interaction. $\hat{h}({\bf r}_i,t)$ in Eq.~(\ref{eq:H1}) is given by,
\begin{equation}
\label{eq:length-gauge}
\hat{h}({\bf r}_i,t) = \frac{\hat{{\bf p}}_i^2}{2}-\sum_\alpha \frac{Z_\alpha}{|{\bf r}_i - {\bf R}_\alpha|}+{\bf r}_i\cdot {\bf E}(t),
\end{equation}
in the length gauge, with $\hat{{\bf p}}_i=-i\nabla_i$, and, 
\begin{equation}
\label{eq:velocity-gauge}
\hat{h}({\bf r}_i,t) = \frac{1}{2}\left[\hat{{\bf p}}_i+{\bf A}(t)\right]^2-\sum_\alpha \frac{Z_\alpha}{|{\bf r}_i - {\bf R}_\alpha|},
\end{equation}
in the velocity gauge, with ${\bf A}(t) = -\int {\bf E}(t)dt$ being the vector potential.

First, in Sec. \ref{sec:TDSE}, we introduce direct solution of the TDSE Eq.~(\ref{eq:TDSE}). This approach provides an exact description for He, Li, and ${\rm H}_2$ and is powerful especially for the investigation of multiphoton ionization by XUV pulses. However, its extension beyond these species is extremely difficult, due to exponential increase in computational cost. Thus, in Sec. \ref{sec:MCSCF}, we discuss a general class of alternative methods that can handle more electrons, which we call multiconfiguration self-consistent-field (MCSCF) approaches. These are extension of those developed in quantum chemistry for the ground-state electronic state, to the dynamics involving excitation and ionization. Variants with different levels of flexibility are reviewed. Section \ref{sec:R-matrix} briefly discusses alternative methods that can describe multielectron atoms. These are time-dependent extension of the $R$-matrix theory developed for electron scattering from an atom and ion.

In {\it ab initio} simulation study of multielectron dynamics, we usually need to (i) prepare the initial state, (ii) propagate the wave function in time, and (iii) read out physically relevant information from the wave function. Furthermore, since ionization is essential, it is one of the major issues how to treat electrons that leave the calculation region. Due to limitations of space, however, let us concentrate on the propagation of the wave function in this review. We briefly note that the initial state can generally be obtained through either propagation in imaginary time or, especially in spectral methods, separate time-independent calculation of the ground state.

Before ending the introductory section, it is worth mentioning other approaches that bypass explicit use of $N$-electron wave function, thus, outside the scope of this review. The time-dependent density functional theory (TDDFT) \cite{Gross1996,TDDFT,Otobe2004PRA,Otobe2008PRB,Telnov2009PRA} favorably scales linearly with $N$. It is, however, difficult to estimate and systematically improve the accuracy of the exchange-correlation potential, whose form beyond the adiabatic limit is not yet known. Whereas an alternative called the time-dependent current-density functional theory \cite{TDDFT} has been proposed, only few approximations for the exchange-correlation vector potential are available \cite{Vignale1996PRL}. Also, these methods deliver only the electron density or current, not the wave function, rendering the extraction of physical observables difficult. Another attractive method that can in principle take account of correlation effects and extract any one- and two-particle observable is the time-dependent two-particle reduced density matrix (TD-2RDM) method \cite{Schaefer-Bung2008PRA,Lackner2015PRA}. A major challenge in this method is how to impose so-called $N$-representability conditions, whose complete list is not known yet. Time-dependent quantum Monte Carlo (TDQMC) method \cite{Christov2007NJP,Christov2011JCP,Oriols} uses de Broglie-Bohm trajectories. The calculation of the quantum potential or guiding waves requires the knowledge of the $N$-electron wave function, thus we need some approximation. It is not obvious how to systematically improve such approximation, although Bohmian trajectories extracted from the wave function calculated with other methods give some insights into strong-field phenomena \cite{Sawada2014PRA,Song2012PRA,Takemoto2011JCP,Wu2013PRAa,Wu2013PRAb,Benseny2014EPJD}.  

\section{Direct Solution of the TDSE}
\label{sec:TDSE}

Direct solution of the time-dependent Schr\"odinger equation (TDSE) has become possible for He, ${\rm H}_2$, and Li. A remarkable advantage of this approach is that it provides, in principle, an exact description. In this section, we review simulations for these three species (The Li case is briefly mentioned in Sec.~\ref{sebsec:TDSE-He}). 

\subsection{He}
\label{sebsec:TDSE-He}

TDSE simulations for He have been developed and applied to study on, e.g., single- and two-photon double ionization \cite{Pindzola1998JPB,Parker2001JPB,Colgan2001JPB,Colgan2002PRL,Laulan2003PRA,Colgan2004JPBa,Ishikawa2005PRA,Foumouo2006PRA,Nikolopoulos2007JPB,Lambropoulos2008NJP,Foumouo2008NJP,Foumouo2010JPB,Feist2008PRA,Palacios2009PRA,Feist2009PRL,Lee2009PRA,Pazourek2011PRA,Zhang2011PRA} as well as single ionization \cite{Laulan2003PRA,Foumouo2006PRA,Palacios2009PRA,Ishikawa2012PRL,Suren2012PRA} including delay in photoemission \cite{Schultze2010Science,Nagele2012PRA}, and also on doubly excited states \cite{Feist2011PRL,Ott2014Nature} and high-field phenomena with a longer wavelength \cite{Parker1996JPB,Parker2000JPB,Parker2006PRL,Armstrong2011NJP,NgokoDjiokap2011PRA}

\subsubsection{Grid method}

Let us first describe a frequently used {\it grid} approach called the time-dependent close coupling (TDCC) method, applied first by Taylor, Parker {\it et al.} \cite{Parker1996JPB,Smyth1998CPC} and later by many others \cite{Pindzola1998JPB,Pindzola1998PRA,Colgan2001JPB,Colgan2002PRL,Colgan2004JPBa,Colgan2004JPB,Pindzola2007JPB,Lee2009PRA,Ishikawa2005PRA,Ishikawa2012PRL,Palacios2009PRA,Feist2008PRA,Schneider2011QDI,Zhang2011PRA}. In the spherical coordinate system, the two-electron wave function $\Psi ({\bf r}_1,{\bf r}_2,t)$ is written as,
\begin{equation}
\label{eq:He-wave-function}
\Psi ({\bf r}_1,{\bf r}_2,t) = \sum_{L,M,l_1,l_2}\frac{P_{l_1,l_2}^{LM}(r_1,r_2,t)}{r_1r_2}\mathcal{Y}_{l_1,l_2}^{LM}(\Omega_1,\Omega_2),
\end{equation}
where $L, M$ denote the total orbital angular and magnetic quantum numbers, respectively, $l_1,l_2$ the angular quantum numbers of the two electrons, $P_{l_1,l_2}^{LM}(r_1,r_2,t)$ the radial wave function, $\Omega_i\, (i=1,2)$ the combined azimuthal and polar angles of the $i$-th electron, and
\begin{equation}
\mathcal{Y}_{l_1l_2}^{LM}(\Omega_1,\Omega_2)=\sum_{m_1,m_2}\langle l_1 m_1 l_2 m_2|LM\rangle Y_{l_1m_1}(\Omega_1)Y_{l_2m_2}(\Omega_2)
\end{equation}
the coupled (or bipolar) spherical harmonics, with $\langle l_1 m_1 l_2 m_2|LM\rangle$ being the Clebsch-Gordan Coefficients. An orbital angular momentum eigenstate of a two-electron system can be specified by a combination of four quantum numbers either $|l_1m_1l_2m_2\rangle $ (i.e., $Y_{l_1m_1}Y_{l_2m_2}$) or $|LMl_1l_2\rangle$, and their mutual conversion is mediated by the Clebsch-Gordan Coefficients. Thus, $\mathcal{Y}_{l_1l_2}^{LM}$ is the explicit form of $|LMl_1l_2\rangle$. In practice, the sums in Eq.\ (\ref{eq:He-wave-function}) are limited to a finite number of partial waves $(L,M,l_1,l_2)$.

If the laser pulse is linearly polarized along the $z$-direction, the value of $M$ does not change throughout the laser-atom interaction. If we further assume that the initial state has $M=0$, as is the case for the ground state He ($L=0, S=0$), we can restrict ourselves to only partial waves with $M=0$. Then, Eq.\ (\ref{eq:He-wave-function}) is simplified to,
\begin{equation}
\label{eq:He-wave-function-M0}
\Psi ({\bf r}_1,{\bf r}_2,t) = \sum_{L,l_1,l_2}\frac{P_{l_1,l_2}^{L}(r_1,r_2,t)}{r_1r_2}\mathcal{Y}_{l_1,l_2}^{L0}(\Omega_1,\Omega_2).
\end{equation}
Whereas Blodgett-Ford {\it et al.} \cite{Blodgett-Ford1993SILAP} earlier used this representation to study He under a breathing mode oscillating field, Parker {\it et al.} \cite{Parker1996JPB} introduced it in the context of intense-field multiphoton ionization for the first time. Also, Pindzola {\it et al.} \cite{Pindzola1996PRAa,Pindzola1996PRAb} applied this approach to electron-hydrogen scattering and later to double ionization of He and ${\rm H}^{-}$ \cite{Pindzola1998JPB,Pindzola1998PRA}.

By substituting Eq.\ (\ref{eq:He-wave-function}) into TDSE, we obtain a set of coupled partial differential equations,
\begin{align} 
&i\frac{\partial}{\partial t}P_{l_1,l_2}^{LM}(r_1,r_2,t) \nonumber \\
&= \sum_{L^\prime,M^\prime}\sum_{l_1^\prime,l_2^\prime}\langle L M l_1 l_2 |\hat{\bf H}| L^\prime M^\prime l_1^\prime l_2^\prime \rangle P_{l_1^\prime,l_2^\prime}^{L^\prime M^\prime}(r_1,r_2,t), \\
&= T_{l_1l_2}(r_1,r_2) P_{l_1,l_2}^{LM}(r_1,r_2,t) \nonumber \\
&+ \sum_{l_1^\prime, l_2^\prime}V_{l_1l_2l_1^\prime l_2^\prime}^L(r_1,r_2)P_{l_1^\prime l_2^\prime}^{LM}(r_1,r_2,t) \nonumber \\
&+ \sum_{L^\prime,M^\prime,l_1^\prime,l_2^\prime} W_{l_1l_2l_1^\prime l_2^\prime}^{LML^\prime M^\prime}(r_1,r_2,t)P_{l_1^\prime l_2^\prime}^{L^\prime M^\prime}(r_1,r_2,t),
\end{align}
called the time-dependent close coupling equations \cite{Pindzola1996PRAa,Pindzola1996PRAb,Pindzola1998PRA}. Here, the operators $T_{l_1l_2}$, $V_{l_1l_2l_1^\prime l_2^\prime}^L$, and $W_{l_1l_2l_1^\prime l_2^\prime}^{LML^\prime M^\prime}$ correspond to the kinetic energy and nuclear Coulomb potential, the electron-electron Coulomb interaction, and the interaction with the time-dependent laser field, respectively. Their explicit forms for $M=0$ can be found, e.g., in \cite{Smyth1998CPC,Pindzola1998PRA,Schneider2011QDI}. Note that $T_{l_1l_2}$ and $V_{l_1l_2l_1^\prime l_2^\prime}^L$ do not depend on $M$.

It is worth mentioning that Colgan {\it et al.} have extended the TDCC method described above to double and triple photoionization of Li \cite{Colgan2004PRL,Colgan2005PRA}. 

\subsubsection{Spectral method}

As an alternate approach, Kamta and Starace \cite{Kamta2001PRL,Kamta2002PRA,NgokoDjiokap2011PRA,Ott2014Nature}, and some others\cite{Laulan2003PRA,Piraux2003EPJD,Laulan2004PRA,Foumouo2006PRA,Foumouo2008NJP,Nikolopoulos2007JPB,Lambropoulos2008NJP} have developed a {\it spectral} method, in which the time-dependent wave function is expanded,
\begin{equation}
\Psi ({\bf r}_1,{\bf r}_2,t) = \sum_{\alpha,L,M}C_{\alpha}^{LM}(t)\Phi_{\alpha}^{LM} ({\bf r}_1,{\bf r}_2),
\end{equation}
in terms of field-free eigenstates $\Phi_{\alpha}^{LM}$ and the expansion coefficients $C_{\alpha}^{LM}(t)$ are propagated in time. They express the eigenstates as,
\begin{align}
\label{eq:He-eigenstates}
\Phi_{\alpha}^{LM} ({\bf r}_1,{\bf r}_2) &= \sum_{l_1,l_2,n_1,n_2} c_{\alpha,n_1,n_2}^{l_1,l_2,L,M} \nonumber \\
&\times \mathcal{A}\, \frac{F_{l_1,n_1}(r_1)}{r_1} \frac{F_{l_2,n_2}(r_2)}{r_2} \mathcal{Y}_{l_1,l_2}^{LM}(\Omega_1,\Omega_2),
\end{align}
where $\mathcal{A}$ denotes the antisymmetrization operator (antisymmetrizer), projecting onto either singlet or triplet states to guarantee the symmetry or antisymmetry of the spatial wave function under the exchange of identical particles. As basis functions $F_{l,n}(r)$, one can choose $B$-spline functions \cite{Boor}, Coulomb Sturmian functions \cite{Piraux2003EPJD}, or radial wave functions of the ${\rm He}^+$ eigenstates \cite{Lambropoulos2008NJP,NgokoDjiokap2011PRA}. For the case of (singly ionized) continuum states, one may also use ${\rm He}^+$ states for the bound electron and $B$-splines for the continuum electron \cite{Lambropoulos2008NJP}. The insertion of Eq.~(\ref{eq:He-eigenstates}) into the TDSE results in a coupled first-order ordinary differential equations for the temporal evolution of the coefficients $C_{\alpha}^{LM}(t)$. These are, more conveniently after conversion to the interaction picture, integrated, e.g., with an explicit scheme of Runge-Kutta type \cite{NumericalRecipes}. Details are described in \cite{Kamta2002PRA,Foumouo2006PRA,Lambropoulos2008NJP,NgokoDjiokap2011PRA}.

\subsubsection{Example: photoionization of an excited helium atom}

As an example, let us consider single-photon single ionization of an excited helium atom by an attosecond XUV pulse. The excited helium has a one-electron excitation character to a good approximation: one of the two electrons is much more deeply bound than the other. Irradiated by an XUV pulse, say, with a photon energy of 73 eV and a peak intensity of $10^{12}\,{\rm W/cm}^2$, predominantly the inner electron absorbs the photon, starts to travel outward, and may collide with the outer electron. We simulate this process with the grid-based TDCC method in the length gauge. 

In order to analyze the correlation-induced response and relaxation dynamics of the remaining ionic subsystem \cite{Suren2012PRA}, we first project out all bound neutral states below the first ionization potential from the instantaneous two-electron wave function $\Psi ({\bf r}_1,{\bf r}_2,t)$ and then project the resulting ionic part $\Psi_{{\rm ion}} ({\bf r}_1,{\bf r}_2,t)$ onto each bound eigenstate $\psi_i$ of ${\rm He}^+$ as,
\begin{equation}
\chi_i ({\bf r},t) = \int \psi_i^* ({\bf r}^\prime)\Psi_{{\rm ion}} ({\bf r},{\bf r}^\prime,t)d^3{\bf r}^\prime,
\end{equation}
interpreted as the continuum wave function correlated to the ionic state $\psi_i$ or the continuum wave packet in ionization channel $i$. The time-dependent population of channel $i$ is given by $2\int |\chi_i ({\bf r},t)|^2 d^3{\bf r}$. We have confirmed that the contribution from doubly excited (autoionizing) states is negligible. 

\begin{figure}[!t]
\centering
\includegraphics[width=9cm]{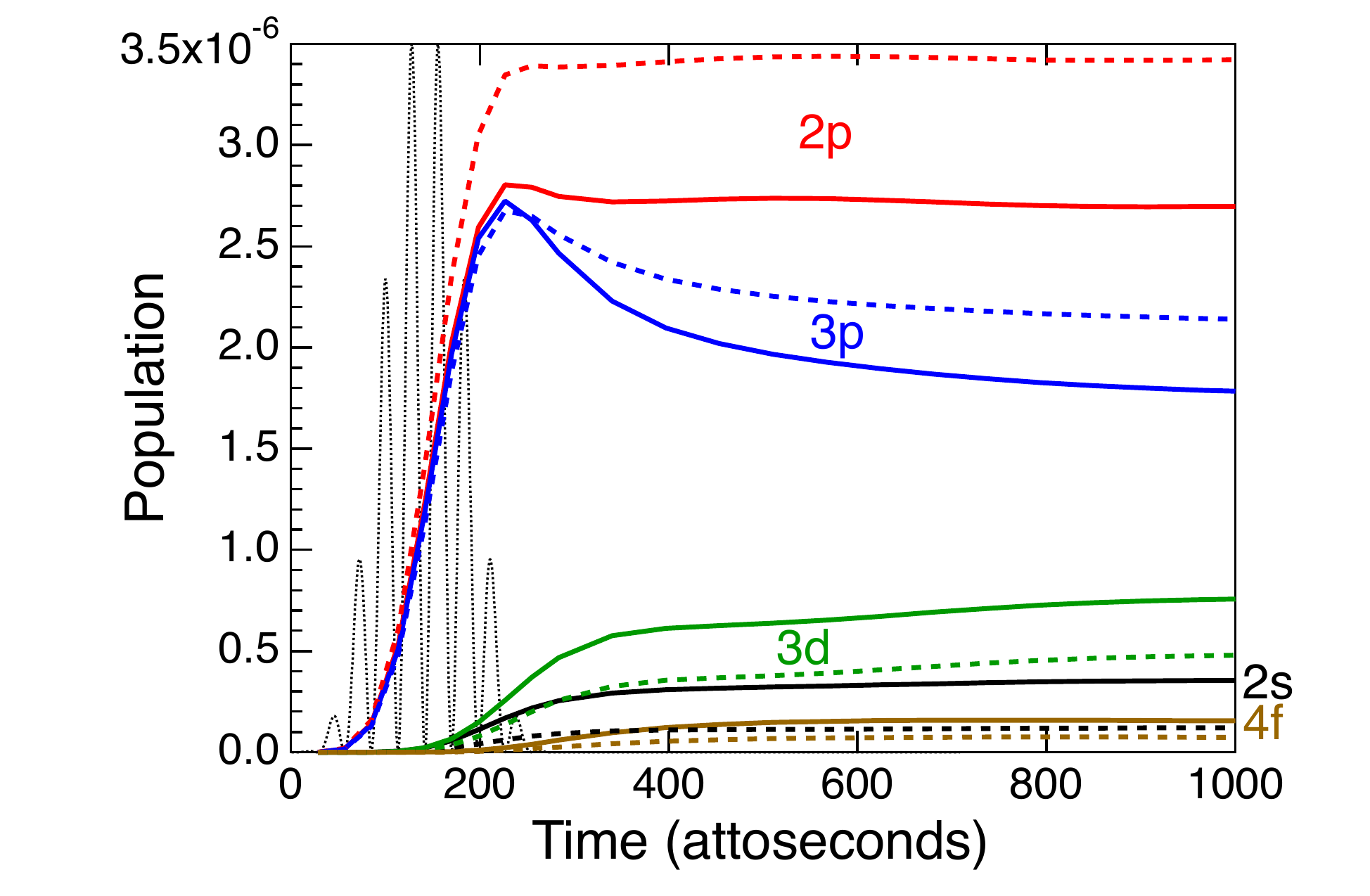}
\caption{Time-dependent populations of several ionic states (channels) for He initially in the $1s2p_z{}^1P$ (solid lines) and $1s2p_x{}^1P$ (dashed lines) states. The thin dotted line represents the magnitude of the laser vector potential in arbitrary units.}
\label{fig:knockup}
\end{figure}

The populations of several ionic channels with the initial state being $1s2p_z{}^1P$ and $1s2p_x{}^1P$ are shown in Fig.~\ref{fig:knockup}. The XUV pulse is assumed to be linearly polarized along the $z$ axis and have a base-to-base pulse duration of 5 cycles, or 280 as (see thin lines). The $1s2p_x{}^1P$ state is composed of $M=\pm 1$, for which the radiation field operator is given by Eq.~(8) of Ref.~\cite{Pindzola1998PRA} 
with $\left(\begin{array}{ccc}L & 1 & L^\prime \\0 & 0 & 0\end{array}\right)$
replaced by
$(-1)^M\left(\begin{array}{ccc}L & 1 & L^\prime \\-M & 0 & M\end{array}\right)$.
The solution is converged with the maximum values of $L, l_1, l_2$ being 2, 5, 5, respectively.

In Fig.~\ref{fig:knockup}, we identify two distinct time scales \cite{Suren2012PRA}. First, during the pulse ($t < 280$ as), instant removal of the $1s$ electron leads to the population of the $2p$ and $3p$ states by shake-up. It should be noticed that the channel population defined above is gauge-dependent {\it during the pulse}, but the qualitative feature is retained even if we use the velocity gauge. Then, {\it after the pulse}, where the results are gauge-independent, population transfer from $2p$ and $3p$ to other ionic states such as $2s$, $3d$, and $4f$ takes place. We refer to these delayed transitions, as the {\em knock-up/knock-down} processes. The motion of the outgoing inner electron through the cloud of the outer remaining electron induces transitions between ionic states \cite{Suren2012PRA}. 
This view is further strengthened if we compare the results for the two different initial states. The transitions from $2p$ and $3p$ to $2s$, $3d$, and $4f$ are clearly reduced by $66\%$, $37\%$, and $52\%$, and larger population remains in $2p$ and $3p$ if He is initially in the $1s2p_x{}^1P$ (dashed lines) state compared with the case where the initial state is $1s2p_z{}^1P$ (solid lines), since the $2p_x$ cloud of the outer electron is distributed along the $x$ axis while the inner electron is ejected along the $z$ axis, resulting in smaller electron-electron interaction. 

\subsection{$\mbox{H}_2$}

Due to the lack of spherical symmetry, TDSE simulations for $\mbox{H}_2$ are more demanding than for its two-electron atomic counterpart He. Let us assume that the internuclear axis lies on the $z$ axis. 

\begin{figure}[!t]
\centering
\includegraphics[width=7cm]{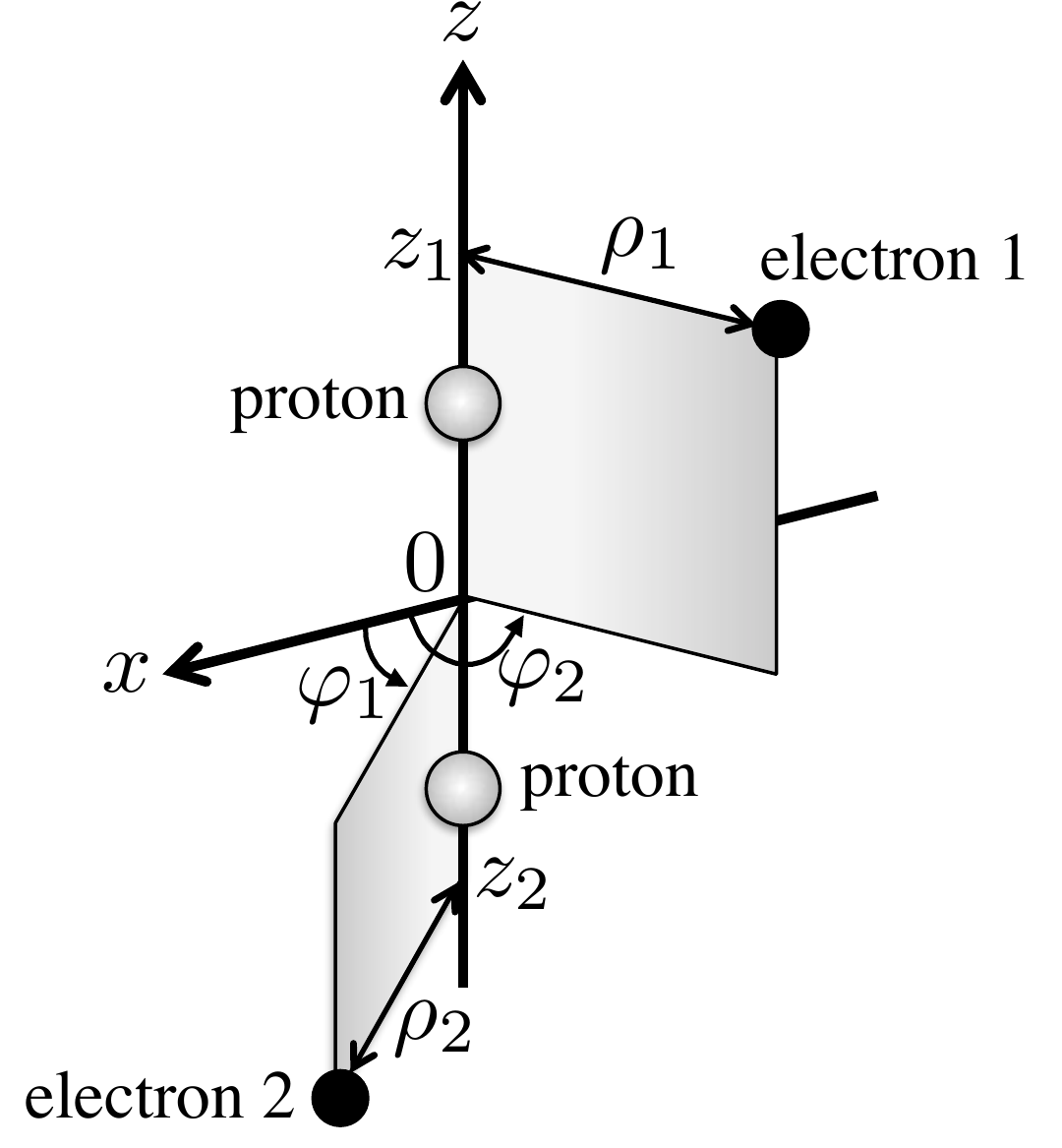}
\caption{Cylindrical coordinates $(\rho_1,z_1,\varphi_1)$ and $(\rho_2,z_2,\varphi_2)$ of the two electrons in ${\rm H}_2$ whose molecular axis lies on the $z$ axis.}
\label{fig:H2-geometry}
\end{figure}

\subsubsection{Grid method}

Kono {\it et al.} \cite{Haruyama2000JCP,Haruyama2002PRA} have developed a grid method with the molecular axis parallel to the laser polarization, by introducing cylindrical coordinates $(\rho, z, \varphi)$ (Fig.~\ref{fig:H2-geometry}). If we define, 
\begin{equation}
\phi = \varphi_1-\varphi_2,\qquad \chi = \frac{\varphi_1+\varphi_2}{2},
\end{equation}
since the $z$ component $M$ ($=0,\pm 1, \pm 2, \cdots$ with $\chi$ between 0 and $2\pi$) of the total angular momentum is conserved, the wave function $\Psi (\{\rho_j\},\{z_j\},\{\varphi_j\},t)$ ($j=1,2$) takes a form $e^{iM\chi}\Phi (\{\rho_j\},\{z_j\},\phi,t)$. Thus, the degrees of freedom can be reduced from six to five. Further, to efficiently handle the long-range nature and singularity of the Coulomb interaction, they have devised the so-called dual transformation method \cite{Kawata1999JCP,Haruyama2000JCP}. In this method, they introduce scaled coordinates $\xi$ and $\zeta$ ($\rho=f(\xi), z=g(\zeta)$) to replace $\rho$ and $z$, respectively, and accordingly transform the wave function and the Hamiltonian so that the transformed wave function vanishes at the Coulomb singular points, equidistant grids near the nucleus in terms of $\xi$ and $\zeta$ generate small grid spacings in terms of $\rho$ and $z$, and differentiation in the transformed Hamiltonian can be well evaluated by the finite difference even near the Coulomb singularity. The transformed TDSE is temporally integrated with the alternating-direction implicit (ADI) scheme \cite{NumericalRecipes}. Using this method, Kono {\it et al.} have simulated $\mbox{H}_2$ in an intense ($I\approx 10^{13}-10^{14}\,\mbox{W/cm}^2$) near-infrared laser pulse to investigate, e.g., formation of localized ionic states ${\rm H}^+{\rm H}^-$ \cite{Haruyama2000JCP,Ohtsuki2001BCSJ} and ionization enhanced by two-electron dynamics \cite{Haruyama2002PRA}.


In order to study mainly double ionization by XUV pulses, Colgan, Pindzola, Robicheaux {\it et al.} \cite{Colgan2004JPB,Colgan2008JPB,Pindzola2009PRA,Lee2010JPB} have extended the TDCC method described in the previous subsection to a molecular hydrogen, for both linear (along the $z$ axis) and circular (in the $xy$ plane) polarizations. In spherical coordinates the total wave function is expanded as, 
\begin{align}
\label{eq:CCWF-H2}
&\Psi ({\bf r}_1,{\bf r}_2,t) = \frac{1}{(2\pi)^2} \nonumber\\
&\times\sum_M\sum_{m_1,m_2}
\frac{P_{m_1m_2}^{M}(r_1,\theta_1,r_2,\theta_2,t)}{r_1r_2\sqrt{\sin\theta_1}\sqrt{\sin\theta_2}}e^{i(m_1\varphi_1+m_2\varphi_2)},
\end{align}
where $P_{m_1m_2}^{M}$ denotes a reduced wave function, and the second sum is taken over $m_1$ and $m_2$ satisfying $M=m_1+m_2$. As is mentioned above, $M$ is conserved for the case of linear polarization along the $z$ axis. By substituting Eq.\ (\ref{eq:CCWF-H2}) into the TDSE, one obtains a set of TDCC equations, whose explicit forms are found in \cite{Colgan2004JPB}. The temporal evolution of $P_{m_1m_2}^{M}$, and thus $\Psi ({\bf r}_1,{\bf r}_2,t)$, can be obtained through the integration of the TDCC equations, e.g., by an implicit algorithm \cite{Pindzola2009PRA}. As variants of this approach, F\o rre {\it et al.} \cite{Simonsen2012PRA} expand the radial and angular parts by $B$ splines \cite{Boor,Bachau2001RPP} and coupled spherical harmonics, respectively, as,
\begin{equation}
\Psi ({\bf r}_1,{\bf r}_2,t) = \sum_{i,j,k}c_{ijk}(t)\frac{B_i(r_1)}{r_1}
\frac{B_j(r_2)}{r_2}\mathcal{Y}_{l_1,l_2}^{LM}(\Omega_1,\Omega_2),
\end{equation}
with $k=\{l_1,l_2,L,M\}$. Schneider {\it et al.} \cite{Guan2010PRA}, on the other hand, use prolate spheroidal coordinates $(\xi,\eta,\varphi)$, in which the ${\rm H}_2^+$ molecule is separable:
\begin{align}
\xi &= \frac{\sqrt{x^2+y^2+(z+\frac{R}{2})^2}+\sqrt{x^2+y^2+(z-\frac{R}{2})^2}}{R}, \\
\eta &= \frac{\sqrt{x^2+y^2+(z+\frac{R}{2})^2}-\sqrt{x^2+y^2+(z-\frac{R}{2})^2}}{R}, \\
\varphi &= \arctan{\frac{y}{x}},
\end{align}
where $R$ denotes the internuclear distance. They expand the total wave function of ${\rm H}_2$ as,
\begin{align}
&\Psi ({\bf r}_1,{\bf r}_2,t) = \frac{1}{2\pi} \nonumber\\
&\times \sum_{m_1,m_2}\Pi_{m_1m_2}(\xi_1,\eta_1,\xi_2,\eta_2,t)e^{i(m_1\varphi_1+m_2\varphi_2)},
\end{align}
and discretize $(\xi,\eta)$ by a finite-element discrete-variable representation. The time evolution of a reduced wave function $\Pi_{m_1m_2}(\xi_1,\eta_1,\xi_2,\eta_2,t)$ is calculated by use of the Arnoldi-Lanczos algorithm \cite{Guan2008PRA}.

\subsubsection{Spectral method}

Alternatively, Saenz {\it et al.} have developed a spectral approach \cite{Awasthi2005JPB}. The field-free eigenstates, including continuum, are obtained from a configuration-interaction calculation \cite{Vanne2004JPB} where the Slater determinants are formed with one-electron ${\rm H}_2^+$ eigenstates expressed in terms of a $B$-spline basis in prolate spheroidal coordinates $(\xi,\eta,\varphi)$. They have developed their method first for ${\rm H}_2$ with parallel orientation of the molecular axis to the laser polarization \cite{Awasthi2005JPB}, but later extended it to an arbitrary molecular orientation. Using the method, Saenz {\it et al.} have studied, e.g., the validity and breakdown of the SAE approximation for ionization and excitation yields \cite{Awasthi2005JPB,Awasthi2010PRA}, $R$- and orientation dependence of strong-field ionization \cite{Vanne2008JMO,Vanne2009PRA,Vanne2010PRA,Forster2014PRA} (up to $\sim 10^{15}\,\mbox{W/cm}^2$ intensity at 800 nm wavelength). Bandrauk {\it et al.} \cite{Dehghanian2010PRA} have also developed a spectral method with eigenstates expressed in terms of Laguerre and Legendre polynomials, and studied enhanced ionization of ${\rm H}_2$ by intense ultrashort laser pulses. 

\subsubsection{Vibrational degree of freedom}
\label{subsubsec:H2-vibration}

Whereas the present review basically focuses on the electron dynamics within the fixed-nuclei approximation, it is worth noting that Bachau, Mart\'{\i}n {\it et al.} \cite{Sanz-Vicario2006PRA,Palacios2006PRL} have developed an elaborate time-dependent close-coupling method that treats not only the electronic but also the vibrational degree of freedom quantum mechanically. In this case, one solves the seven-dimensional TDSE,
\begin{equation}
\left(\hat{H}_0({\bf r}_1,{\bf r}_2,R)+V(t)-i\frac{\partial}{\partial t}\right)\Psi ({\bf r}_1,{\bf r}_2,R,t)=0,
\end{equation}
where $\hat{H}_0$ denotes the field-free Hamiltonian of $\mbox{H}_2$, and $V(t)$ the laser-$\mbox{H}_2$ interaction Hamiltonian (these authors usually use the velocity gauge). Assuming negligible nonadiabatic couplings, i.e., Born-Oppenheimer (BO) approximation, we expand $\Psi ({\bf r}_1,{\bf r}_2,R,t)$ with fully correlated adiabatic BO vibronic stationary states of the form $\psi ({\bf r}_1,{\bf r}_2,R)\chi (R)$ with $\psi$ and $\chi$ being the electronic and nuclear wave functions, respectively. These eigenstates include the bound, the resonant doubly excited, and the nonresonant singly ionized continuum of $\mbox{H}_2$. The technical details that make use of a $B$-spline basis in spherical coordinates are found in \cite{Martin1999JPB,Sanz-Vicario2006PRA,Bachau2001RPP}. By construction, this method does not account for double ionization. Bachau, Mart\'{\i}n {\it et al.} have developed their method initially for linear polarization along the $z$ direction, i.e., parallel to the molecular axis, but recently extended it to circular polarization whose electric field is in the $yz$ plane \cite{Perez-Torres2014PRA}. They have intensively studied various aspects of dissociative photoionization of $\mbox{H}_2$ and $\mbox{D}_2$ molecules by attosecond and femtosecond XUV pulses, e.g., autoionization \cite{Sanz-Vicario2006PRA}, control by pulse duration \cite{Palacios2006PRL}, electron localization \cite{Sansone2010Nature}, and circular dichroism in molecular-frame photoelectron angular distributions \cite{Perez-Torres2014PRA}. 


\section{Multiconfiguration Self-Consistent-Field (MCSCF) Approach}
\label{sec:MCSCF}

\begin{figure}[!t]
\centering
\includegraphics[width=8.3cm]{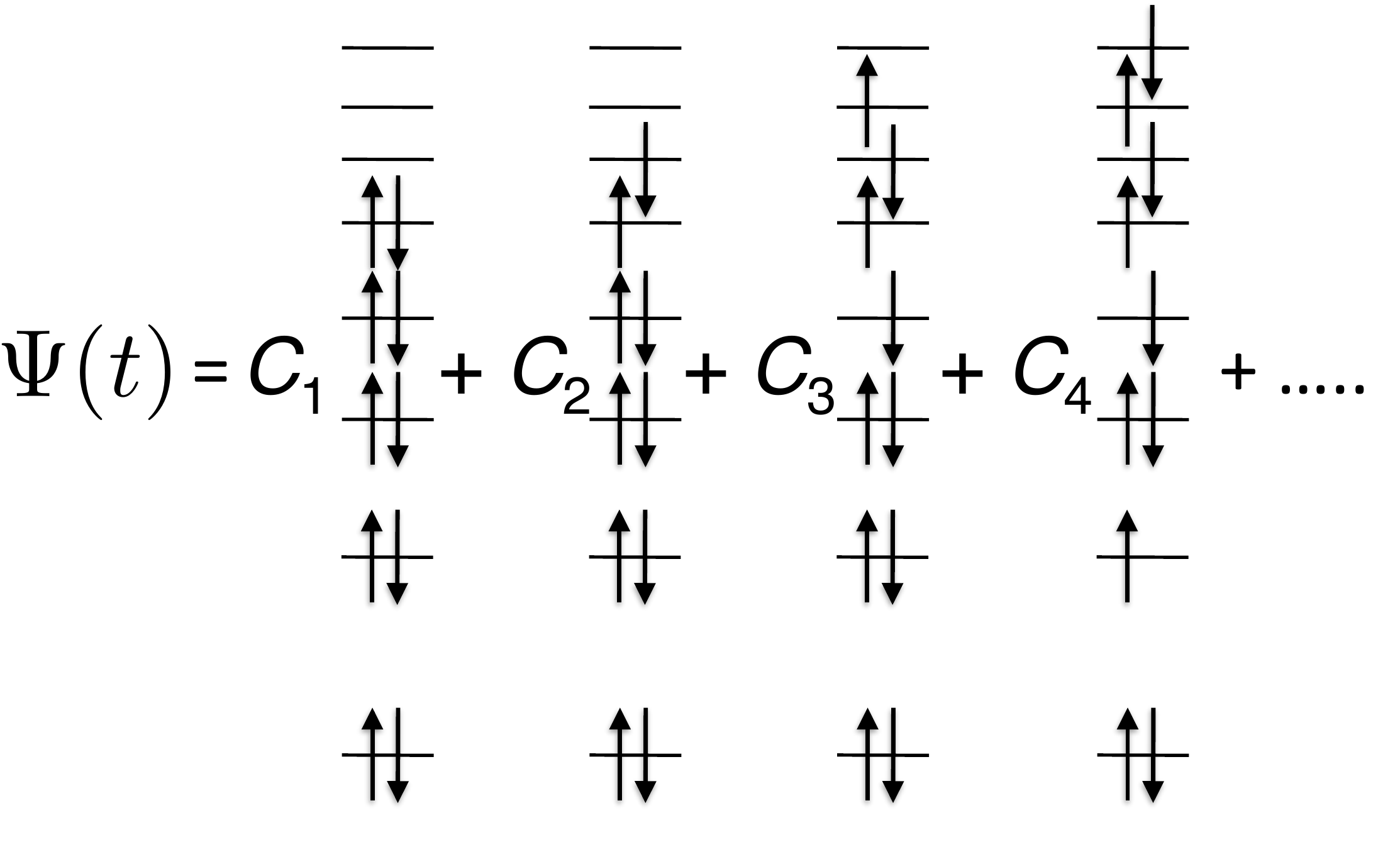}
\caption{Schematic representation of the multiconfiguration expansion Eq.~(\ref{eq:general-mcwf}). Each term on the right-hand side corresponds to a configuration $\Phi_1, \Phi_2, \cdots$. The CI coefficients $C_1, C_2, \cdots$ are usually assumed to vary in time, whereas spatial molecular orbitals, represented by horizontal bars, can be taken as either time-independent or -dependent.}
\label{fig:MC}
\end{figure}

Although the TDSE approach (Sec.~\ref{sec:TDSE}) provides an exact theoretical framework, this method is practically unfeasible for multi electron systems beyond He, ${\rm H}_2$, and Li, especially in an intense long-wavelength laser field. In order to handle multi electron dynamics, time-dependent multiconfiguration self-consistent-field (MCSCF) methods have actively been developed. The idea is to express the total wave function $\Psi (t)$ as a superposition of different Slater determinants or configuration state functions (CSF) \cite{Nguyen-Dang2009CP,Nguyen-Dang2013JCP,Miranda2011JCPa,Sato2013PRA,Sato2015PRA} (Fig.~\ref{fig:MC}):
\begin{equation}
\label{eq:general-mcwf}
\Psi (t) = \sum_{\bf I}^{\sf P} \Phi_{\bf I}(t)C_{\bf I}(t),
\end{equation}
where expansion coefficients $\{C_{\bf I}\}$ are called configuration interaction (CI) coefficients, and bases $\{\Phi\}$ are the Slater determinants \cite{Szabo1996,Helgaker}
\begin{align}
\Phi_{\bf I}(t) = \frac{1}{\sqrt{N!}}\left|\begin{array}{cccc}
\chi_{j_1}({\bf r}_1) & \chi_{j_2}({\bf r}_1) & \cdots & \chi_{j_N}({\bf r}_1) \\
\chi_{j_1}({\bf r}_2) & \chi_{j_2}({\bf r}_2) & \cdots & \chi_{j_N}({\bf r}_2) \\
\vdots & \vdots & \ddots & \vdots \\
\chi_{j_1}({\bf r}_N) & \chi_{j_2}({\bf r}_N) & \cdots & \chi_{j_N}({\bf r}_N)
\end{array}\right|
\nonumber\\
(j_1,\cdots,j_N\in {\bf I})
\end{align}
built from $N$ spin orbitals $\chi_j$ out of $2n$ spin orbitals $\{\phi_p;p=1,2,\cdots,n\}\otimes\{\alpha,\beta\}$  (in the spin-restricted treatment) with $\{\phi_p\}$ being spatial orbital functions and $\alpha (\beta)$ the up- (down-) spin eigenfunction. The summation in Eq.~(\ref{eq:general-mcwf}) with respect to configurations {\bf I} runs through the
element of a CI space ${\sf P}$, consisting of
a given set of determinants. The general multiconfiguration ansatz Eq.~(\ref{eq:general-mcwf}) can represent a very wide spectrum of methods; whereas $\{C_I\}$ are usually taken as time-dependent, they can also be fixed \cite{Miranda2011JCPa}. $\{\phi_p\}$, and thus $\{\Phi\}$, can be considered either time-dependent or independent. The sum is taken over either the full-CI space ${\sf P}_{\rm FCI}$ composed of all the possible configurations ${\bf I}$ to distribute $N$ electrons among the $2n$ spin orbitals or any arbitrary subspace ${\sf P}$ of ${\sf P}_{\rm FCI}$. Orthonormal spatial orbitals are often employed, but this choice is not mandatory.

The time-dependent variational principle (TDVP) or the Dirac-Frenkel variational principle \cite{Frenkel,Loewdin1972CPL,Moccia1973IJQC} requires the action integral,
\begin{equation}
\label{eq:action-integral}
S[\Psi] = \int_{t_0}^{t_1}\langle\Psi | \left(\hat{H}-i\frac{\partial}{\partial t}\right)|\Psi\rangle,
\end{equation}
to be stationary, i.e.,
\begin{equation}
\label{eq:TDVP}
\delta S = \delta\langle\Psi|\hat{H}|\Psi\rangle-i\left(\langle\delta\Psi | \frac{\partial\Psi}{\partial t}\rangle
-\langle\frac{\partial\Psi}{\partial t}|\delta\Psi \rangle\right) = 0,
\end{equation}
with respect to the variations $\delta\Psi$ of the total wave function permitted within the given multiconfiguration ansatz. By substituting Eq.~(\ref{eq:general-mcwf}) into Eq.~(\ref{eq:TDVP}) and after laborious algebra, one can derive the equations of motion (EOM) for the CI coefficients and orbital functions.

\begin{figure}[!t]
\centering
\includegraphics[width=9cm]{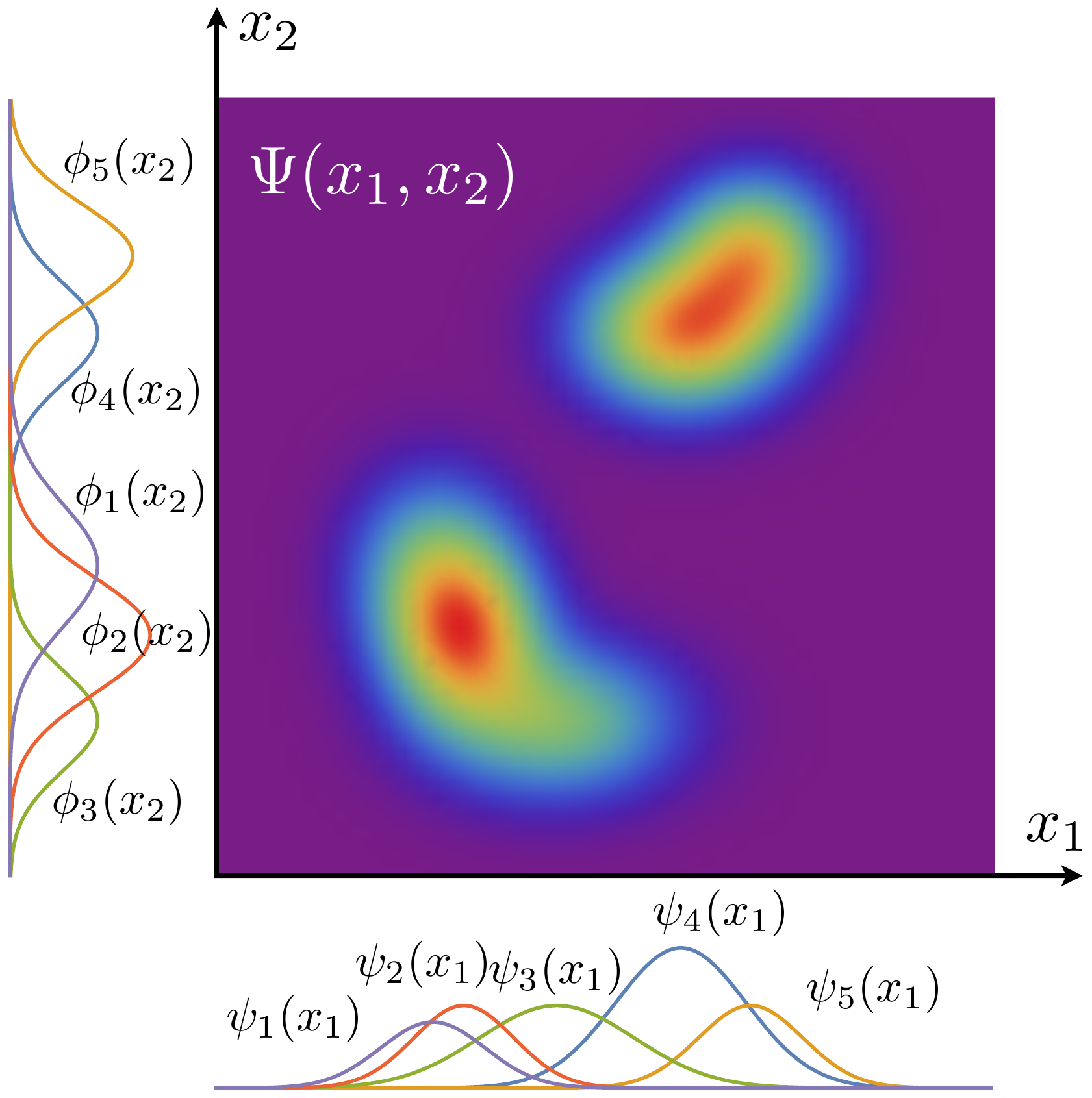}
\caption{Correlated two-particle model wave function $\Psi (x_1,x_2)$ expressed as multiple configurations $\sum_{i=1}^{5} \psi_i(x_1)\phi_i(x_2)$ with single-particle orbital functions $\psi_i(x)$ and $\phi_i(x)$. See text for details.}
\label{fig:MC-image}
\end{figure}

\begin{figure}[!t]
\centering
\includegraphics[width=8.7cm]{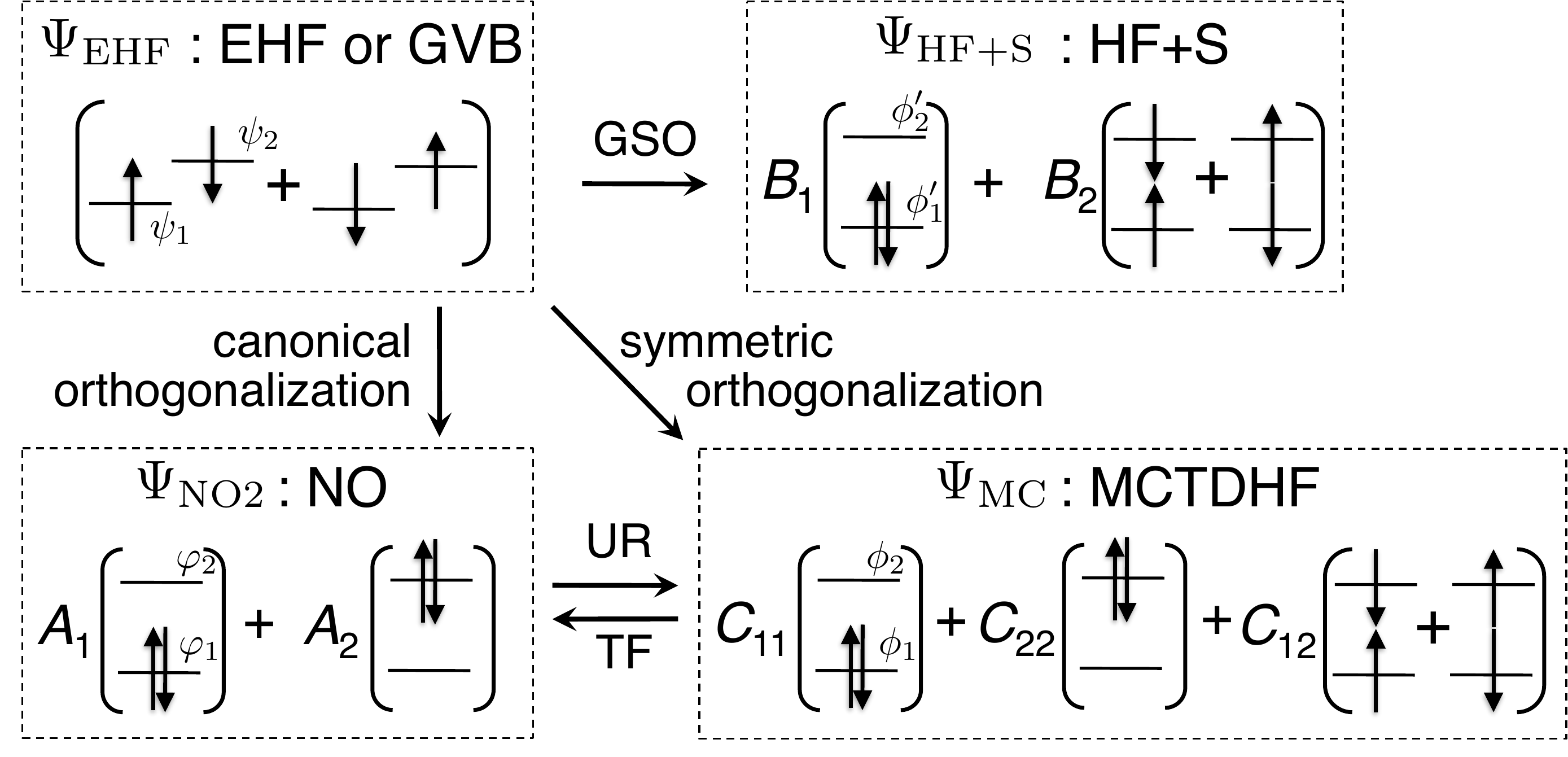}
\caption{Different representations of the identical two-electron wave function and orbital transformations from one to another. GSO: Gram-Schmidt orthogonalization, TF: Takagi's factorization, UR: unitary rotation.}
\label{fig:transformation}
\end{figure}

The computational gain thanks to the use of a limited number of orbital functions are concisely explained in Ref.~\cite{Caillat2005PRA}. Let us consider a one-dimensional two-particle system and try to approximate its model wave function $\Psi (x_1,x_2)$ as shown in Fig.~\ref{fig:MC-image} by,
\begin{equation}
\label{eq:expansion-model}
\Psi (x_1,x_2) \approx \sum_{i=1}^{i_{\rm max}} \psi_i(x_1)\phi_i(x_2),
\end{equation}
where we neglect the antisymmetrization for simplicity. Since $\Psi (x_1,x_2)$ is correlated, it cannot be well described by the product $\psi(x_1)\phi(x_2)$ ($i_{\rm max}=1$). With an appropriate number of $i_{\rm max}$, however, the expansion Eq.~(\ref{eq:expansion-model}) can efficiently cover the major part of the wave function. On the other hand, TDSE simulations as discussed in Sec.~\ref{sec:TDSE} explicitly treat the entire $(x_1,x_2)$ space, most of which is hardly occupied by the electrons. Thus, the TDSE approach is computationally much more demanding. Also, from Fig.~\ref{fig:MC-image}, one can qualitatively understand that TD-MCSCF methods with time-dependent orbital functions, which move with the electron cloud, can efficiently represent the total wave function by a smaller number of orbital functions than those with time-independent ones.

Before reviewing representative examples, let us emphasize, especially to readers with a physics background, that it is not very meaningful to discuss each spatial orbital $\phi_p$ as if it were a physical entity. One can see this as follows by considering a two-electron system, say, He (Fig.~\ref{fig:transformation}) \cite{Sato2014JPB}. To account for one electron ejected by a strong laser pulse and the other electron remaining in the ion, it would be reasonable to express the wave function as,
\begin{equation}
\label{eq:wf-EHF}
\Psi_{\rm EHF} = \frac{1}{2}\left[\psi_1({\bf r}_1,t)\psi_2({\bf r}_2,t)+\psi_2({\bf r}_1,t)\psi_1({\bf r}_2,t)\right]\Theta_S,
\end{equation}
with $\psi_1$ and $\psi_2$ being nonorthogonal, where $\Theta_S$ denotes a singlet spin function. This form is called the generalized valence bond (GVB) or extended Hartree-Fock (EHF) wave function and was used to explain the mechanism of NSDI \cite{Dahlen2001PRA,Nguyen2006PRA}. The canonical orthogonalization transforms Eq.~(\ref{eq:wf-EHF}) into an expression in terms of orthonormal orbitals $\{\varphi_1,\varphi_2\}$:
\begin{equation}
\label{eq:wf-NO-2}
\Psi_{\rm NO2} = \sum_{i=1}^2A_i(t)||\varphi_i({\bf r}_1,t)\bar{\varphi}_i({\bf r}_2,t)||,
\end{equation}
where $||\varphi_i\bar{\varphi}_j||$ denotes a normalized Slater determinant built from the direct product of a spatial orbital function $\varphi_i ({\bf r})$ and an up spin eigenfunction and the direct product of a spatial orbital $\varphi_j ({\bf r})$ and an down spin. Equation (\ref{eq:wf-NO-2}) corresponds to a two-orbital case of natural orbital (NO) expansion. Any arbitrary rotation,
\begin{equation}
\left(\begin{array}{c}\phi_1 \\ \phi_2\end{array}\right) = {\rm U}\left(\begin{array}{c}\varphi_1 \\ \varphi_2\end{array}\right),
\end{equation}
by a unitary matrix ${\rm U}$ yields the multiconfiguration time-dependent Hartree-Fock (MCTDHF) form in terms of orthonormal orbitals $\{\phi_1,\phi_2\}$:
\begin{equation}
\label{eq:wf-MC-2}
\Psi_{\rm MC} = \sum_{i,j=1}^2C_{ij}(t)||\phi_i({\bf r}_1,t)\bar{\phi}_j({\bf r}_2,t)||,
\end{equation}
where the matrix ${\rm C}$ with elements $C_{ij}$ is complex symmetric. Inversely, $\Psi_{\rm MC}$ can be transformed into $\Psi_{\rm NO2}$ through Takagi's factorization \cite{BunseGerstner1988JCAM} (this holds true for any number of orbitals). Moreover, by means of Gram-Schmidt orthonormalization, one can also rewrite Eq.~(\ref{eq:wf-EHF}) as,
\begin{align}
\label{eq:wf-HFS-2}
\Psi_{\rm HF+S} &= B_1(t)||\phi^\prime_1({\bf r}_1,t)\bar{\phi}^\prime_1({\bf r}_2,t)|| \nonumber\\
&+B_2(t)[||\phi^\prime_1({\bf r}_1,t)\bar{\phi}^\prime_2({\bf r}_2,t)||+||\phi^\prime_2({\bf r}_1,t)\bar{\phi}^\prime_1({\bf r}_2,t)||].
\end{align}
Hence, the equivalent two-electron wave function can be expanded in different ways Eqs.\ (\ref{eq:wf-EHF}), (\ref{eq:wf-NO-2}), (\ref{eq:wf-MC-2}), and (\ref{eq:wf-HFS-2}) (Fig.~\ref{fig:transformation}). This observation emphasizes that orbital functions are a mathematical tool, i.e., a kind of single-particle basis functions, to construct the multielectron wave function, rather than a physical entity.

An obvious extension of Eq.~(\ref{eq:wf-NO-2}) is an expansion of the two-electron wave function with more than two NOs,
\begin{equation}
\label{eq:wf-NO-n}
\Psi_{\rm NO} = \sum_{i=1}^{n}A_i(t)||\varphi_i({\bf r}_1,t)\bar{\varphi}_i({\bf r}_2,t)||.
\end{equation}
The {\it time-dependent natural orbital (TD-NO)} method \cite{Sato2014JPB} directly propagates $A_i(t)$ and $\varphi_i({\bf r},t)$ to simulate laser-driven two-electron systems. We need many NOs to quantitatively describe correlation-induced phenomena such as NSDI. The {\it time-dependent renormalized natural orbital theory (TDRNOT)} \cite{Brics2013PRA,Rapp2014PRAa,Brics2014PRAb} propagates, instead of NOs, renormalized natural orbitals (RNOs) $\tilde{\varphi_i}$ defined as,
\begin{equation}
\tilde{\varphi_i}({\bf r},t) = |A_i(t)|\varphi_i({\bf r},t).
\end{equation}
The equations of motion for RNOs can be found in \cite{Rapp2014PRA}.

\subsection{Time-Dependent Configuration Interaction (TDCI) Method}
\label{subsec:TDCI}

In the TDCI approach, the spatial orbital functions $\phi_p$ are time-independent, and the total wave function is expressed by the truncated CI expansion:
\begin{equation}
\label{eq:CI-expansion}
\Psi (t) = \Phi_0C_0(t) + \sum_{ia}\Phi_i^a C_i^a(t)+  \sum_{ijab}\Phi_{ij}^{ab}C_{ij}^{ab}(t) + \cdots,
\end{equation}
comprised of the closed-shell HF determinant $\Phi_0$ built from the $N/2$ lowest orbitals, singly excited determinants $\Phi_i^a$ with $\phi_i$ in $\Phi_0$ replaced by $\phi_a$, 
\begin{equation}
|\Phi_i^a\rangle = \frac{1}{\sqrt{2}} \sum_{\sigma \in \{\uparrow, \downarrow\}} \hat{a}_{a\sigma}^\dag\hat{a}_{i\sigma}|\Phi_0\rangle,
\end{equation}
similarly defined doubly excited determinants $\Phi_{ij}^{ab}$, etc., with $\uparrow (\downarrow)$ representing the up (down) spin. The expansion Eq.~(\ref{eq:CI-expansion}) is truncated at a given order. Approaches that include up to single excitation, double excitation, $\cdots$, are called TDCI singles (TDCIS), TDCI singles and doubles (TDCISD), $\cdots$, respectively.

The orbital functions are obtained from the ground state Hartree-Fock method. Whereas the configurations $\Phi_0, \Phi_i^a, \Phi_{ij}^{ab}, \cdots$ are fixed in time, the time-dependent CI coefficients $C_0, C_i^a, C_{ij}^{ab}, \cdots$ account for the laser-driven dynamics. The equations of motion for the latter can be derived through the substitution of Eq.~(\ref{eq:CI-expansion}) into the time-dependent variational principle. In the case of TDCI, which uses time-independent orbitals, the same EOMs can be obtained by inserting Eq.~(\ref{eq:CI-expansion}) directly into TDSE as well. The resulting EOMs are given for TDCIS and the length gauge by,
\begin{align}
i\frac{\partial}{\partial t}C_0(t) &= {\bf E}(t)\cdot\sum_{a,i}\langle\Phi_0|\hat{\bf r}|\Phi_i^a\rangle C_i^a(t),\\
i\frac{\partial}{\partial t}C_i^a(t) &=(\varepsilon_a - \varepsilon_i) C_i^a(t)
+ \sum_{b,j}\langle\Phi_i^a|\hat{H}^\prime|\Phi_j^b\rangle C_j^b(t) \nonumber\\
&+ {\bf E}(t)\cdot \left( \langle\Phi_i^a|\hat{\bf r}|\Phi_0\rangle C_0(t)  + \sum_{b,j}\langle\Phi_i^a|\hat{\bf r}|\Phi_j^b\rangle C_j^b(t) \right), \label{eq:TDCIS-EOM-2}
\end{align}
with $\hat{\bf r}=\sum_{i=1}^N {\bf r}_i$ being the dipole operator. $\hat{H}_{HF}$ denotes the mean-field Hamiltonian of a one-electron nature that defines the HF ground state $\Phi_0$ and orbital functions $\phi_p$ with their orbital energies $\varepsilon_p$ and that includes the electron-electron mean-field potential $\hat{V}_{MF}$ composed of the Coulomb and exchange operators. $\hat{H}^\prime = \hat{H}_2-\hat{V}_{MF}$ accounts for the electron-electron correlation beyond the mean-field contribution. The second term on the right-hand side of Eq.~(\ref{eq:TDCIS-EOM-2}) describes the coupling between configurations singly excited from different orbitals, thus can be viewed as interchannel interactions. It should be, however, noticed that this term is present because each $\Phi_i^a$ is not an eigenstate of the total field-free Hamiltonian $\hat{H}_{HF}+\hat{H}^\prime$ even within the CIS ansatz. $\hat{H}^\prime$ does not couple the ground state and the singly excited states since $\langle\Phi_0|\hat{H}^\prime|\Phi_i^a\rangle=0$ (Brillouin's theorem \cite{McWeeny}).
The detailed description of their implementation can be found in \cite{Greenman2010PRA}.  

In TDCIS, only one electron can be fully active at once; it is implicitly assumed that high-field phenomena are predominantly single-electron processes. In this sense, this approach can be considered as an extended ab-initio formulation of SAE. However, the electron can originate from any occupied orbital. Thus, multiple-orbital (multichannel) effects are taken into account. 

In the context of high-field phenomena and attosecond physics, the TDCIS method was introduced \cite{Rohringer2006PRA,Pabst2013EPJST} and implemented \cite{Greenman2010PRA} by Santra {\it et al.}. Reference \cite{Rohringer2006PRA} also discusses an interesting reformulation with a conceptual advantage in terms of a time-dependent orbital,
\begin{equation}
\chi_i(t) = \frac{1}{\sqrt{2}}\sum_a C_i^a(t)\phi_a
\end{equation}
that assembles all the single excitations from the occupied orbital $\phi_i$. These authors later included spin-orbit splitting for the occupied orbitals \cite{Rohringer2009PRA,Pabst2012PRAa}. The TDCIS has been applied to both perturbative and nonperturbative multiphoton processes such as decoherence in attosecond photoionization \cite{Pabst2011PRL}, attosecond transient-absorption spectroscopy \cite{Pabst2012PRAa}, the Cooper minimum \cite{Pabst2012PRAb} and multielectron effects in the giant enhancement \cite{Pabst2013PRL} in HHG spectra, two-photon ionization of ${\rm Ne}^{8+}$ by intense ultrafast x rays \cite{Sytcheva2012PRA}, and adiabaticity and diabaticity in strong-field ionization \cite{Karamatskou2013PRA}.

One of the drawbacks of TDCIS is the lack of the size extensivity \cite{Helgaker}; the separate treatment of two subsystems with TDCIS, which involves up to double excitation in whole, is not consistent with that of the whole system with the same method. Also, as is usual for approaches involving truncated expansion Eq.\ (\ref{eq:CI-expansion}) with temporally fixed orbital functions, the TDCIS equations are not gauge invariant (see Subsection \ref{subsec:gauge-dependence}). Moreover, its applicability is limited to systems whose initial state (ground state) is correctly described by the HF method.

\subsection{Multiconfiguration Time-Dependent Hartree-Fock (MCTDHF) Method}
\label{subsec:MCTDHF}

Let us take both the CI coefficients and orthonormal orbital functions as time-dependent variational degrees of freedom and take the sum in Eq.~(\ref{eq:general-mcwf}) over the full-CI space ${\sf P}_{\rm FCI}$:
\begin{equation}
\label{eq:MCTDHF-wf}
\Psi (t) = \sum_{\bf I}^{{\sf P}_{\rm FCI}} \Phi_{\bf I}(t)C_{\bf I}(t).
\end{equation}
The sum is taken over all the possible ways to distribute $N$ electrons among the $2n$ spin orbitals. One can use a more intuitive, symbolical notation,
\begin{equation}
\Psi_{\rm MCTDHF} : \{\phi_1(t)\phi_2(t)\cdots\phi_n(t)\}^N,
\end{equation}
representing the $N$-electron full-CI wave function with $n$ time-dependent orbitals.
In the community of high-field phenomena and attosecond physics, the term MCTDHF usually refers to this most comprehensive variant, whereas this term has also been used for non full-CI expansion in quantum chemistry. In the limit of a large number of $n$, the MCTDHF method can deliver the exact solutions of TDSE in principle. The EOMs for the CI coefficients $C_{\bf I}(t)$ and orbital functions $\{\phi_i (t)\}$ are derived in Refs.~\cite{Kato2004CPL,Caillat2005PRA} and discussed in Sec.~\ref{subsec:CASSCF}.

\begin{figure}[!t]
\centering
\includegraphics[width=9cm]{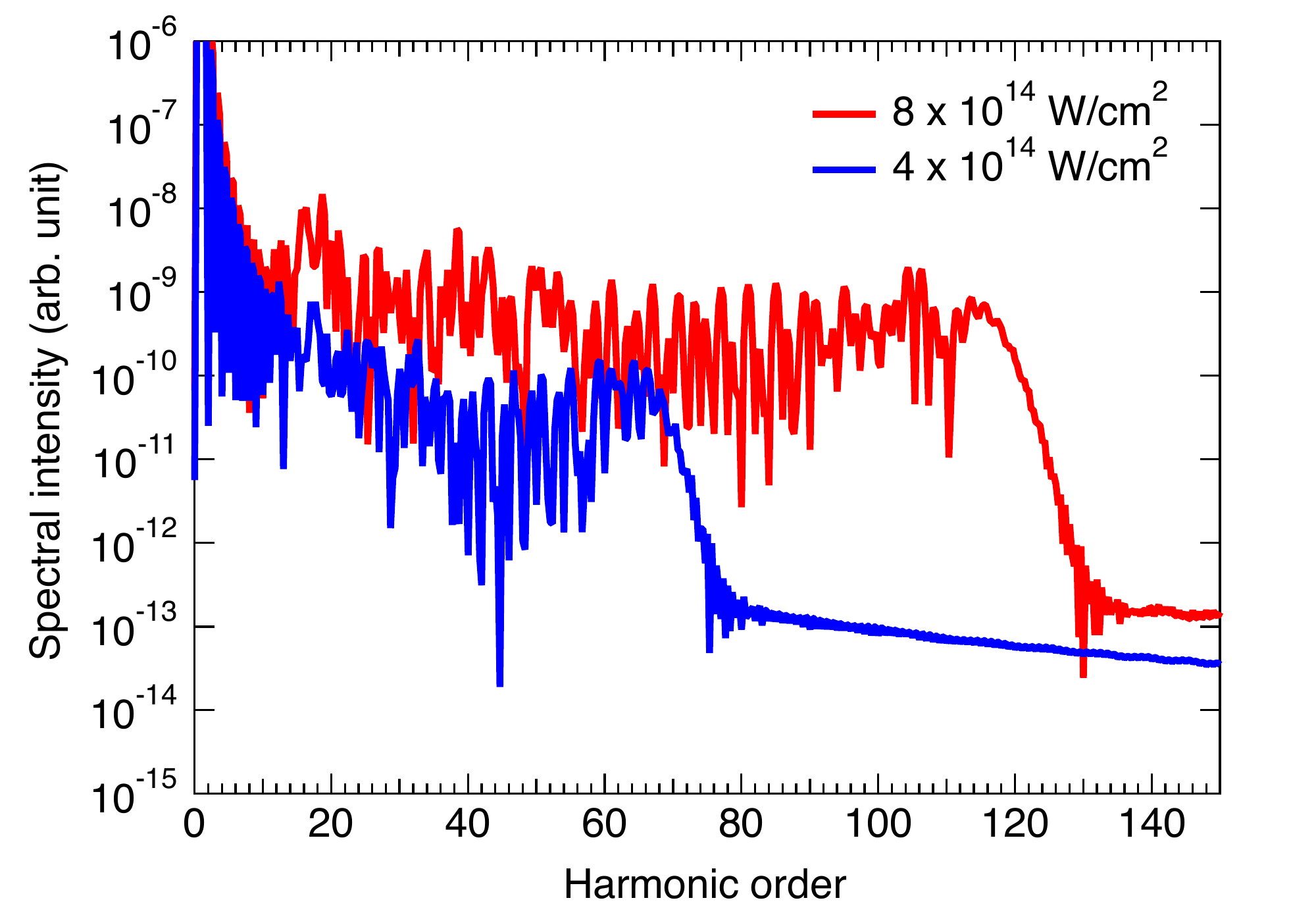}
\caption{High-harmonic spectra from He calculated with the MCTDHF method for the case of 800 nm wavelength and 4 and $8\times 10^{14}\,{\rm W/cm}^2$ peak intensity.}
\label{fig:HeHHG}
\end{figure}

As an example, in Fig.~\ref{fig:HeHHG} we show the high-harmonic spectra from He driven by a near-infrared (NIR) laser pulse with a wavelength of 800 nm and a peak intensity of 4 and $8\times 10^{14}\,{\rm W/cm}^2$, calculated with the MCTDHF method. The pulse is composed of single-cycle turn-on, constant intensity for a single cycle, and single-cycle turn-off. The simulation (in the length gauge) is sufficiently converged with two orbitals and the maximum angular momentum of 79 for each orbital. The cutoff energy $E_c$ calculated from the cutoff law,
\begin{equation}
\label{eq:cutoff}
E_c = I_p + 3.17U_p,
\end{equation}
with $I_p$ and $U_p$ being the ionization potential and ponderomotive energy, respectively, corresponds to the 65th and 114th order for each of the two intensities, consistent with the spectra in Fig.~\ref{fig:HeHHG}.
It would be extremely difficult to obtain such spectra for NIR fundamental wavelengths by means of direct solution of the TDSE.

\subsection{Time-Dependent Complete Active-Space Self-Consistent Field (TD-CASSCF) Method}
\label{subsec:CASSCF}

\begin{figure}[!t]
\centering
\includegraphics[width=9cm]{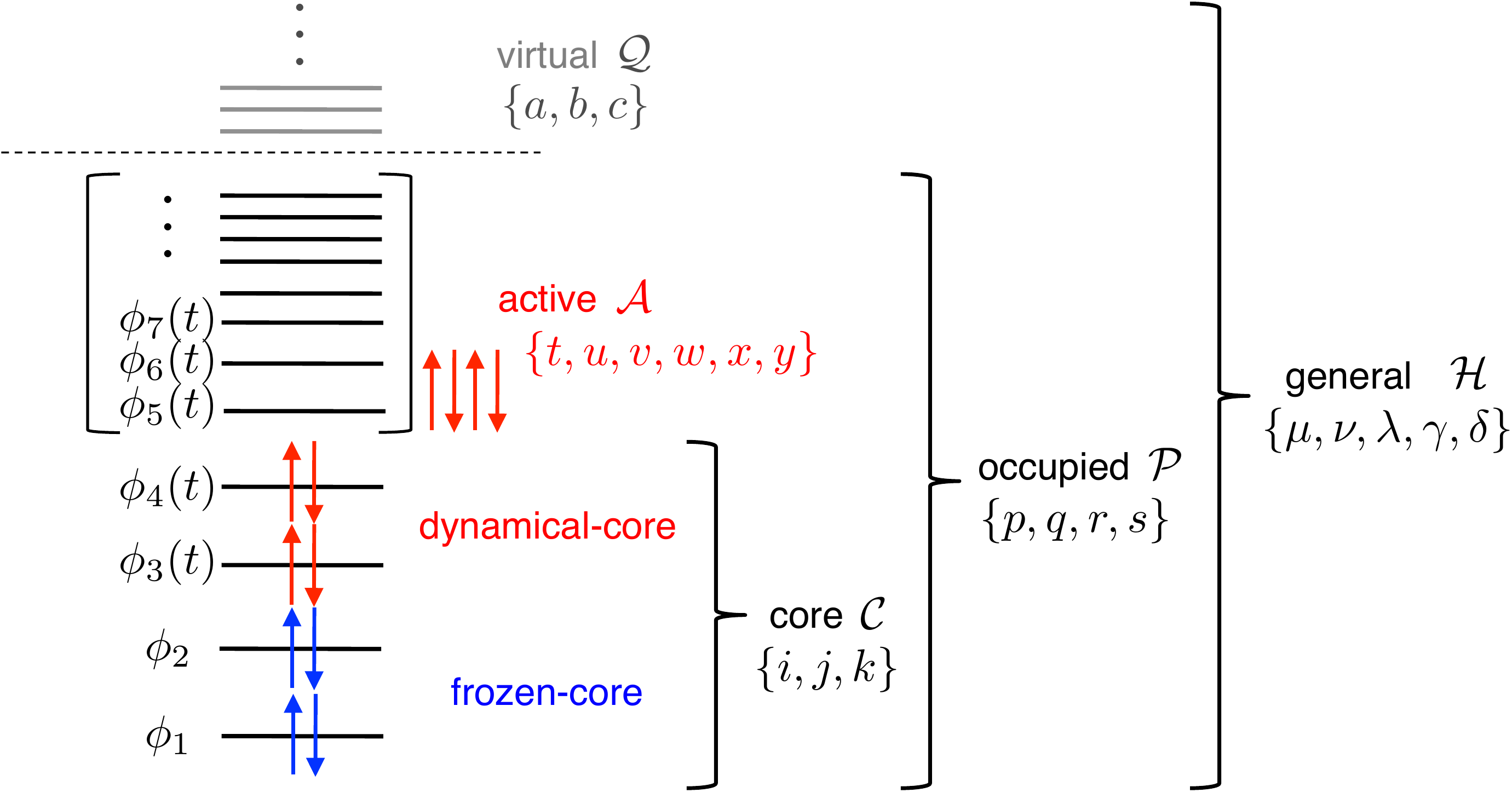}
\caption{Schematic explanation of the TD-CASSCF concept, illustrating a twelve-electron system with two frozen-core, two dynamical-core, and eight active orbitals. The classification of orbitals and the indices we use are also shown.}
\label{fig:CASSCF-concept}
\end{figure}

The MCTDHF method, though powerful, is difficult to apply to large systems, since the number of the configurations involved (CI dimension), and thus its computational time, scales factorially with the number of electrons $N$. Sato and Ishikawa \cite{Sato2013PRA} have proposed and formulated a more flexible method based on the CASSCF concept. This method introduces {\it core} ($\mathcal{C}$) and {\it active} ($\mathcal{A}$) orbital subspaces (Fig.\ \ref{fig:CASSCF-concept}). In an intense long-wavelength laser field, one reasonably expects that only high-lying electrons are strongly driven whereas deeply bound core electrons remain nonionized. Accordingly, we assume that the $n_C$ core orbitals are doubly occupied all the time. On the other hand, we consider all the possible distributions of $N_A(=N-2n_C)$ electrons among $n_A$ active orbitals. The CASSCF wave function can be symbolically expressed as,
\begin{align}
\Psi_{\rm CASSCF}& : \phi_1^2(t)\phi_1^2(t)\cdots\phi_{n_C}^2(t) \nonumber \\
&\times \{\phi_{n_C+1}(t)\phi_{n_C+2}(t)\cdots\phi_{n_C+n_A(t)}\}^{N_A}.
\end{align}
It should be noticed that not only the active orbitals but also the core orbitals, though constrained to the closed-shell structure, vary in time, in general, responding to the field formed by the laser and the other electrons. Alternatively, it is also possible to divide the core space further into fixed {\it frozen core} and time-dependent {\it dynamical core} subspaces. The TD-CASSCF approach includes time-dependent Hartree-Fock (TDHF) \cite{Pindzola1991PRL} and MCTDHF as special cases (see below). 

Hereafter, we use orbital indices $\{i,j,k\}$ for core ($\mathcal{C}$),
$\{t,u,v,w,x,y\}$ for active ($\mathcal{A}$), $\{p,q,r,s\}$ for occupied
($\mathcal{P}=\mathcal{C}+\mathcal{A}$), $\{a,b,c\}$ for virtual ($\mathcal{Q}$), and
$\{\mu,\nu,\lambda,\gamma,\delta\}$ for general ($\mathcal{H}=\mathcal{P}+\mathcal{Q}$)
orbitals (Fig.\ \ref{fig:CASSCF-concept}). For the discussion of the EOMs, it is convenient to use the second quantization formalism, in which the Hamiltonian is expressed as,
\begin{equation}
\label{eq:Hamiltonian-sq}
\hat{H} = \sum_{\mu\nu} h^\mu_\nu \hat{E}^\mu_\nu
+ \frac{1}{2} \sum_{\mu\nu\lambda\gamma} g^{\mu\lambda}_{\nu\gamma} \hat{E}^{\mu\lambda}_{\nu\gamma},
\end{equation}
with
$
\hat{E}^\mu_\nu = \sum_\sigma
\hat{a}^\dagger_{\mu\sigma}\hat{a}_{\nu\sigma}
$, 
$
\hat{E}^{\mu\lambda}_{\nu\gamma} = \sum_{\sigma\tau}
\hat{a}^\dagger_{\mu\sigma}\hat{a}^\dagger_{\lambda\tau}
\hat{a}_{\gamma\tau}\hat{a}_{\nu\sigma}
$, and
\begin{eqnarray}
\label{eq:ham1e}
h^\mu_\nu = \int d{\bf r} \phi^*_\mu({\bf r}) 
h\left({\bf r}, {\bf p}\right)
\phi_\nu({\bf r}),
\end{eqnarray}
\begin{eqnarray}
\label{eq:ham2e}
g^{\mu\lambda}_{\nu\gamma} = \int\!\!\!\int \! d{\bf r}_1 d{\bf r}_2
\frac{
\phi^*_\mu({\bf r}_1) \phi_\nu({\bf r}_1)
\phi^*_\lambda({\bf r}_2) \phi_\gamma({\bf r}_2)
}{|{\bf r}_1 - {\bf r}_2|}.
\end{eqnarray}
The wave function $\Phi_{\bf I}(t)$ for configuration ${\bf I}$ is given by,
\begin{equation}
\label{eq:configuration}
\vert \Phi_{\bf I} (t) \rangle = \prod_{i\in\mathcal{C}} \hat{a}^\dagger_{i\uparrow}
\hat{a}^\dagger_{i\downarrow} 
\prod_\sigma\prod_{t \in \mathcal{A}}
(\hat{a}^\dagger_{t\sigma})^{I_{t\sigma}}\vert vac \rangle,
\end{equation}
where $\vert vac \rangle$ denotes the vacuum state, and $I_{t\sigma} = \{0,1\}$ is an index that specifies the configuration, satisfying $\sum_\sigma\sum_{t \in \mathcal{A}}
I_{t\sigma} = N_\textrm{A}$. It should be noticed that the annihilation and creation operators, $\hat{a}_{\mu\sigma}$ and $\hat{a}^\dagger_{\mu\sigma}$, respectively, are time-dependent in general.

Assuming the orthonormality of orbitals, let us also introduce an anti-Hermitian transformation matrix $X$ whose element is defined as \cite{Miranda2011JCPa,Sato2013PRA},
\begin{equation}
\label{eq:dermo}
X^\mu_\nu = \langle\phi_\mu|\dot{\phi_\nu}\rangle.
\end{equation}
Then, the temporal variation $|\dot{\Psi}\rangle$ of the total wave function is written as the sum of the contributions from the evolution of the CI coefficients and that of the orbitals:
\begin{equation}
\label{eq:der}
|\dot\Psi\rangle = 
\sum_{\bf I} |\Phi_{\bf I}\rangle\dot{C}_{\bf I} + \hat{X}|\Psi\rangle,
\end{equation}
where,
\begin{equation}
\hat{X}=\sum_{\mu\nu}X^\mu_\nu\hat{E}^\mu_\nu.
\end{equation}

The TDVP leads to the equations of motion \cite{Sato2013PRA,Sato2015PRA},
\begin{align}
\label{eq:g-tdci}
\dot{C}_{\bf I} &= - i\langle\Phi_{\bf I}|
\hat{H} |\Psi\rangle
-
\langle\Phi_{\bf I}|
\hat{X} |\Psi\rangle,\\
\label{eq:g-tdmo}
\langle\Psi|
\hat{E}^\mu_\nu\hat{{\sf Q}}\hat{X} -
\hat{X}\hat{{\sf Q}}\hat{E}^\mu_\nu
|\Psi\rangle &= -i
\langle\Psi|
\hat{E}^\mu_\nu\hat{{\sf Q}}\hat{H} -
\hat{H}\hat{{\sf Q}}\hat{E}^\mu_\nu
|\Psi\rangle,
\end{align}
where $\hat{{\sf Q}}=\hat{1}-\hat{{\sf P}}$ with $\hat{{\sf P}} = \sum_{\bf I}^{\sf P} \vert {\bf I}\rangle\langle{\bf I}\vert$ being the projector onto the CI space ${\sf P}$. It should be noted that {\it Eqs.~(\ref{eq:g-tdci}) and (\ref{eq:g-tdmo}) are valid for not only CASSCF but also general MCSCF wave functions with arbitrary CI spaces {\sf P}}, including MCTDHF (Sec.~\ref{subsec:MCTDHF}) and time-dependent occupation-restricted multiple-active-space (Sec.~\ref{subsec:TD-ORMAS}) methods.

Numerical implementation requires the transformation of Eq.~(\ref{eq:g-tdmo}) into equations of motion in terms of each orbital function. For that purpose, we have analyzed $\hat{E}^\mu_\nu$ and $X^\mu_\nu$. The orbital rotation operators $\hat{E}^\mu_\nu$ are categorized, according to the classification of the relevant orbitals, as,
\begin{equation}
\left\{\hat{E}^\mu_\nu\right\} = 
\left\{\hat{E}^i_j, \hat{E}^i_t, \hat{E}^t_i, \hat{E}^t_u, \hat{E}^p_a, \hat{E}^a_p, \hat{E}^a_b\right\}.
\end{equation}
By noting the projector $\hat{{\sf Q}}$ in Eq.~(\ref{eq:g-tdmo}), we can classify these into the following two, for the case of TD-CASSCF, MCTDHF, and TDHF:
 
\begin{enumerate}
\item {\it Redundant} ($\hat{E}^i_j, \hat{E}^t_u, \hat{E}^a_b$).
Both 
$\hat{E}^\mu_\nu \vert \Psi \rangle$ and
$\hat{E}^\nu_\mu \vert \Psi \rangle$ lie inside ${\sf P}$ or vanish.
In this case, Eq.~(\ref{eq:g-tdmo}) reduces to a trivial identity $0=0$, and the corresponding $X^\mu_\nu$ may be matrix elements of an arbitrary one-electron anti-Hermitian operator $\hat{\theta}(t)$ \cite{Caillat2005PRA}, 
\begin{eqnarray}
\label{eq:x_red}
X^\mu_\nu = \langle \phi_\mu \vert \hat{\theta}(t) \vert \phi_\nu
\rangle, \hspace{.5em} \hat{\theta}^\dagger(t) = -\hat{\theta}(t).
\end{eqnarray}
This redundancy originates from the fact that the unitary rotation inside the core ($\mathcal{C}$), active ($\mathcal{A}$), or virtual ($\mathcal{Q}$) orbitals does not alter the space covered by the given MC ansatz Eq.~(\ref{eq:general-mcwf}). This observation reemphasizes that molecular orbitals are mathematical instruments rather than physical entities. The simplest choice is $\hat{\theta}(t)=0$, thus $X^\mu_\nu = 0$.

\item {\it Non-redundant uncoupled} ($\hat{E}_i^t, \hat{E}_t^i, \hat{E}_p^a, \hat{E}_a^p$). 
At least one of $\hat{E}^\mu_\nu \vert \Psi \rangle$ and
$\hat{E}^\nu_\mu \vert \Psi \rangle$ do not vanish, and $\hat{E}^\mu_\nu \vert \Psi \rangle$ and
$\hat{E}^\nu_\mu \vert \Psi \rangle$ lie, if non-vanishing, outside ${\sf P}$.
This type of rotations do not contribute to the CI equations,
Eq.~(\ref{eq:g-tdci}), through the second term. Nevertheless, they contribute to the orbital EOMs 
Eq.~(\ref{eq:g-tdmo}), which reduces to a simpler expression
\cite{Sato2013PRA}, 
\begin{equation}
\label{eq:rmat_inter}
\langle \Psi \vert \left[\hat{E}^\mu_\nu, \hat{E}^\gamma_\lambda\right] \vert
\Psi \rangle X^\gamma_\lambda = -{\rm i}
\langle \Psi \vert \left[\hat{E}^\mu_\nu, \hat{H}\right] \vert
\Psi \rangle.
\end{equation}
Explicit formulas for $X_i^t, X_t^i, X_p^a$ and $X_a^p$ are given in Eqs.~(33)-(36) of Ref.~\cite{Sato2013PRA}, with $R_\nu^\mu=iX_\nu^\mu$.

\end{enumerate}

The orbital EOMs explicitly in terms of orbital functions, required for numerical implementation, are given by, 
%
\begin{eqnarray}
\label{eq:tdmoc}
\vert\dot{\phi}_i\rangle &=&
- {\rm i}\hat{Q} \hat{F} \vert\phi_i\rangle
+ \sum_p \vert \phi_p \rangle X^p_i, \quad\mbox{(core)}\\
\label{eq:tdmoa}
\vert\dot{\phi}_t\rangle &=&
- {\rm i}\hat{Q} \hat{F}_t \vert\phi_t\rangle
+ \sum_p \vert \phi_p \rangle X^p_t, \quad\mbox{(active)}
\end{eqnarray}
%
for core and active orbitals, respectively, where,
\begin{eqnarray}\label{eq:projq} 
\hat{Q} = \hat{1} - \sum_p |\phi_p\rangle\langle\phi_p|,
\end{eqnarray}
is the orbital projector onto the $\mathcal{Q}$ space, and,
%
\begin{eqnarray}
\label{eq:fockc}
\hat{F} |\phi_i\rangle &=& 
\hat{f} |\phi_i\rangle + \sum_{tu} D^t_u \hat{G}^u_t |\phi_i\rangle, \\
\label{eq:focka}
\hat{F}_t \vert\phi_t\rangle &=&
\hat{f}|\phi_t\rangle + \sum_{uvwx}
\hat{W}^v_w \vert\phi_u\rangle P^{uw}_{xv} \left(D^{-1}\right)^x_t,\\
\label{eq:cfock}
\hat{f} |\phi_p\rangle &=& 
\hat{h} |\phi_p\rangle +
2 \sum_j \hat{G}^j_j |\phi_p\rangle,\\
\hat{G}^t_u|\phi_p\rangle &=& \hat{W}^t_u|\phi_p\rangle -
1/2\hat{W}^t_p|\phi_u\rangle,
\end{eqnarray}
with, 
\begin{equation}
\label{eq:meanfield}
 W^p_q({\bf r}) =
 \int d{\bf r}^\prime \frac{\phi^*_p({\bf r}^\prime) \phi_q({\bf r}^\prime)}
 {|{\bf r} - {\bf r}^\prime|}.
\end{equation}
being the general mean-field operator, and $D^t_u \equiv \langle\Psi_\textrm{A}|\hat{E}^u_t |\Psi_\textrm{A}\rangle$ and 
$P^{tv}_{uw} \equiv \langle\Psi_\textrm{A}|\hat{E}^{uw}_{tv}|\Psi_\textrm{A}\rangle$ being
one- and two-electron reduced density matrix (RDM) elements,
respectively, defined within the active space.
The second terms on the right-hand sides of Eqs.~(\ref{eq:tdmoc}) and (\ref{eq:tdmoa}) represents, aside from the redundancy within the same orbital subspace, the coupling between the core and active orbitals.
We need {\it not} explicitly work with virtual orbitals thanks to $\hat{Q}$ \cite{Caillat2005PRA}.

The wave function Eq.\ (\ref{eq:general-mcwf}) with configurations Eq.\ (\ref{eq:configuration}) can be rewritten as,
\begin{equation}
\label{eq:mcscf_2q}
|\Psi\rangle = \hat{\Phi}_\textrm{C} |\Psi_\textrm{A}\rangle,
\end{equation}
with the active part,
\begin{equation}
|\Psi_\textrm{A}\rangle = \sum_{\bf I}^{\sf P} |{\bf I}\rangle C_{\bf I},
\end{equation}
where,
\begin{equation}
\label{eq:det_act}
\vert {\bf I} \rangle = \prod_\sigma\prod_{t \in \mathcal{A}}
(\hat{a}^\dagger_{t\sigma})^{I_{t\sigma}}\vert vac \rangle,
\end{equation}
and the core generator,
\begin{equation}
\hat{\Phi}_\textrm{C} \equiv \prod_{i\in\mathcal{C}}
\hat{a}^\dagger_{i\uparrow}
\hat{a}^\dagger_{i\downarrow}.
\end{equation}
As is already mentioned above, MCTDHF and TDHF are special cases of the TD-CASSCF method. The former corresponds to $n_C=0$, for which $\hat{\Phi}_\textrm{C}$ becomes the identity operator, and the latter to $n_A=0$ and $\vert \Psi_\textrm{A}\rangle=\vert vac \rangle$. Therefore, the equations of motion for the CI coefficients and orbitals presented in this subsection are valid also for MCTDHF and TDHF.

The separation of the core wave function in Eq.~(\ref{eq:mcscf_2q}) allows us to reduce the equations of motion for the CI coefficients, Eq.~(\ref{eq:g-tdci}) to those in terms of the active-part wave function $\Psi_A$ \cite{Sato2013PRA,Sato2015PRA}, i.e., effectively an $N_\textrm{A}$-electron problem
\begin{equation}
\label{eq:tdci}
\dot{C}_{\bf I} = -i
\langle{\bf I} \vert \hat{H}_\textrm{A} - E_\textrm{A}\hat{1} \vert \Psi_\textrm{A} \rangle -
\langle{\bf I} \vert \hat{X} \vert \Psi_\textrm{A} \rangle,
\end{equation}
\begin{equation}
\label{eq:hama}
\hat{H}_\textrm{A} =
\sum_{tu} f^t_u \hat{E}^t_u + \frac{1}{2} \sum_{tuvw} g^{tv}_{uw}
\hat{E}^{tv}_{uw},
\end{equation}
where $\hat{1}$ is the identity operator, $E_\textrm{A} \equiv
\langle\Psi_\textrm{A}|\hat{H}_\textrm{A}|\Psi_\textrm{A}\rangle$,
and 
\begin{eqnarray}
\label{eq:fmat}
f^t_u &=& \langle\phi_t| \cdot \hat{f}|\phi_u\rangle, \\
\label{eq:gmat}
g^{tv}_{uw} &=& \langle\phi_t| \cdot \hat{W}^v_w|\phi_u\rangle.
\end{eqnarray}
Here, without loss of generality, we have adopted a particular phase choice $\langle \Psi \vert \dot{\Psi} \rangle = 0$. For another, more common choice ${\rm i}\langle\Psi|\dot{\Psi}\rangle = \langle\Psi|\hat{H}|\Psi\rangle$, Eq.~(\ref{eq:tdci}) is replaced by,
\begin{equation}\label{eq:tdci-another}
\dot{C}_{\bf I} = -i
\langle{\bf I} \vert \hat{H}_\textrm{A} + E_\textrm{C}\hat{1} \vert \Psi_\textrm{A} \rangle -
\langle{\bf I} \vert \hat{X} \vert \Psi_\textrm{A} \rangle,
\end{equation}
with $E_\textrm{C} = 2\sum_j f^j_j$. Whereas both options, just shifting the origin of energy, are equivalent, the former is numerically more stable \cite{Sato2013PRA}.

The TD-CASSCF method is gauge invariant (see Subsection \ref{subsec:gauge-dependence}) and size extensive. Accuracy can be systematically controlled between TDHF and MCTDHF. The account of correlation can also be flexibly controlled. Moreover, correlation (beyond the HF mean-field potential) in the ground state can be taken into account, unlike in the case of the TDCI method. The number of the configurations involved (CI dimension) scales factorially with the number of active electrons $N_A$, significantly reduced from that in the MCTDHF approach, which scales factorially with the total number of electrons $N$. The detailed discussion on the computational cost of the TD-CASSCF method can be found in Ref.~\cite{Sato2013PRA}.

\subsection{Time-Dependent Occupation-Restricted Multiple-Active-Space (TD-ORMAS) Method}
\label{subsec:TD-ORMAS}

The classification into core and active orbital subspaces introduced in the TD-CASSCF reduces the CI dimension compared with that in the MCTDHF. Nevertheless, the computational cost increases exponentially with $N_A$, thus large-active-space calculations will become prohibitive. Also, we cannot take a large core subspace if we are to properly describe inner-shell excitation and ionization by short wave-length pulses. We can further decrease the CI dimension by resorting to {\it non-complete} CI expansion. 

The ORMAS model \cite{Ivanic2003JCP-a,Sato2015PRA} offers a highly flexible framework in this direction. In this model, we further subdivide 
the active orbital space $\mathcal{A}$ into a
given number, $G$, of subgroups,
\begin{equation}
\mathcal{A} = \sum_{g=1}^G \mathcal{A}_g,
\end{equation}
where,
\begin{equation}
\mathcal{A}_g =
\left\{\phi^{(g)}_1,\phi^{(g)}_2,\cdot\cdot\cdot,\phi^{(g)}_{n_g}\right\} \quad (1 \le g \le G),
\end{equation}
with $n_\textrm{A} = \sum_{g=1}^G n_g$.
The minimum and maximum numbers of electrons that can be assigned to each subgroup are specified as,
\begin{equation}
\label{eq:occb}
N^\textrm{min}_g \leq N_g \leq N^\textrm{max}_g \quad (1 \le g \le G),
\end{equation}
with $N_A=\sum_{g=1}^{G} N_g$.

\begin{figure}[!t]
\centering
\includegraphics[width=8.3cm]{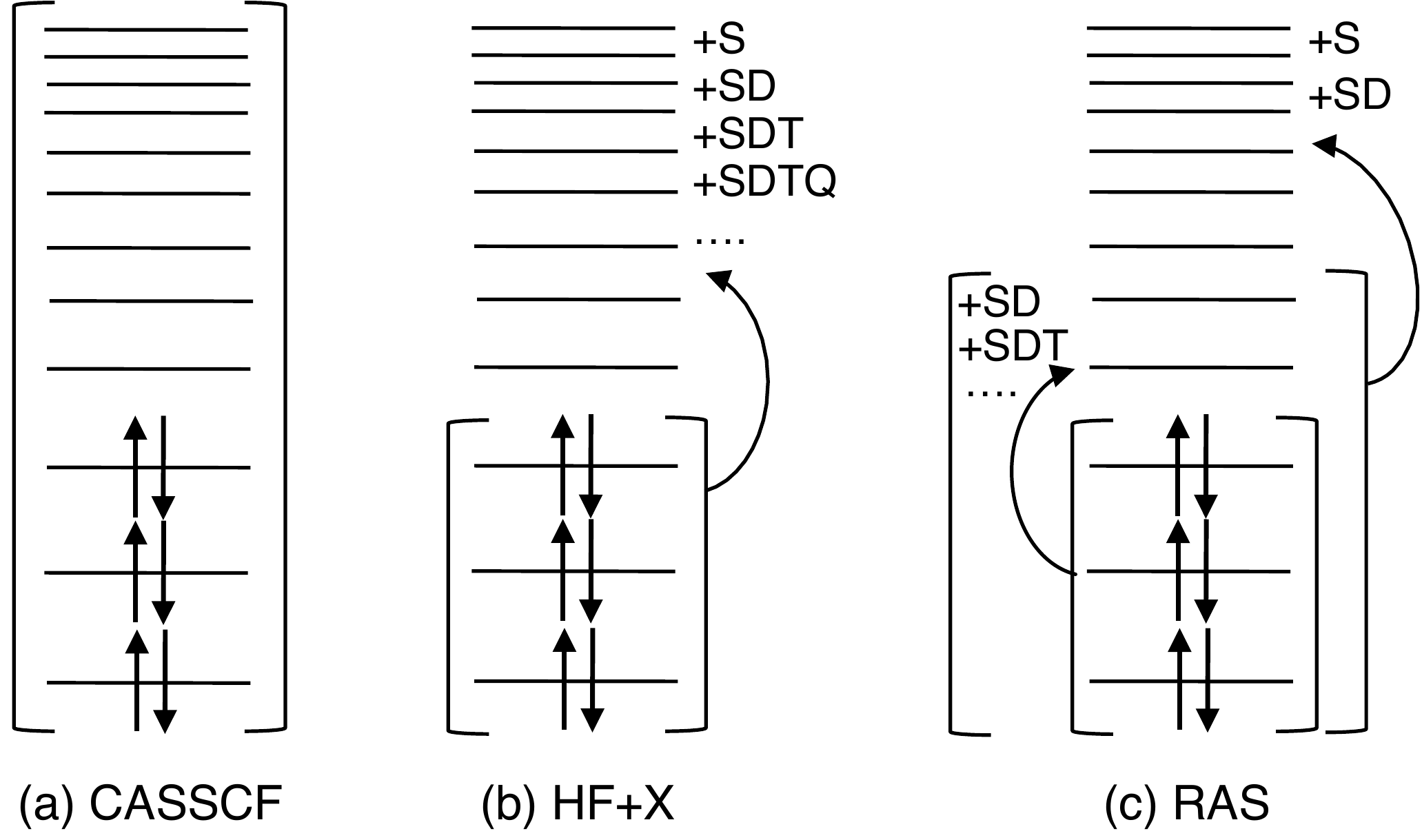}
\caption{Schematic examples of the ORMAS-CI space for six active electrons and twelve active orbitals. The vertical arrows represent electrons in the HF configurations, to be distributed following the respective ORMAS restriction. The curved upward arrows represent the excitation from one to the other subgroup. }
\label{fig:ormas}
\end{figure}

This ansatz involves a wide variety of CI expansions, some of whose examples are shown in Fig.~\ref{fig:ormas}:
\begin{enumerate}
\renewcommand{\labelenumi}{\arabic{enumi})}

\item CASSCF: $G=1, N^1_\textrm{min}=N^1_\textrm{max}=N_\textrm{A}$ [Fig.~\ref{fig:ormas} (a)].

\item HF+X: $G=2, N_\textrm{A}-L \leq N_1 \leq N_\textrm{A}$, $0 \leq N_2 \leq L$ for a given $L$, $n_1 = N_\textrm{A}/2$ [Fig.~\ref{fig:ormas}~(b)]. The CI space is composed of the HF reference space ($\mathcal{A}_1$) and up to $L$-fold excitation from $\mathcal{A}_1$ to $\mathcal{A}_2$. Let us denote the corresponding method, explicitly considered by Miyagi and Madsen \cite{Miyagi2013PRA, Miyagi2014PRA, Bauch2014PRA}, as HF+X for brevity, with X = S, SD, SDT, etc. It should be noted that this expansion is distinct from TDCI methods, in that the orbital functions are time-dependent. 


\item RAS: $G=3$, and the maximum number of holes $M_\text{hole}$ in $\mathcal{A}_1$ and the maximum number of electrons $M_\text{elec}$ in
$\mathcal{A}_3$ are specified, whereas $N_2$ is unconstrained, i.e.,
      \begin{equation}
      \label{eq:or_ras}
       N_\textrm{A} - M_\textrm{hole} \leq N_1 \leq N_\textrm{A}, 
	\hspace{.5em} 0 \leq N_3 \leq M_\textrm{elec}.
      \end{equation}
This model is called {\it restricted active space} (RAS) model proposed 
      by Olsen {\it et al} \cite{Olsen1988JCP}.


\end{enumerate}
Note that Miyagi and Madsen \cite{Miyagi2014PRA} use a term ``TD-RASSCF'' for their method that uses the second type of CI spaces above but that this method is actually {\it not} based on the RAS scheme of Ref.~\cite{Olsen1988JCP}, introduced much earlier. For consistency with the terminology widely used in the stationary electronic structure theory and to promote collaboration between atomic physicists and quantum chemists, in this review, we refer to the latter method as RAS,
which includes the former as a special case.

The equations (\ref{eq:g-tdci}), (\ref{eq:g-tdmo}), (\ref{eq:tdmoc}), (\ref{eq:tdmoa}), and (\ref{eq:tdci}) of motion themselves hold also for the TD-ORMAS approach. For the case of TD-CASSCF, including MCTDHF and TDHF, the orbital rotation operators $\hat{E}_\nu^\mu$ are classified as either redundant or non-redundant coupled, as discussed in the previous subsection. General TD-ORMAS methods, on the other hand, require a third category as follows:
\begin{enumerate}

\item {\it Redundant}.  $\{\hat{E}^i_j, \hat{E}^a_b\}$ as well as active intra-group rotations $\{\hat{E}_u^t;\phi_t,\phi_u\in\mathcal{A}_g\}$ belong to this. Equation (\ref{eq:g-tdmo}) is reduced to Eq.~(\ref{eq:x_red}).

\item {\it Non-redundant uncoupled}. $\{\hat{E}_i^t, \hat{E}_t^i, \hat{E}_p^a, \hat{E}_a^p\}$ belong to this, as in the case of TD-CASSCF. Equation (\ref{eq:g-tdmo}) is simplified to Eq.~(\ref{eq:rmat_inter}).

\item {\it Non-redundant coupled}.
Either $\hat{E}^{\mu}_{\nu} \vert \Psi \rangle$ or $\hat{E}^{\nu}_{\mu} \vert \Psi \rangle$ 
lies across ${\sf P}$ and ${\sf Q}$. Active inter-group rotations $\{E^t_u; \phi_t \in
\mathcal{A}_g, \phi_u \in \mathcal{A}_{g^\prime}, g \ne
g^\prime\}$ belong to this, in general. These contribute to both the CI and orbital
EOMs. We need to work directly with Eq.~(\ref{eq:g-tdmo}). The coupled equations to be solved for  $\{X^t_u; \phi_t \in
\mathcal{A}_g, \phi_u \in \mathcal{A}_{g^\prime}, g \ne
g^\prime\}$ are given in Eq.~(57) of Ref.~\cite{Sato2015PRA}.

\end{enumerate}

Numerical implementation of the TD-ORMAS method is described in Ref.~\cite{Sato2015PRA}. It is worth noting that the orbital EOMs Eqs.~(\ref{eq:tdmoc}) and (\ref{eq:tdmoa}), and CI EOMs Eq.~(\ref{eq:tdci}), derived by Sato and Ishikawa in Ref.~\cite{Sato2015PRA} are valid even for {\it general} MCSCF wave functions with {\it arbitrary} CI spaces {\sf P} if we treat all the active-active rotations $E^t_u$ as non-redundant coupled and solve Eq.~(57) of Ref.~\cite{Sato2015PRA} for $X^t_u$. The TD-ORMAS method is, however, advantageous from the viewpoint of the efficiency and stability of time propagation, since it can limit non-redundant coupled rotations only to active inter-group rotations. Equations of motion for general MCSCF wave functions have recently been derived independently also by Haxton and McCurdy \cite{Haxton2015PRA}.


\subsection{Remark on the Gauge Dependence}
\label{subsec:gauge-dependence}

The TDSE Eq.~(\ref{eq:TDSE}) can be represented in either the length Eq.~(\ref{eq:length-gauge}) or the velocity gauge Eq.~(\ref{eq:velocity-gauge}). The gauge principle states that all physical observables are gauge invariant \cite{Bandrauk2013JPB}. The wave functions $\Psi_{\rm L} (t)$ and $\Psi_{\rm V} (t)$ expressed in the length and velocity gauges, respectively, are related to each other through the gauge transformation,
\begin{equation}
\label{eq:gauge-transformation}
\Psi_{\rm V}(t) = \hat{\mathcal{U}}(t) \Psi_{\rm L}(t),
\end{equation}
where the transformation operator,
\begin{equation}
\label{eq:gauge-transformation-operator}
\hat{\mathcal{U}}(t) = \exp \left[-i{\bf A}(t)\cdot \sum_{i=1}^{N}{\bf r}_i\right]
\end{equation}
is unitary. One can easily verify this, by substituting Eq.~(\ref{eq:gauge-transformation}) into the TDSE with Eq.~(\ref{eq:velocity-gauge}) and finding that $\Psi_{\rm L}$ indeed satisfies the TDSE with Eq.~(\ref{eq:length-gauge}).

Let us denote the orbital functions calculated with a given multiconfiguration ansatz Eq.~(\ref{eq:general-mcwf}) within the length gauge by $\{\phi_p^{\rm L}({\bf r})\}$. Equation (\ref{eq:gauge-transformation}) is fulfilled if one constructs the wave function $\Psi_{\rm V}(t)$ of the same ansatz with the CI coefficients unchanged using the orbital functions $\{\phi_p^{\rm V}({\bf r})\}$ defined by,
\begin{equation}
\label{eq:orbital-gauge-transformation}
\phi_p^{\rm V}({\bf r}) = \exp \left[-i{\bf A}(t)\cdot {\bf r}\right] \phi_p^{\rm L}({\bf r}).
\end{equation}
Since this requires the orbital functions to be time-dependent, TD-MCSCF methods with time-independent orbital functions such as TDCI discussed in Subsec.~\ref{subsec:TDCI} are, in general, not gauge invariant, i.e., the results obtained within the length gauge are not equal to those within the velocity gauge. This is because $\Psi_{\rm V}(t)$ does not necessarily belong to the subspace of the Hilbert space spanned by $\{\Phi_{\bf I}\}$, in which $\Psi_{\rm L}(t)$ is optimized.

It should be noticed that the length- and velocity-gauge Hamiltonians $\hat{H}_{\rm L}(t)$ with Eq.~(\ref{eq:length-gauge}) and $\hat{H}_{\rm V}(t)$ with Eq.~(\ref{eq:velocity-gauge}), respectively, are related by \cite{Bandrauk2013JPB},
\begin{equation}
\hat{H}_{\rm V} = \hat{\mathcal{U}}\hat{H}_{\rm L}\hat{\mathcal{U}}^\dag+i\frac{d\hat{\mathcal{U}}}{dt}\hat{\mathcal{U}}^\dag.
\end{equation}
Then, since the gauge-transformation operator $\hat{\mathcal{U}}(t)$ is unitary [Eq.~(\ref{eq:gauge-transformation-operator})], one can see that the formulas for the TDVP Eq.~(\ref{eq:TDVP}) in the two representations are equivalent. This guarantees that the wave function transformed via Eq.~(\ref{eq:orbital-gauge-transformation}) from the wave function satisfying the length-gauge TDVP satisfies the velocity-gauge TDVP. Therefore, TD-MCSCF methods with orbital functions varying in time are gauge invariant in general \cite{Sato2013PRA,Miyagi2014PRA,Sato2015PRA}

\subsection{Time-Dependent Coupled-Cluster method}

A somewhat different direction of particular interest is time-dependent extensions \cite{Huber2011JCP,Kvaal2012JCP} of the so-called coupled cluster method, though they have not been applied to laser-induced ionization dynamics yet, and further theoretical sophistications appear to be required. Whereas coupled cluster is often abbreviated as CC, we do not use it in this review to avoid confusion with close coupling in Sec.~\ref{sec:TDSE}.

Let us first assume that orbital functions are fixed in time, as implemented by Huber and Klamroth \cite{Huber2011JCP} to study laser-driven excitation of small molecules. We write the total wave function as,
\begin{equation}
\label{eq:cluster-expansion}
|\Psi (t)\rangle = e^{\hat{T}}|\Psi_0\rangle,
\end{equation}
where $|\Psi_0\rangle$ denotes a reference Slater determinant, taken to be the HF ground state, and the cluster operator $\hat{T}$ is defined as,
\begin{equation}
\label{eq:cluster-operator}
\hat{T} = \sum_{ia} \tau_i^a(t)\hat{a}_{a}^\dag\hat{a}_{i}+  \sum_{ijab} \tau_{ij}^{ab}(t) \hat{a}_{b}^\dag \hat{a}_{j} \hat{a}_{a}^\dag \hat{a}_{i} + \cdots,
\end{equation}
which is truncated at a given term in practice. The time-dependent cluster amplitudes $\tau_i^a, \tau_{ij}^{ab}, \cdots$ correspond to $C_i^a, C_{ij}^{ab}, \cdots$ in Eq.~(\ref{eq:CI-expansion}) in the first order, whereas Eq.~(\ref{eq:cluster-expansion}) involves excitation of any arbitrary order even if we truncate Eq.~(\ref{eq:cluster-operator}), e.g., at the second term. As a consequence, the exponential form recovers size extensivity which the TDCI method (Sec.~\ref{subsec:TDCI}) lacks.
If we include up to the second term in Eq.~(\ref{eq:cluster-operator}) (coupled cluster singles and doubles) and substitute the cluster expansion Eq.~(\ref{eq:cluster-expansion}) into the TDSE, we obtain, after some algebra, the EOMs for the amplitudes $\tau_i^a(t)$ and $\tau_{ij}^{ab}(t)$, whose explicit forms are given in Ref.~\cite{Huber2011JCP}.

In order to simulate dynamics involving ionization without the need of a huge basis set, it would be advantageous to allow orbital functions to vary in time (cf. difference between TDCI and HF+X). Kvaal has developed orbital adaptive time-dependent coupled-cluster method \cite{Kvaal2012JCP}, though Ref.~\cite{Kvaal2012JCP} assumes a time-independent Hamiltonian, thus no external field. This method is based on  {\it bivariational} (rather than variational) principle; $|\Psi\rangle$ and $\langle \Psi^\prime |$ are independently variated, and constructed based on different sets of molecular orbitals $\{\varphi_p\}$ and $\{\tilde{\varphi}_p\}$, respectively, that are not individually orthogonal but satisfy biorthogonality $\langle \tilde{\varphi}_p | \varphi_q \rangle = \delta_{pq}$. The EOMs for the variational degrees of freedom, i.e., 
$\{\varphi_p\}$, $\{\tilde{\varphi}_p\}$, and their corresponding cluster amplitudes are derived from the principle that the action integral is stationary under all possible variations, as detailed in Ref.~\cite{Kvaal2012JCP}. It seems that the initial ground state cannot be obtained by imaginary time relaxation.

\section{$R$-Matrix Approach}
\label{sec:R-matrix}

\begin{figure}[!t]
\centering
\includegraphics[width=9cm]{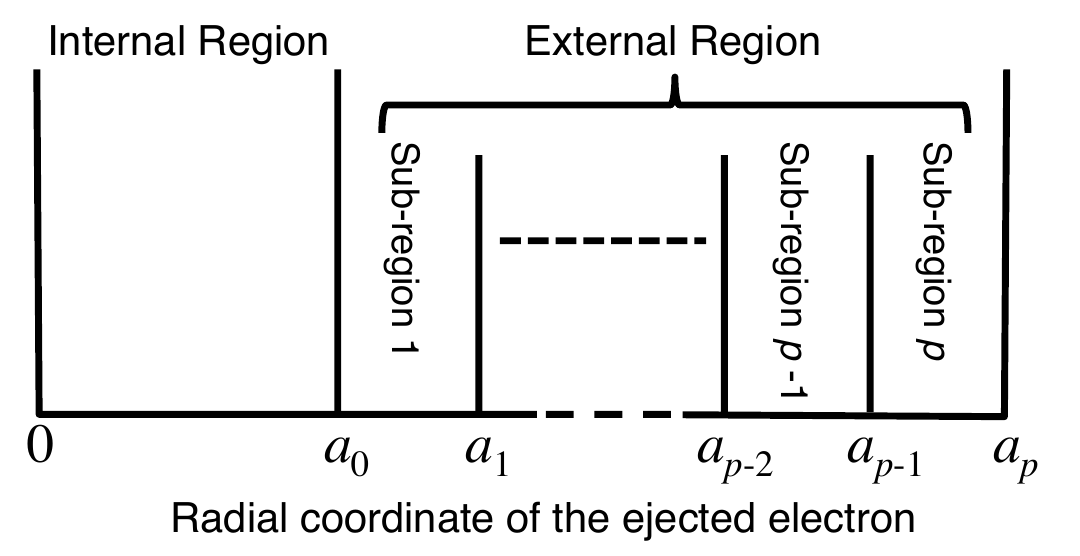}
\caption{Partitioning of the radial coordinate of the ejected electron in the $R$-matrix approach.}
\label{fig:R-matrix}
\end{figure}

In this section we briefly mention approaches based on the $R$-matrix theory originally developed for electron-atom and electron-ion collision processes. The length gauge is preferred in these approaches \cite{Hutchinson2010JPB}.

In the time-dependent $R$-matrix (TDRM) approach, originally proposed by Burke and Burke \cite{Burke1997JPB} and further developed by van der Hart, Lysaght, {\it et al.} \cite{Lysaght2008PRL,Lysaght2009PRA}, single ionization at most of an $(N+1)$-electron atom is considered, and the time-dependent wave function $\Psi ({\bf r}_1,\cdots,{\bf r}_{N+1},t)$ is expanded as \cite{Burke1997JPB},
\begin{align}
&\Psi ({\bf r}_1,\cdots,{\bf r}_{N+1},t) \nonumber \\
&= \mathcal{A}\sum_i
\Phi_i ({\bf r}_1,\cdots,{\bf r}_{N} ; \Omega_{N+1}\sigma_{N+1})r_{N+1}^{-1}\psi_{i}(r_{N+1},t),
\end{align}
where $\Phi_i$ denote time-independent channel functions formed by coupling the residual $N$-electron state with the angular and spin eigenfunctions of the ejected electron, and $\psi_{i}$ the time-dependent reduced radial functions that describe the motion of the ejected electron in the $i$-th channel. 

By introducing an equally spaced temporal mesh,
\begin{equation}
t_q = q\Delta t, \quad q = 0,1,2,\cdots,
\end{equation}
with $\Delta t$ being the time step, one can express the temporal evolution of the wave function $\Psi (t)$ from $t=t_q$ to $t_{q+1}$ as,
\begin{equation}
\label{eq:TDRM-implicit}
[\hat{H}(t_{q+1/2})-E]\Psi (t_{q+1}) = \Theta (t_q),
\end{equation}
with $t_{q+1/2} = t_q + \Delta t/2$, $E\equiv 2i\Delta t^{-1}$ and,
\begin{equation}
\Theta (t_q) = -[\hat{H} (t_{q+1/2}) + E]\Psi (t_q),
\end{equation}
where terms of $O(\Delta t^3)$ are neglected. 

In order to solve the implicit step Eq.~(\ref{eq:TDRM-implicit}), we partition the radial coordinate $r_{N+1}$ of the ejected electron into an internal and external regions with the boundary radius $a_0$ (Fig.~\ref{fig:R-matrix}). In the internal region, exchange and electron-electron correlation effects between the ejected and the remaining electrons are important, whereas in the external region the ejected electron moves in the local long-range multipole potential of the residual $N$-electron system and the laser field. The external region is further partitioned into $p$ subregions with the boundary radii $(a_0 < ) a_1 < \cdots < a_p$ (Fig.~\ref{fig:R-matrix}). The outermost boundary $a_p$ is set so large that $\psi_{i}(a_p,t)=0$. The matrix equations that couple $\psi_{i}(a_s,t)$ and $\psi_{i}^\prime (a_s,t)$ $(s=0,\cdots, p)$, where $\psi_{i}^\prime \equiv \frac{\partial }{\partial r}\psi_{i}$, between neighboring boundaries are obtained and solved as described in Refs.~\cite{Burke1997JPB,Lysaght2009PRA}. Whereas the TDRM theory was originally formulated in terms of the $R$-matrix basis functions \cite{Lysaght2009PRA}, also a mixed finite-difference and $R$-matrix basis set technique has recently been developed, in which finite-difference methods are adopted for the external region \cite{Lysaght-QDI}.

Guan {\it et al.} \cite{Guan2007PRA,Bartschat-QDI} have developed a time-dependent extension of the $B$-spline $R$-matrix (BSR) method \cite{Zatsarinny2000JPB,Zatsarinny2004JPB,Zatsarinny2006CPC}, for multiphoton single and double ionization of atoms. The time-dependent wave function $\Psi$ is expanded as,
\begin{equation}
\Psi ({\bf r}_1,\cdots,{\bf r}_{N+1},t) = \sum_q C_q(t) \Phi_q ({\bf r}_1,\cdots,{\bf r}_{N+1}),
\end{equation}
in terms of a known set of time-independent $(N+1)$-electron states $\Phi_q$ and expansion coefficients $C_q(t)$ describing the dynamics. $\Phi_q$ are formed from appropriately symmetrized products of atomic orbitals that are not necessarily orthogonal, by means of the BSR method. The temporal evolution of $C_q(t)$ is obtained by the Arnoldi-Lanczos approach \cite{Park1986JCP,Schneider2005JNCS,Saad}


\section{Conclusion}

Today, thanks to elaboration of quantum chemistry methods, not only specialized theoreticians but also experimentalists can routinely calculate ground-state properties of large systems containing tens to hundreds of electrons. In marked contrast, approaches for the dynamics in a strong, time-dependent external field is still in the early stage of development. Major difficulty stems from spatially extended continuum electrons and dynamical electronic correlation more prominent than the static one in the ground state. In this paper, we have reviewed, if not exhaustively, major promising activities to tackle this grand challenge. Whereas its extension beyond three electrons would not be realistic, full TDSE simulation (Sec.~\ref{sec:TDSE}) has been a powerful method to investigate the electronic correlation in photo excitation and ionization, and will continue to be a stringent benchmark for any new approach. Further development and sophistication in both MCSCF (Sec.~\ref{sec:MCSCF}) and $R$-matrix (Sec.~\ref{sec:R-matrix}) approaches may someday realize routine calculations for large complex atoms and molecules. In the present review, we have concentrated ourselves on the electron dynamics within the fixed-nuclei approximation except for in Sec.~\ref{subsubsec:H2-vibration}. Comparison with experiments will involve nuclear motion. The existence of different dissociation and Coulomb explosion channels may imply that nuclear dynamics also should be, at least partially, treated quantum mechanically, for which another endeavor such as in Ref.~\cite{Kato2009JCP} will be required. Vast unexplored terrain lies before us.


%

%

\section*{Acknowledgment}
We would like to thank Joachim Burgd\"order, Johannes Feist, Stefan Nagele, Renate Pazourek, Iva B\ifmmode \check{r}\else \v{r}\fi{}ezinov\'a, Fabian Lackner, Lars Bojer Madsen, Haruhide Miyagi, Shungo Miyabe, Ryoto Sawada, Iliya Olegovich Tikhomirov, Yoshihisa Futakuchi, Misha Ivanov, Suren Sukiasyan, and Daniel Haxton for fruitful discussions. We are also grateful to support by Grant-in-Aid for Scientific Research (Grants No. 23750007, No. 23656043, No. 25286064, No. 26390076, and No. 26600111) from the Japan Society for the Promotion of Science (JSPS), Advanced Photon Science Alliance (APSA) project commissioned by the Ministry of Education, Culture, Sports, Science and Technology (MEXT) of Japan, the Center of Innovation Program from the Japan Science and Technology Agency (JST), and the Cooperative Research Program of ``Network Joint Research Center for Materials and Devices'' (Japan).

\section*{Note added in proof}

We refer the readers to Ref.\, \cite{Mercouris1994PRA} and Ref.\, \cite{Mercouris2010AQC} and references therein for the state-specific expansion approach by Mercouris, Komninos, Nicolaides {\it et al.}, which we could not discuss in this review.



\bibliography{ishikenbib}

\providecommand{\noopsort}[1]{}\providecommand{\singleletter}[1]{#1}%
\begin{thebibliography}{161}%
\makeatletter
\providecommand \@ifxundefined [1]{%
 \@ifx{#1\undefined}
}%
\providecommand \@ifnum [1]{%
 \ifnum #1\expandafter \@firstoftwo
 \else \expandafter \@secondoftwo
 \fi
}%
\providecommand \@ifx [1]{%
 \ifx #1\expandafter \@firstoftwo
 \else \expandafter \@secondoftwo
 \fi
}%
\providecommand \natexlab [1]{#1}%
\providecommand \enquote  [1]{``#1''}%
\providecommand \bibnamefont  [1]{#1}%
\providecommand \bibfnamefont [1]{#1}%
\providecommand \citenamefont [1]{#1}%
\providecommand \href@noop [0]{\@secondoftwo}%
\providecommand \href [0]{\begingroup \@sanitize@url \@href}%
\providecommand \@href[1]{\@@startlink{#1}\@@href}%
\providecommand \@@href[1]{\endgroup#1\@@endlink}%
\providecommand \@sanitize@url [0]{\catcode `\\12\catcode `\$12\catcode
  `\&12\catcode `\#12\catcode `\^12\catcode `\_12\catcode `\%12\relax}%
\providecommand \@@startlink[1]{}%
\providecommand \@@endlink[0]{}%
\providecommand \url  [0]{\begingroup\@sanitize@url \@url }%
\providecommand \@url [1]{\endgroup\@href {#1}{\urlprefix }}%
\providecommand \urlprefix  [0]{URL }%
\providecommand \Eprint [0]{\href }%
\providecommand \doibase [0]{http://dx.doi.org/}%
\providecommand \selectlanguage [0]{\@gobble}%
\providecommand \bibinfo  [0]{\@secondoftwo}%
\providecommand \bibfield  [0]{\@secondoftwo}%
\providecommand \translation [1]{[#1]}%
\providecommand \BibitemOpen [0]{}%
\providecommand \bibitemStop [0]{}%
\providecommand \bibitemNoStop [0]{.\EOS\space}%
\providecommand \EOS [0]{\spacefactor3000\relax}%
\providecommand \BibitemShut  [1]{\csname bibitem#1\endcsname}%
\let\auto@bib@innerbib\@empty
\bibitem [{\citenamefont {Protopapas}\ \emph {et~al.}(1997)\citenamefont
  {Protopapas}, \citenamefont {Keitel},\ and\ \citenamefont
  {Knight}}]{Protopapas1997RPP}%
  \BibitemOpen
  \bibfield  {author} {\bibinfo {author} {\bibfnamefont {M.}~\bibnamefont
  {Protopapas}}, \bibinfo {author} {\bibfnamefont {C.~H.}\ \bibnamefont
  {Keitel}}, \ and\ \bibinfo {author} {\bibfnamefont {P.~L.}\ \bibnamefont
  {Knight}},\ }\href {http://stacks.iop.org/0034-4885/60/i=4/a=001} {\bibfield
  {journal} {\bibinfo  {journal} {Rep. Prog. Phys.}\ }\textbf {\bibinfo
  {volume} {60}},\ \bibinfo {pages} {389} (\bibinfo {year} {1997})}\BibitemShut
  {NoStop}%
\bibitem [{\citenamefont {Brabec}\ and\ \citenamefont
  {Krausz}(2000)}]{Brabec2000RMP}%
  \BibitemOpen
  \bibfield  {author} {\bibinfo {author} {\bibfnamefont {T.}~\bibnamefont
  {Brabec}}\ and\ \bibinfo {author} {\bibfnamefont {F.}~\bibnamefont
  {Krausz}},\ }\href {\doibase 10.1103/RevModPhys.72.545} {\bibfield  {journal}
  {\bibinfo  {journal} {Rev. Mod. Phys.}\ }\textbf {\bibinfo {volume} {72}},\
  \bibinfo {pages} {545} (\bibinfo {year} {2000})}\BibitemShut {NoStop}%
\bibitem [{\citenamefont {Popmintchev}\ \emph {et~al.}(2012)\citenamefont
  {Popmintchev}, \citenamefont {Chen}, \citenamefont {Popmintchev},
  \citenamefont {Arpin}, \citenamefont {Brown}, \citenamefont {Alisauskas},
  \citenamefont {Andriukaitis}, \citenamefont {Balciunas}, \citenamefont
  {Mucke}, \citenamefont {Pugzlys}, \citenamefont {Baltu{\v s}ka},
  \citenamefont {Shim}, \citenamefont {Schrauth}, \citenamefont {Gaeta},
  \citenamefont {Hernandez-Garcia}, \citenamefont {Plaja}, \citenamefont
  {Becker}, \citenamefont {Jaro{\'{n}}-Becker}, \citenamefont {Murnane},\ and\
  \citenamefont {Kapteyn}}]{Popmintchev2012Nature}%
  \BibitemOpen
  \bibfield  {author} {\bibinfo {author} {\bibfnamefont {T.}~\bibnamefont
  {Popmintchev}}, \bibinfo {author} {\bibfnamefont {M.~C.}\ \bibnamefont
  {Chen}}, \bibinfo {author} {\bibfnamefont {D.}~\bibnamefont {Popmintchev}},
  \bibinfo {author} {\bibfnamefont {P.}~\bibnamefont {Arpin}}, \bibinfo
  {author} {\bibfnamefont {S.}~\bibnamefont {Brown}}, \bibinfo {author}
  {\bibfnamefont {S.}~\bibnamefont {Alisauskas}}, \bibinfo {author}
  {\bibfnamefont {G.}~\bibnamefont {Andriukaitis}}, \bibinfo {author}
  {\bibfnamefont {T.}~\bibnamefont {Balciunas}}, \bibinfo {author}
  {\bibfnamefont {O.~D.}\ \bibnamefont {Mucke}}, \bibinfo {author}
  {\bibfnamefont {A.}~\bibnamefont {Pugzlys}}, \bibinfo {author} {\bibfnamefont
  {A.}~\bibnamefont {Baltu{\v s}ka}}, \bibinfo {author} {\bibfnamefont
  {B.}~\bibnamefont {Shim}}, \bibinfo {author} {\bibfnamefont {S.~E.}\
  \bibnamefont {Schrauth}}, \bibinfo {author} {\bibfnamefont {A.}~\bibnamefont
  {Gaeta}}, \bibinfo {author} {\bibfnamefont {C.}~\bibnamefont
  {Hernandez-Garcia}}, \bibinfo {author} {\bibfnamefont {L.}~\bibnamefont
  {Plaja}}, \bibinfo {author} {\bibfnamefont {A.}~\bibnamefont {Becker}},
  \bibinfo {author} {\bibfnamefont {A.}~\bibnamefont {Jaro{\'{n}}-Becker}},
  \bibinfo {author} {\bibfnamefont {M.~M.}\ \bibnamefont {Murnane}}, \ and\
  \bibinfo {author} {\bibfnamefont {H.~C.}\ \bibnamefont {Kapteyn}},\
  }\href@noop {} {\bibfield  {journal} {\bibinfo  {journal} {Science}\ }\textbf
  {\bibinfo {volume} {336}},\ \bibinfo {pages} {1287} (\bibinfo {year}
  {2012})}\BibitemShut {NoStop}%
\bibitem [{\citenamefont {Chang}(2011)}]{Chang2011}%
  \BibitemOpen
  \bibfield  {author} {\bibinfo {author} {\bibfnamefont {Z.}~\bibnamefont
  {Chang}},\ }\href@noop {} {\emph {\bibinfo {title} {Fundamentals of
  Attosecond Optics}}}\ (\bibinfo  {publisher} {CRC Press},\ \bibinfo {address}
  {Boca Raton, FL},\ \bibinfo {year} {2011})\BibitemShut {NoStop}%
\bibitem [{\citenamefont {Plaja}\ \emph {et~al.}(2013)\citenamefont {Plaja},
  \citenamefont {Torres},\ and\ \citenamefont {Za\"{\i}r}}]{AttosecondPhysics}%
  \BibitemOpen
  \bibinfo {editor} {\bibfnamefont {L.}~\bibnamefont {Plaja}}, \bibinfo
  {editor} {\bibfnamefont {R.}~\bibnamefont {Torres}}, \ and\ \bibinfo {editor}
  {\bibfnamefont {A.}~\bibnamefont {Za\"{\i}r}},\ eds.,\ \href@noop {} {\emph
  {\bibinfo {title} {Attosecond Physics}}},\ \bibinfo {series} {Springer Series
  in Optical Sciences}, Vol.\ \bibinfo {volume} {177}\ (\bibinfo  {publisher}
  {Springer},\ \bibinfo {address} {Berlin},\ \bibinfo {year}
  {2013})\BibitemShut {NoStop}%
\bibitem [{\citenamefont {Itatani}\ \emph {et~al.}(2004)\citenamefont
  {Itatani}, \citenamefont {Levesque}, \citenamefont {Zeidler}, \citenamefont
  {Niikura}, \citenamefont {P\'epin}, \citenamefont {Kieffer}, \citenamefont
  {Corkum},\ and\ \citenamefont {Villeneuve}}]{Itatani2004Nature}%
  \BibitemOpen
  \bibfield  {author} {\bibinfo {author} {\bibfnamefont {J.}~\bibnamefont
  {Itatani}}, \bibinfo {author} {\bibfnamefont {J.}~\bibnamefont {Levesque}},
  \bibinfo {author} {\bibfnamefont {D.}~\bibnamefont {Zeidler}}, \bibinfo
  {author} {\bibfnamefont {H.}~\bibnamefont {Niikura}}, \bibinfo {author}
  {\bibfnamefont {H.}~\bibnamefont {P\'epin}}, \bibinfo {author} {\bibfnamefont
  {J.~C.}\ \bibnamefont {Kieffer}}, \bibinfo {author} {\bibfnamefont {P.~B.}\
  \bibnamefont {Corkum}}, \ and\ \bibinfo {author} {\bibfnamefont {D.~M.}\
  \bibnamefont {Villeneuve}},\ }\href@noop {} {\bibfield  {journal} {\bibinfo
  {journal} {Nature}\ }\textbf {\bibinfo {volume} {432}},\ \bibinfo {pages}
  {867} (\bibinfo {year} {2004})}\BibitemShut {NoStop}%
\bibitem [{\citenamefont {Haessler}\ \emph {et~al.}(2010)\citenamefont
  {Haessler}, \citenamefont {Caillat}, \citenamefont {Boutu}, \citenamefont
  {Giovanetti-Teixeira}, \citenamefont {Ruchon}, \citenamefont {Auguste},
  \citenamefont {Diveki}, \citenamefont {Breger}, \citenamefont {Maquet},
  \citenamefont {Carr{\'e}}, \citenamefont {Ta{\"\i}eb},\ and\ \citenamefont
  {Sali{\`e}res}}]{Haessler2010NatPhys}%
  \BibitemOpen
  \bibfield  {author} {\bibinfo {author} {\bibfnamefont {S.}~\bibnamefont
  {Haessler}}, \bibinfo {author} {\bibfnamefont {J.}~\bibnamefont {Caillat}},
  \bibinfo {author} {\bibfnamefont {W.}~\bibnamefont {Boutu}}, \bibinfo
  {author} {\bibfnamefont {C.}~\bibnamefont {Giovanetti-Teixeira}}, \bibinfo
  {author} {\bibfnamefont {T.}~\bibnamefont {Ruchon}}, \bibinfo {author}
  {\bibfnamefont {T.}~\bibnamefont {Auguste}}, \bibinfo {author} {\bibfnamefont
  {Z.}~\bibnamefont {Diveki}}, \bibinfo {author} {\bibfnamefont
  {P.}~\bibnamefont {Breger}}, \bibinfo {author} {\bibfnamefont
  {A.}~\bibnamefont {Maquet}}, \bibinfo {author} {\bibfnamefont
  {B.}~\bibnamefont {Carr{\'e}}}, \bibinfo {author} {\bibfnamefont
  {R.}~\bibnamefont {Ta{\"\i}eb}}, \ and\ \bibinfo {author} {\bibfnamefont
  {P.}~\bibnamefont {Sali{\`e}res}},\ }\href@noop {} {\bibfield  {journal}
  {\bibinfo  {journal} {Nature Phys.}\ }\textbf {\bibinfo {volume} {6}},\
  \bibinfo {pages} {200} (\bibinfo {year} {2010})}\BibitemShut {NoStop}%
\bibitem [{\citenamefont {Sali{\`e}res}\ \emph {et~al.}(2012)\citenamefont
  {Sali{\`e}res}, \citenamefont {Maquet}, \citenamefont {Haessler},
  \citenamefont {Caillat},\ and\ \citenamefont {Ta{\"\i}eb}}]{Salieres2012RPP}%
  \BibitemOpen
  \bibfield  {author} {\bibinfo {author} {\bibfnamefont {P.}~\bibnamefont
  {Sali{\`e}res}}, \bibinfo {author} {\bibfnamefont {A.}~\bibnamefont
  {Maquet}}, \bibinfo {author} {\bibfnamefont {S.}~\bibnamefont {Haessler}},
  \bibinfo {author} {\bibfnamefont {J.}~\bibnamefont {Caillat}}, \ and\
  \bibinfo {author} {\bibfnamefont {R.}~\bibnamefont {Ta{\"\i}eb}},\ }\href
  {http://stacks.iop.org/0034-4885/75/i=6/a=062401} {\bibfield  {journal}
  {\bibinfo  {journal} {Rep. Prog. Phys.}\ }\textbf {\bibinfo {volume} {75}},\
  \bibinfo {pages} {062401} (\bibinfo {year} {2012})}\BibitemShut {NoStop}%
\bibitem [{\citenamefont {Agostini}\ and\ \citenamefont
  {DiMauro}(2004)}]{Agostini2004RPP}%
  \BibitemOpen
  \bibfield  {author} {\bibinfo {author} {\bibfnamefont {P.}~\bibnamefont
  {Agostini}}\ and\ \bibinfo {author} {\bibfnamefont {L.~F.}\ \bibnamefont
  {DiMauro}},\ }\href {http://stacks.iop.org/0034-4885/67/i=6/a=R01} {\bibfield
   {journal} {\bibinfo  {journal} {Rep. Prog. Phys.}\ }\textbf {\bibinfo
  {volume} {67}},\ \bibinfo {pages} {813} (\bibinfo {year} {2004})}\BibitemShut
  {NoStop}%
\bibitem [{\citenamefont {Krausz}\ and\ \citenamefont
  {Ivanov}(2009)}]{Krausz2009RMP}%
  \BibitemOpen
  \bibfield  {author} {\bibinfo {author} {\bibfnamefont {F.}~\bibnamefont
  {Krausz}}\ and\ \bibinfo {author} {\bibfnamefont {M.}~\bibnamefont
  {Ivanov}},\ }\href {\doibase 10.1103/RevModPhys.81.163} {\bibfield  {journal}
  {\bibinfo  {journal} {Rev. Mod. Phys.}\ }\textbf {\bibinfo {volume} {81}},\
  \bibinfo {pages} {163} (\bibinfo {year} {2009})}\BibitemShut {NoStop}%
\bibitem [{\citenamefont {Gallmann}\ \emph {et~al.}(2012)\citenamefont
  {Gallmann}, \citenamefont {Cirelli},\ and\ \citenamefont
  {Keller}}]{Gallmann2012ARPC}%
  \BibitemOpen
  \bibfield  {author} {\bibinfo {author} {\bibfnamefont {L.}~\bibnamefont
  {Gallmann}}, \bibinfo {author} {\bibfnamefont {C.}~\bibnamefont {Cirelli}}, \
  and\ \bibinfo {author} {\bibfnamefont {U.}~\bibnamefont {Keller}},\ }\href
  {\doibase 10.1146/annurev-physchem-032511-143702} {\bibfield  {journal}
  {\bibinfo  {journal} {Annu. Rev. Phys. Chem.}\ }\textbf {\bibinfo {volume}
  {63}},\ \bibinfo {pages} {447} (\bibinfo {year} {2012})},\ \bibinfo {note}
  {pMID: 22404594},\ \Eprint
  {http://arxiv.org/abs/http://dx.doi.org/10.1146/annurev-physchem-032511-143702}
  {http://dx.doi.org/10.1146/annurev-physchem-032511-143702} \BibitemShut
  {NoStop}%
\bibitem [{\citenamefont {Sekikawa}\ \emph {et~al.}(2004)\citenamefont
  {Sekikawa}, \citenamefont {Kosuge}, \citenamefont {Kanai},\ and\
  \citenamefont {Watanabe}}]{Sekikawa2004Nature}%
  \BibitemOpen
  \bibfield  {author} {\bibinfo {author} {\bibfnamefont {T.}~\bibnamefont
  {Sekikawa}}, \bibinfo {author} {\bibfnamefont {A.}~\bibnamefont {Kosuge}},
  \bibinfo {author} {\bibfnamefont {T.}~\bibnamefont {Kanai}}, \ and\ \bibinfo
  {author} {\bibfnamefont {S.}~\bibnamefont {Watanabe}},\ }\href@noop {}
  {\bibfield  {journal} {\bibinfo  {journal} {Nature}\ }\textbf {\bibinfo
  {volume} {432}},\ \bibinfo {pages} {605} (\bibinfo {year}
  {2004})}\BibitemShut {NoStop}%
\bibitem [{\citenamefont {Nabekawa}\ \emph {et~al.}(2005)\citenamefont
  {Nabekawa}, \citenamefont {Hasegawa}, \citenamefont {Takahashi},\ and\
  \citenamefont {Midorikawa}}]{Nabekawa2005PRL}%
  \BibitemOpen
  \bibfield  {author} {\bibinfo {author} {\bibfnamefont {Y.}~\bibnamefont
  {Nabekawa}}, \bibinfo {author} {\bibfnamefont {H.}~\bibnamefont {Hasegawa}},
  \bibinfo {author} {\bibfnamefont {E.~J.}\ \bibnamefont {Takahashi}}, \ and\
  \bibinfo {author} {\bibfnamefont {K.}~\bibnamefont {Midorikawa}},\ }\href
  {\doibase 10.1103/PhysRevLett.94.043001} {\bibfield  {journal} {\bibinfo
  {journal} {Phys. Rev. Lett.}\ }\textbf {\bibinfo {volume} {94}},\ \bibinfo
  {pages} {043001} (\bibinfo {year} {2005})}\BibitemShut {NoStop}%
\bibitem [{\citenamefont {Gordon}\ \emph {et~al.}(2006)\citenamefont {Gordon},
  \citenamefont {K\"artner}, \citenamefont {Rohringer},\ and\ \citenamefont
  {Santra}}]{Gordon2006PRL}%
  \BibitemOpen
  \bibfield  {author} {\bibinfo {author} {\bibfnamefont {A.}~\bibnamefont
  {Gordon}}, \bibinfo {author} {\bibfnamefont {F.~X.}\ \bibnamefont
  {K\"artner}}, \bibinfo {author} {\bibfnamefont {N.}~\bibnamefont
  {Rohringer}}, \ and\ \bibinfo {author} {\bibfnamefont {R.}~\bibnamefont
  {Santra}},\ }\href {\doibase 10.1103/PhysRevLett.96.223902} {\bibfield
  {journal} {\bibinfo  {journal} {Phys. Rev. Lett.}\ }\textbf {\bibinfo
  {volume} {96}},\ \bibinfo {pages} {223902} (\bibinfo {year}
  {2006})}\BibitemShut {NoStop}%
\bibitem [{\citenamefont {Rohringer}\ and\ \citenamefont
  {Santra}(2009)}]{Rohringer2009PRA}%
  \BibitemOpen
  \bibfield  {author} {\bibinfo {author} {\bibfnamefont {N.}~\bibnamefont
  {Rohringer}}\ and\ \bibinfo {author} {\bibfnamefont {R.}~\bibnamefont
  {Santra}},\ }\href {\doibase 10.1103/PhysRevA.79.053402} {\bibfield
  {journal} {\bibinfo  {journal} {Phys. Rev. A}\ }\textbf {\bibinfo {volume}
  {79}},\ \bibinfo {pages} {053402} (\bibinfo {year} {2009})}\BibitemShut
  {NoStop}%
\bibitem [{\citenamefont {Smirnova}\ \emph
  {et~al.}(2009{\natexlab{a}})\citenamefont {Smirnova}, \citenamefont
  {Mairesse}, \citenamefont {Patchkovskii}, \citenamefont {Dudovich},
  \citenamefont {Villeneuve}, \citenamefont {Corkum},\ and\ \citenamefont
  {Ivanov}}]{Smirnova2009Nature}%
  \BibitemOpen
  \bibfield  {author} {\bibinfo {author} {\bibfnamefont {O.}~\bibnamefont
  {Smirnova}}, \bibinfo {author} {\bibfnamefont {Y.}~\bibnamefont {Mairesse}},
  \bibinfo {author} {\bibfnamefont {S.}~\bibnamefont {Patchkovskii}}, \bibinfo
  {author} {\bibfnamefont {N.}~\bibnamefont {Dudovich}}, \bibinfo {author}
  {\bibfnamefont {D.}~\bibnamefont {Villeneuve}}, \bibinfo {author}
  {\bibfnamefont {P.}~\bibnamefont {Corkum}}, \ and\ \bibinfo {author}
  {\bibfnamefont {M.~Y.}\ \bibnamefont {Ivanov}},\ }\href@noop {} {\bibfield
  {journal} {\bibinfo  {journal} {Nature}\ }\textbf {\bibinfo {volume} {460}},\
  \bibinfo {pages} {972} (\bibinfo {year} {2009}{\natexlab{a}})}\BibitemShut
  {NoStop}%
\bibitem [{\citenamefont {Smirnova}\ \emph
  {et~al.}(2009{\natexlab{b}})\citenamefont {Smirnova}, \citenamefont
  {Patchkovskii}, \citenamefont {Mairesse}, \citenamefont {Dudovich},\ and\
  \citenamefont {Ivanov}}]{Smirnova2009PNAS}%
  \BibitemOpen
  \bibfield  {author} {\bibinfo {author} {\bibfnamefont {O.}~\bibnamefont
  {Smirnova}}, \bibinfo {author} {\bibfnamefont {S.}~\bibnamefont
  {Patchkovskii}}, \bibinfo {author} {\bibfnamefont {Y.}~\bibnamefont
  {Mairesse}}, \bibinfo {author} {\bibfnamefont {N.}~\bibnamefont {Dudovich}},
  \ and\ \bibinfo {author} {\bibfnamefont {M.~Y.}\ \bibnamefont {Ivanov}},\
  }\href {\doibase 10.1073/pnas.0907434106} {\bibfield  {journal} {\bibinfo
  {journal} {Proc. Natl. Acad. Sci. U.S.A.}\ }\textbf {\bibinfo {volume}
  {106}},\ \bibinfo {pages} {16556} (\bibinfo {year} {2009}{\natexlab{b}})},\
  \Eprint
  {http://arxiv.org/abs/http://www.pnas.org/content/106/39/16556.full.pdf+html}
  {http://www.pnas.org/content/106/39/16556.full.pdf+html} \BibitemShut
  {NoStop}%
\bibitem [{\citenamefont {Akagi}\ \emph {et~al.}(2009)\citenamefont {Akagi},
  \citenamefont {Otobe}, \citenamefont {Staudte}, \citenamefont {Shiner},
  \citenamefont {Turner}, \citenamefont {D{\"o}rner}, \citenamefont
  {Villeneuve},\ and\ \citenamefont {Corkum}}]{Akagi2009Science}%
  \BibitemOpen
  \bibfield  {author} {\bibinfo {author} {\bibfnamefont {H.}~\bibnamefont
  {Akagi}}, \bibinfo {author} {\bibfnamefont {T.}~\bibnamefont {Otobe}},
  \bibinfo {author} {\bibfnamefont {A.}~\bibnamefont {Staudte}}, \bibinfo
  {author} {\bibfnamefont {A.}~\bibnamefont {Shiner}}, \bibinfo {author}
  {\bibfnamefont {F.}~\bibnamefont {Turner}}, \bibinfo {author} {\bibfnamefont
  {R.}~\bibnamefont {D{\"o}rner}}, \bibinfo {author} {\bibfnamefont {D.~M.}\
  \bibnamefont {Villeneuve}}, \ and\ \bibinfo {author} {\bibfnamefont {P.~B.}\
  \bibnamefont {Corkum}},\ }\href {\doibase 10.1126/science.1175253} {\bibfield
   {journal} {\bibinfo  {journal} {Science}\ }\textbf {\bibinfo {volume}
  {325}},\ \bibinfo {pages} {1364} (\bibinfo {year} {2009})},\ \Eprint
  {http://arxiv.org/abs/http://www.sciencemag.org/content/325/5946/1364.full.pdf}
  {http://www.sciencemag.org/content/325/5946/1364.full.pdf} \BibitemShut
  {NoStop}%
\bibitem [{\citenamefont {Boguslavskiy}\ \emph {et~al.}(2012)\citenamefont
  {Boguslavskiy}, \citenamefont {Mikosch}, \citenamefont {Gijsbertsen},
  \citenamefont {Spanner}, \citenamefont {Patchkovskii}, \citenamefont {Gador},
  \citenamefont {Vrakking},\ and\ \citenamefont
  {Stolow}}]{Boguslavskiy2012Science}%
  \BibitemOpen
  \bibfield  {author} {\bibinfo {author} {\bibfnamefont {A.~E.}\ \bibnamefont
  {Boguslavskiy}}, \bibinfo {author} {\bibfnamefont {J.}~\bibnamefont
  {Mikosch}}, \bibinfo {author} {\bibfnamefont {A.}~\bibnamefont
  {Gijsbertsen}}, \bibinfo {author} {\bibfnamefont {M.}~\bibnamefont
  {Spanner}}, \bibinfo {author} {\bibfnamefont {S.}~\bibnamefont
  {Patchkovskii}}, \bibinfo {author} {\bibfnamefont {N.}~\bibnamefont {Gador}},
  \bibinfo {author} {\bibfnamefont {M.~J.~J.}\ \bibnamefont {Vrakking}}, \ and\
  \bibinfo {author} {\bibfnamefont {A.}~\bibnamefont {Stolow}},\ }\href
  {\doibase 10.1126/science.1212896} {\bibfield  {journal} {\bibinfo  {journal}
  {Science}\ }\textbf {\bibinfo {volume} {335}},\ \bibinfo {pages} {1336}
  (\bibinfo {year} {2012})},\ \Eprint
  {http://arxiv.org/abs/http://www.sciencemag.org/content/335/6074/1336.full.pdf}
  {http://www.sciencemag.org/content/335/6074/1336.full.pdf} \BibitemShut
  {NoStop}%
\bibitem [{\citenamefont {Gross}\ \emph {et~al.}(1996)\citenamefont {Gross},
  \citenamefont {Dobson},\ and\ \citenamefont {Petersilka}}]{Gross1996}%
  \BibitemOpen
  \bibfield  {author} {\bibinfo {author} {\bibfnamefont {E.}~\bibnamefont
  {Gross}}, \bibinfo {author} {\bibfnamefont {J.}~\bibnamefont {Dobson}}, \
  and\ \bibinfo {author} {\bibfnamefont {M.}~\bibnamefont {Petersilka}},\ }in\
  \href {\doibase 10.1007/BFb0016643} {{\emph
  {\bibinfo {booktitle} {Density Functional Theory II}}}},\ \bibinfo {series}
  {Topics in Current Chemistry}, Vol.\ \bibinfo {volume} {181},\ \bibinfo
  {editor} {edited by\ \bibinfo {editor} {\bibfnamefont {R.}~\bibnamefont
  {Nalewajski}}}\ (\bibinfo  {publisher} {Springer Berlin Heidelberg},\
  \bibinfo {year} {1996})\ pp.\ \bibinfo {pages} {81--172}\BibitemShut
  {NoStop}%
\bibitem [{\citenamefont {Ullrich}(2012)}]{TDDFT}%
  \BibitemOpen
  \bibfield  {author} {\bibinfo {author} {\bibfnamefont {C.~A.}\ \bibnamefont
  {Ullrich}},\ }\href@noop {} {\emph {\bibinfo {title} {Time-Dependent
  Density-Functional Theory: Concepts and Applications}}},\ Oxford Graduate
  Texts\ (\bibinfo  {publisher} {Oxford University Press},\ \bibinfo {address}
  {Oxford},\ \bibinfo {year} {2012})\BibitemShut {NoStop}%
\bibitem [{\citenamefont {Otobe}\ \emph {et~al.}(2004)\citenamefont {Otobe},
  \citenamefont {Yabana},\ and\ \citenamefont {Iwata}}]{Otobe2004PRA}%
  \BibitemOpen
  \bibfield  {author} {\bibinfo {author} {\bibfnamefont {T.}~\bibnamefont
  {Otobe}}, \bibinfo {author} {\bibfnamefont {K.}~\bibnamefont {Yabana}}, \
  and\ \bibinfo {author} {\bibfnamefont {J.-I.}\ \bibnamefont {Iwata}},\ }\href
  {\doibase 10.1103/PhysRevA.69.053404} {\bibfield  {journal} {\bibinfo
  {journal} {Phys. Rev. A}\ }\textbf {\bibinfo {volume} {69}},\ \bibinfo
  {pages} {053404} (\bibinfo {year} {2004})}\BibitemShut {NoStop}%
\bibitem [{\citenamefont {Otobe}\ \emph {et~al.}(2008)\citenamefont {Otobe},
  \citenamefont {Yamagiwa}, \citenamefont {Iwata}, \citenamefont {Yabana},
  \citenamefont {Nakatsukasa},\ and\ \citenamefont {Bertsch}}]{Otobe2008PRB}%
  \BibitemOpen
  \bibfield  {author} {\bibinfo {author} {\bibfnamefont {T.}~\bibnamefont
  {Otobe}}, \bibinfo {author} {\bibfnamefont {M.}~\bibnamefont {Yamagiwa}},
  \bibinfo {author} {\bibfnamefont {J.-I.}\ \bibnamefont {Iwata}}, \bibinfo
  {author} {\bibfnamefont {K.}~\bibnamefont {Yabana}}, \bibinfo {author}
  {\bibfnamefont {T.}~\bibnamefont {Nakatsukasa}}, \ and\ \bibinfo {author}
  {\bibfnamefont {G.~F.}\ \bibnamefont {Bertsch}},\ }\href {\doibase
  10.1103/PhysRevB.77.165104} {\bibfield  {journal} {\bibinfo  {journal} {Phys.
  Rev. B}\ }\textbf {\bibinfo {volume} {77}},\ \bibinfo {pages} {165104}
  (\bibinfo {year} {2008})}\BibitemShut {NoStop}%
\bibitem [{\citenamefont {Telnov}\ and\ \citenamefont
  {Chu}(2009)}]{Telnov2009PRA}%
  \BibitemOpen
  \bibfield  {author} {\bibinfo {author} {\bibfnamefont {D.~A.}\ \bibnamefont
  {Telnov}}\ and\ \bibinfo {author} {\bibfnamefont {S.-I.}\ \bibnamefont
  {Chu}},\ }\href {\doibase 10.1103/PhysRevA.80.043412} {\bibfield  {journal}
  {\bibinfo  {journal} {Phys. Rev. A}\ }\textbf {\bibinfo {volume} {80}},\
  \bibinfo {pages} {043412} (\bibinfo {year} {2009})}\BibitemShut {NoStop}%
\bibitem [{\citenamefont {Vignale}\ and\ \citenamefont
  {Kohn}(1996)}]{Vignale1996PRL}%
  \BibitemOpen
  \bibfield  {author} {\bibinfo {author} {\bibfnamefont {G.}~\bibnamefont
  {Vignale}}\ and\ \bibinfo {author} {\bibfnamefont {W.}~\bibnamefont {Kohn}},\
  }\href {\doibase 10.1103/PhysRevLett.77.2037} {\bibfield  {journal} {\bibinfo
   {journal} {Phys. Rev. Lett.}\ }\textbf {\bibinfo {volume} {77}},\ \bibinfo
  {pages} {2037} (\bibinfo {year} {1996})}\BibitemShut {NoStop}%
\bibitem [{\citenamefont {Sch\"afer-Bung}\ and\ \citenamefont
  {Nest}(2008)}]{Schaefer-Bung2008PRA}%
  \BibitemOpen
  \bibfield  {author} {\bibinfo {author} {\bibfnamefont {B.}~\bibnamefont
  {Sch\"afer-Bung}}\ and\ \bibinfo {author} {\bibfnamefont {M.}~\bibnamefont
  {Nest}},\ }\href {\doibase 10.1103/PhysRevA.78.012512} {\bibfield  {journal}
  {\bibinfo  {journal} {Phys. Rev. A}\ }\textbf {\bibinfo {volume} {78}},\
  \bibinfo {pages} {012512} (\bibinfo {year} {2008})}\BibitemShut {NoStop}%
\bibitem [{\citenamefont {Lackner}\ \emph {et~al.}(2015)\citenamefont
  {Lackner}, \citenamefont {B\ifmmode~\check{r}\else \v{r}\fi{}ezinov\'a},
  \citenamefont {Sato}, \citenamefont {Ishikawa},\ and\ \citenamefont
  {Burgd\"orfer}}]{Lackner2015PRA}%
  \BibitemOpen
  \bibfield  {author} {\bibinfo {author} {\bibfnamefont {F.}~\bibnamefont
  {Lackner}}, \bibinfo {author} {\bibfnamefont {I.}~\bibnamefont
  {B\ifmmode~\check{r}\else \v{r}\fi{}ezinov\'a}}, \bibinfo {author}
  {\bibfnamefont {T.}~\bibnamefont {Sato}}, \bibinfo {author} {\bibfnamefont
  {K.~L.}\ \bibnamefont {Ishikawa}}, \ and\ \bibinfo {author} {\bibfnamefont
  {J.}~\bibnamefont {Burgd\"orfer}},\ }\href {\doibase
  10.1103/PhysRevA.91.023412} {\bibfield  {journal} {\bibinfo  {journal} {Phys.
  Rev. A}\ }\textbf {\bibinfo {volume} {91}},\ \bibinfo {pages} {023412}
  (\bibinfo {year} {2015})}\BibitemShut {NoStop}%
\bibitem [{\citenamefont {Christov}(2007)}]{Christov2007NJP}%
  \BibitemOpen
  \bibfield  {author} {\bibinfo {author} {\bibfnamefont {I.~P.}\ \bibnamefont
  {Christov}},\ }\href {http://stacks.iop.org/1367-2630/9/i=3/a=070} {\bibfield
   {journal} {\bibinfo  {journal} {New J. Phys.}\ }\textbf {\bibinfo {volume}
  {9}},\ \bibinfo {pages} {70} (\bibinfo {year} {2007})}\BibitemShut {NoStop}%
\bibitem [{\citenamefont {Christov}(2011)}]{Christov2011JCP}%
  \BibitemOpen
  \bibfield  {author} {\bibinfo {author} {\bibfnamefont {I.~P.}\ \bibnamefont
  {Christov}},\ }\href {\doibase http://dx.doi.org/10.1063/1.3615061}
  {\bibfield  {journal} {\bibinfo  {journal} {J. Chem. Phys.}\ }\textbf
  {\bibinfo {volume} {135}},\ \bibinfo {eid} {044120} (\bibinfo {year}
  {2011})}\BibitemShut {NoStop}%
\bibitem [{\citenamefont {Oriols}\ and\ \citenamefont
  {Mompart}(2012)}]{Oriols}%
  \BibitemOpen
  \bibinfo {editor} {\bibfnamefont {X.}~\bibnamefont {Oriols}}\ and\ \bibinfo
  {editor} {\bibfnamefont {J.}~\bibnamefont {Mompart}},\ eds.,\ \href@noop {}
  {\emph {\bibinfo {title} {Applied Bohmian Mechanics: From Nanoscale Systems
  to Cosmology}}}\ (\bibinfo  {publisher} {Pan Stanford Publishing},\ \bibinfo
  {address} {Singapore},\ \bibinfo {year} {2012})\BibitemShut {NoStop}%
\bibitem [{\citenamefont {Sawada}\ \emph {et~al.}(2014)\citenamefont {Sawada},
  \citenamefont {Sato},\ and\ \citenamefont {Ishikawa}}]{Sawada2014PRA}%
  \BibitemOpen
  \bibfield  {author} {\bibinfo {author} {\bibfnamefont {R.}~\bibnamefont
  {Sawada}}, \bibinfo {author} {\bibfnamefont {T.}~\bibnamefont {Sato}}, \ and\
  \bibinfo {author} {\bibfnamefont {K.~L.}\ \bibnamefont {Ishikawa}},\ }\href
  {\doibase 10.1103/PhysRevA.90.023404} {\bibfield  {journal} {\bibinfo
  {journal} {Phys. Rev. A}\ }\textbf {\bibinfo {volume} {90}},\ \bibinfo
  {pages} {023404} (\bibinfo {year} {2014})}\BibitemShut {NoStop}%
\bibitem [{\citenamefont {Song}\ \emph {et~al.}(2012)\citenamefont {Song},
  \citenamefont {Guo}, \citenamefont {Li}, \citenamefont {Chen}, \citenamefont
  {Zeng},\ and\ \citenamefont {Yang}}]{Song2012PRA}%
  \BibitemOpen
  \bibfield  {author} {\bibinfo {author} {\bibfnamefont {Y.}~\bibnamefont
  {Song}}, \bibinfo {author} {\bibfnamefont {F.-M.}\ \bibnamefont {Guo}},
  \bibinfo {author} {\bibfnamefont {S.-Y.}\ \bibnamefont {Li}}, \bibinfo
  {author} {\bibfnamefont {J.-G.}\ \bibnamefont {Chen}}, \bibinfo {author}
  {\bibfnamefont {S.-L.}\ \bibnamefont {Zeng}}, \ and\ \bibinfo {author}
  {\bibfnamefont {Y.-J.}\ \bibnamefont {Yang}},\ }\href {\doibase
  10.1103/PhysRevA.86.033424} {\bibfield  {journal} {\bibinfo  {journal} {Phys.
  Rev. A}\ }\textbf {\bibinfo {volume} {86}},\ \bibinfo {pages} {033424}
  (\bibinfo {year} {2012})}\BibitemShut {NoStop}%
\bibitem [{\citenamefont {Takemoto}\ and\ \citenamefont
  {Becker}(2011)}]{Takemoto2011JCP}%
  \BibitemOpen
  \bibfield  {author} {\bibinfo {author} {\bibfnamefont {N.}~\bibnamefont
  {Takemoto}}\ and\ \bibinfo {author} {\bibfnamefont {A.}~\bibnamefont
  {Becker}},\ }\href {\doibase http://dx.doi.org/10.1063/1.3553178} {\bibfield
  {journal} {\bibinfo  {journal} {J. Chem. Phys.}\ }\textbf {\bibinfo {volume}
  {134}},\ \bibinfo {eid} {074309} (\bibinfo {year} {2011})}\BibitemShut
  {NoStop}%
\bibitem [{\citenamefont {Wu}\ \emph {et~al.}(2013{\natexlab{a}})\citenamefont
  {Wu}, \citenamefont {Augstein},\ and\ \citenamefont {Figueira~de
  Morisson~Faria}}]{Wu2013PRAa}%
  \BibitemOpen
  \bibfield  {author} {\bibinfo {author} {\bibfnamefont {J.}~\bibnamefont
  {Wu}}, \bibinfo {author} {\bibfnamefont {B.~B.}\ \bibnamefont {Augstein}}, \
  and\ \bibinfo {author} {\bibfnamefont {C.}~\bibnamefont {Figueira~de
  Morisson~Faria}},\ }\href {\doibase 10.1103/PhysRevA.88.023415} {\bibfield
  {journal} {\bibinfo  {journal} {Phys. Rev. A}\ }\textbf {\bibinfo {volume}
  {88}},\ \bibinfo {pages} {023415} (\bibinfo {year}
  {2013}{\natexlab{a}})}\BibitemShut {NoStop}%
\bibitem [{\citenamefont {Wu}\ \emph {et~al.}(2013{\natexlab{b}})\citenamefont
  {Wu}, \citenamefont {Augstein},\ and\ \citenamefont {Figueira~de
  Morisson~Faria}}]{Wu2013PRAb}%
  \BibitemOpen
  \bibfield  {author} {\bibinfo {author} {\bibfnamefont {J.}~\bibnamefont
  {Wu}}, \bibinfo {author} {\bibfnamefont {B.~B.}\ \bibnamefont {Augstein}}, \
  and\ \bibinfo {author} {\bibfnamefont {C.}~\bibnamefont {Figueira~de
  Morisson~Faria}},\ }\href {\doibase 10.1103/PhysRevA.88.063416} {\bibfield
  {journal} {\bibinfo  {journal} {Phys. Rev. A}\ }\textbf {\bibinfo {volume}
  {88}},\ \bibinfo {pages} {063416} (\bibinfo {year}
  {2013}{\natexlab{b}})}\BibitemShut {NoStop}%
\bibitem [{\citenamefont {Benseny}\ \emph {et~al.}(2014)\citenamefont
  {Benseny}, \citenamefont {Albareda}, \citenamefont {Sanz}, \citenamefont
  {Mompart},\ and\ \citenamefont {Oriols}}]{Benseny2014EPJD}%
  \BibitemOpen
  \bibfield  {author} {\bibinfo {author} {\bibfnamefont {A.}~\bibnamefont
  {Benseny}}, \bibinfo {author} {\bibfnamefont {G.}~\bibnamefont {Albareda}},
  \bibinfo {author} {\bibfnamefont {A.~S.}\ \bibnamefont {Sanz}}, \bibinfo
  {author} {\bibfnamefont {J.}~\bibnamefont {Mompart}}, \ and\ \bibinfo
  {author} {\bibfnamefont {X.}~\bibnamefont {Oriols}},\ }\href {\doibase
  10.1140/epjd/e2014-50222-4} {\bibfield  {journal} {\bibinfo  {journal} {Eur.
  Phys. J. D}\ }\textbf {\bibinfo {volume} {68}},\ \bibinfo {eid} {286}
  (\bibinfo {year} {2014}),\ 10.1140/epjd/e2014-50222-4}\BibitemShut {NoStop}%
\bibitem [{\citenamefont {Pindzola}\ and\ \citenamefont
  {Robicheaux}(1998{\natexlab{a}})}]{Pindzola1998JPB}%
  \BibitemOpen
  \bibfield  {author} {\bibinfo {author} {\bibfnamefont {M.~S.}\ \bibnamefont
  {Pindzola}}\ and\ \bibinfo {author} {\bibfnamefont {F.}~\bibnamefont
  {Robicheaux}},\ }\href {http://stacks.iop.org/0953-4075/31/i=19/a=010}
  {\bibfield  {journal} {\bibinfo  {journal} {J. Phys. B}\ }\textbf {\bibinfo
  {volume} {31}},\ \bibinfo {pages} {L823} (\bibinfo {year}
  {1998}{\natexlab{a}})}\BibitemShut {NoStop}%
\bibitem [{\citenamefont {Parker}\ \emph {et~al.}(2001)\citenamefont {Parker},
  \citenamefont {Moore}, \citenamefont {Meharg}, \citenamefont {Dundas},\ and\
  \citenamefont {Taylor}}]{Parker2001JPB}%
  \BibitemOpen
  \bibfield  {author} {\bibinfo {author} {\bibfnamefont {J.~S.}\ \bibnamefont
  {Parker}}, \bibinfo {author} {\bibfnamefont {L.~R.}\ \bibnamefont {Moore}},
  \bibinfo {author} {\bibfnamefont {K.~J.}\ \bibnamefont {Meharg}}, \bibinfo
  {author} {\bibfnamefont {D.}~\bibnamefont {Dundas}}, \ and\ \bibinfo {author}
  {\bibfnamefont {K.~T.}\ \bibnamefont {Taylor}},\ }\href
  {http://stacks.iop.org/0953-4075/34/i=3/a=103} {\bibfield  {journal}
  {\bibinfo  {journal} {J. Phys. B}\ }\textbf {\bibinfo {volume} {34}},\
  \bibinfo {pages} {L69} (\bibinfo {year} {2001})}\BibitemShut {NoStop}%
\bibitem [{\citenamefont {Colgan}\ \emph {et~al.}(2001)\citenamefont {Colgan},
  \citenamefont {Pindzola},\ and\ \citenamefont {Robicheaux}}]{Colgan2001JPB}%
  \BibitemOpen
  \bibfield  {author} {\bibinfo {author} {\bibfnamefont {J.}~\bibnamefont
  {Colgan}}, \bibinfo {author} {\bibfnamefont {M.~S.}\ \bibnamefont
  {Pindzola}}, \ and\ \bibinfo {author} {\bibfnamefont {F.}~\bibnamefont
  {Robicheaux}},\ }\href {http://stacks.iop.org/0953-4075/34/i=15/a=101}
  {\bibfield  {journal} {\bibinfo  {journal} {J. Phys. B}\ }\textbf {\bibinfo
  {volume} {34}},\ \bibinfo {pages} {L457} (\bibinfo {year}
  {2001})}\BibitemShut {NoStop}%
\bibitem [{\citenamefont {Colgan}\ and\ \citenamefont
  {Pindzola}(2002)}]{Colgan2002PRL}%
  \BibitemOpen
  \bibfield  {author} {\bibinfo {author} {\bibfnamefont {J.}~\bibnamefont
  {Colgan}}\ and\ \bibinfo {author} {\bibfnamefont {M.~S.}\ \bibnamefont
  {Pindzola}},\ }\href {\doibase 10.1103/PhysRevLett.88.173002} {\bibfield
  {journal} {\bibinfo  {journal} {Phys. Rev. Lett.}\ }\textbf {\bibinfo
  {volume} {88}},\ \bibinfo {pages} {173002} (\bibinfo {year}
  {2002})}\BibitemShut {NoStop}%
\bibitem [{\citenamefont {Laulan}\ and\ \citenamefont
  {Bachau}(2003)}]{Laulan2003PRA}%
  \BibitemOpen
  \bibfield  {author} {\bibinfo {author} {\bibfnamefont {S.}~\bibnamefont
  {Laulan}}\ and\ \bibinfo {author} {\bibfnamefont {H.}~\bibnamefont
  {Bachau}},\ }\href {\doibase 10.1103/PhysRevA.68.013409} {\bibfield
  {journal} {\bibinfo  {journal} {Phys. Rev. A}\ }\textbf {\bibinfo {volume}
  {68}},\ \bibinfo {pages} {013409} (\bibinfo {year} {2003})}\BibitemShut
  {NoStop}%
\bibitem [{\citenamefont {Colgan}\ and\ \citenamefont
  {Pindzola}(2004)}]{Colgan2004JPBa}%
  \BibitemOpen
  \bibfield  {author} {\bibinfo {author} {\bibfnamefont {J.}~\bibnamefont
  {Colgan}}\ and\ \bibinfo {author} {\bibfnamefont {M.~S.}\ \bibnamefont
  {Pindzola}},\ }\href {http://stacks.iop.org/0953-4075/37/i=6/a=001}
  {\bibfield  {journal} {\bibinfo  {journal} {J. Phys. B}\ }\textbf {\bibinfo
  {volume} {37}},\ \bibinfo {pages} {1153} (\bibinfo {year}
  {2004})}\BibitemShut {NoStop}%
\bibitem [{\citenamefont {Ishikawa}\ and\ \citenamefont
  {Midorikawa}(2005)}]{Ishikawa2005PRA}%
  \BibitemOpen
  \bibfield  {author} {\bibinfo {author} {\bibfnamefont {K.~L.}\ \bibnamefont
  {Ishikawa}}\ and\ \bibinfo {author} {\bibfnamefont {K.}~\bibnamefont
  {Midorikawa}},\ }\href {\doibase 10.1103/PhysRevA.72.013407} {\bibfield
  {journal} {\bibinfo  {journal} {Phys. Rev. A}\ }\textbf {\bibinfo {volume}
  {72}},\ \bibinfo {pages} {013407} (\bibinfo {year} {2005})}\BibitemShut
  {NoStop}%
\bibitem [{\citenamefont {Foumouo}\ \emph {et~al.}(2006)\citenamefont
  {Foumouo}, \citenamefont {Kamta}, \citenamefont {Edah},\ and\ \citenamefont
  {Piraux}}]{Foumouo2006PRA}%
  \BibitemOpen
  \bibfield  {author} {\bibinfo {author} {\bibfnamefont {E.}~\bibnamefont
  {Foumouo}}, \bibinfo {author} {\bibfnamefont {G.~L.}\ \bibnamefont {Kamta}},
  \bibinfo {author} {\bibfnamefont {G.}~\bibnamefont {Edah}}, \ and\ \bibinfo
  {author} {\bibfnamefont {B.}~\bibnamefont {Piraux}},\ }\href {\doibase
  10.1103/PhysRevA.74.063409} {\bibfield  {journal} {\bibinfo  {journal} {Phys.
  Rev. A}\ }\textbf {\bibinfo {volume} {74}},\ \bibinfo {pages} {063409}
  (\bibinfo {year} {2006})}\BibitemShut {NoStop}%
\bibitem [{\citenamefont {Nikolopoulos}\ and\ \citenamefont
  {Lambropoulos}(2007)}]{Nikolopoulos2007JPB}%
  \BibitemOpen
  \bibfield  {author} {\bibinfo {author} {\bibfnamefont {L.~A.~A.}\
  \bibnamefont {Nikolopoulos}}\ and\ \bibinfo {author} {\bibfnamefont
  {P.}~\bibnamefont {Lambropoulos}},\ }\href
  {http://stacks.iop.org/0953-4075/40/i=7/a=004} {\bibfield  {journal}
  {\bibinfo  {journal} {J. Phys. B}\ }\textbf {\bibinfo {volume} {40}},\
  \bibinfo {pages} {1347} (\bibinfo {year} {2007})}\BibitemShut {NoStop}%
\bibitem [{\citenamefont {Lambropoulos}\ and\ \citenamefont
  {Nikolopoulos}(2008)}]{Lambropoulos2008NJP}%
  \BibitemOpen
  \bibfield  {author} {\bibinfo {author} {\bibfnamefont {P.}~\bibnamefont
  {Lambropoulos}}\ and\ \bibinfo {author} {\bibfnamefont {L.~A.~A.}\
  \bibnamefont {Nikolopoulos}},\ }\href
  {http://stacks.iop.org/1367-2630/10/i=2/a=025012} {\bibfield  {journal}
  {\bibinfo  {journal} {New J. Phys.}\ }\textbf {\bibinfo {volume} {10}},\
  \bibinfo {pages} {025012} (\bibinfo {year} {2008})}\BibitemShut {NoStop}%
\bibitem [{\citenamefont {Foumouo}\ \emph {et~al.}(2008)\citenamefont
  {Foumouo}, \citenamefont {Antoine}, \citenamefont {Bachau},\ and\
  \citenamefont {Piraux}}]{Foumouo2008NJP}%
  \BibitemOpen
  \bibfield  {author} {\bibinfo {author} {\bibfnamefont {E.}~\bibnamefont
  {Foumouo}}, \bibinfo {author} {\bibfnamefont {P.}~\bibnamefont {Antoine}},
  \bibinfo {author} {\bibfnamefont {H.}~\bibnamefont {Bachau}}, \ and\ \bibinfo
  {author} {\bibfnamefont {B.}~\bibnamefont {Piraux}},\ }\href
  {http://stacks.iop.org/1367-2630/10/i=2/a=025017} {\bibfield  {journal}
  {\bibinfo  {journal} {New J. Phys.}\ }\textbf {\bibinfo {volume} {10}},\
  \bibinfo {pages} {025017} (\bibinfo {year} {2008})}\BibitemShut {NoStop}%
\bibitem [{\citenamefont {Foumouo}\ \emph {et~al.}(2010)\citenamefont
  {Foumouo}, \citenamefont {Hamido}, \citenamefont {Antoine}, \citenamefont
  {Piraux}, \citenamefont {Bachau},\ and\ \citenamefont
  {Shakeshaft}}]{Foumouo2010JPB}%
  \BibitemOpen
  \bibfield  {author} {\bibinfo {author} {\bibfnamefont {E.}~\bibnamefont
  {Foumouo}}, \bibinfo {author} {\bibfnamefont {A.}~\bibnamefont {Hamido}},
  \bibinfo {author} {\bibfnamefont {P.}~\bibnamefont {Antoine}}, \bibinfo
  {author} {\bibfnamefont {B.}~\bibnamefont {Piraux}}, \bibinfo {author}
  {\bibfnamefont {H.}~\bibnamefont {Bachau}}, \ and\ \bibinfo {author}
  {\bibfnamefont {R.}~\bibnamefont {Shakeshaft}},\ }\href
  {http://stacks.iop.org/0953-4075/43/i=9/a=091001} {\bibfield  {journal}
  {\bibinfo  {journal} {J. Phys. B}\ }\textbf {\bibinfo {volume} {43}},\
  \bibinfo {pages} {091001} (\bibinfo {year} {2010})}\BibitemShut {NoStop}%
\bibitem [{\citenamefont {Feist}\ \emph {et~al.}(2008)\citenamefont {Feist},
  \citenamefont {Nagele}, \citenamefont {Pazourek}, \citenamefont {Persson},
  \citenamefont {Schneider}, \citenamefont {Collins},\ and\ \citenamefont
  {Burgd\"orfer}}]{Feist2008PRA}%
  \BibitemOpen
  \bibfield  {author} {\bibinfo {author} {\bibfnamefont {J.}~\bibnamefont
  {Feist}}, \bibinfo {author} {\bibfnamefont {S.}~\bibnamefont {Nagele}},
  \bibinfo {author} {\bibfnamefont {R.}~\bibnamefont {Pazourek}}, \bibinfo
  {author} {\bibfnamefont {E.}~\bibnamefont {Persson}}, \bibinfo {author}
  {\bibfnamefont {B.~I.}\ \bibnamefont {Schneider}}, \bibinfo {author}
  {\bibfnamefont {L.~A.}\ \bibnamefont {Collins}}, \ and\ \bibinfo {author}
  {\bibfnamefont {J.}~\bibnamefont {Burgd\"orfer}},\ }\href {\doibase
  10.1103/PhysRevA.77.043420} {\bibfield  {journal} {\bibinfo  {journal} {Phys.
  Rev. A}\ }\textbf {\bibinfo {volume} {77}},\ \bibinfo {pages} {043420}
  (\bibinfo {year} {2008})}\BibitemShut {NoStop}%
\bibitem [{\citenamefont {Palacios}\ \emph {et~al.}(2009)\citenamefont
  {Palacios}, \citenamefont {Rescigno},\ and\ \citenamefont
  {McCurdy}}]{Palacios2009PRA}%
  \BibitemOpen
  \bibfield  {author} {\bibinfo {author} {\bibfnamefont {A.}~\bibnamefont
  {Palacios}}, \bibinfo {author} {\bibfnamefont {T.~N.}\ \bibnamefont
  {Rescigno}}, \ and\ \bibinfo {author} {\bibfnamefont {C.~W.}\ \bibnamefont
  {McCurdy}},\ }\href {\doibase 10.1103/PhysRevA.79.033402} {\bibfield
  {journal} {\bibinfo  {journal} {Phys. Rev. A}\ }\textbf {\bibinfo {volume}
  {79}},\ \bibinfo {pages} {033402} (\bibinfo {year} {2009})}\BibitemShut
  {NoStop}%
\bibitem [{\citenamefont {Feist}\ \emph {et~al.}(2009)\citenamefont {Feist},
  \citenamefont {Nagele}, \citenamefont {Pazourek}, \citenamefont {Persson},
  \citenamefont {Schneider}, \citenamefont {Collins},\ and\ \citenamefont
  {Burgd\"orfer}}]{Feist2009PRL}%
  \BibitemOpen
  \bibfield  {author} {\bibinfo {author} {\bibfnamefont {J.}~\bibnamefont
  {Feist}}, \bibinfo {author} {\bibfnamefont {S.}~\bibnamefont {Nagele}},
  \bibinfo {author} {\bibfnamefont {R.}~\bibnamefont {Pazourek}}, \bibinfo
  {author} {\bibfnamefont {E.}~\bibnamefont {Persson}}, \bibinfo {author}
  {\bibfnamefont {B.~I.}\ \bibnamefont {Schneider}}, \bibinfo {author}
  {\bibfnamefont {L.~A.}\ \bibnamefont {Collins}}, \ and\ \bibinfo {author}
  {\bibfnamefont {J.}~\bibnamefont {Burgd\"orfer}},\ }\href {\doibase
  10.1103/PhysRevLett.103.063002} {\bibfield  {journal} {\bibinfo  {journal}
  {Phys. Rev. Lett.}\ }\textbf {\bibinfo {volume} {103}},\ \bibinfo {pages}
  {063002} (\bibinfo {year} {2009})}\BibitemShut {NoStop}%
\bibitem [{\citenamefont {Lee}\ \emph {et~al.}(2009)\citenamefont {Lee},
  \citenamefont {Pindzola},\ and\ \citenamefont {Robicheaux}}]{Lee2009PRA}%
  \BibitemOpen
  \bibfield  {author} {\bibinfo {author} {\bibfnamefont {T.-G.}\ \bibnamefont
  {Lee}}, \bibinfo {author} {\bibfnamefont {M.~S.}\ \bibnamefont {Pindzola}}, \
  and\ \bibinfo {author} {\bibfnamefont {F.}~\bibnamefont {Robicheaux}},\
  }\href {\doibase 10.1103/PhysRevA.79.053420} {\bibfield  {journal} {\bibinfo
  {journal} {Phys. Rev. A}\ }\textbf {\bibinfo {volume} {79}},\ \bibinfo
  {pages} {053420} (\bibinfo {year} {2009})}\BibitemShut {NoStop}%
\bibitem [{\citenamefont {Pazourek}\ \emph {et~al.}(2011)\citenamefont
  {Pazourek}, \citenamefont {Feist}, \citenamefont {Nagele}, \citenamefont
  {Persson}, \citenamefont {Schneider}, \citenamefont {Collins},\ and\
  \citenamefont {Burgd\"orfer}}]{Pazourek2011PRA}%
  \BibitemOpen
  \bibfield  {author} {\bibinfo {author} {\bibfnamefont {R.}~\bibnamefont
  {Pazourek}}, \bibinfo {author} {\bibfnamefont {J.}~\bibnamefont {Feist}},
  \bibinfo {author} {\bibfnamefont {S.}~\bibnamefont {Nagele}}, \bibinfo
  {author} {\bibfnamefont {E.}~\bibnamefont {Persson}}, \bibinfo {author}
  {\bibfnamefont {B.~I.}\ \bibnamefont {Schneider}}, \bibinfo {author}
  {\bibfnamefont {L.~A.}\ \bibnamefont {Collins}}, \ and\ \bibinfo {author}
  {\bibfnamefont {J.}~\bibnamefont {Burgd\"orfer}},\ }\href {\doibase
  10.1103/PhysRevA.83.053418} {\bibfield  {journal} {\bibinfo  {journal} {Phys.
  Rev. A}\ }\textbf {\bibinfo {volume} {83}},\ \bibinfo {pages} {053418}
  (\bibinfo {year} {2011})}\BibitemShut {NoStop}%
\bibitem [{\citenamefont {Zhang}\ \emph {et~al.}(2011)\citenamefont {Zhang},
  \citenamefont {Peng}, \citenamefont {Xu}, \citenamefont {Starace},
  \citenamefont {Morishita},\ and\ \citenamefont {Gong}}]{Zhang2011PRA}%
  \BibitemOpen
  \bibfield  {author} {\bibinfo {author} {\bibfnamefont {Z.}~\bibnamefont
  {Zhang}}, \bibinfo {author} {\bibfnamefont {L.-Y.}\ \bibnamefont {Peng}},
  \bibinfo {author} {\bibfnamefont {M.-H.}\ \bibnamefont {Xu}}, \bibinfo
  {author} {\bibfnamefont {A.~F.}\ \bibnamefont {Starace}}, \bibinfo {author}
  {\bibfnamefont {T.}~\bibnamefont {Morishita}}, \ and\ \bibinfo {author}
  {\bibfnamefont {Q.}~\bibnamefont {Gong}},\ }\href {\doibase
  10.1103/PhysRevA.84.043409} {\bibfield  {journal} {\bibinfo  {journal} {Phys.
  Rev. A}\ }\textbf {\bibinfo {volume} {84}},\ \bibinfo {pages} {043409}
  (\bibinfo {year} {2011})}\BibitemShut {NoStop}%
\bibitem [{\citenamefont {Ishikawa}\ and\ \citenamefont
  {Ueda}(2012)}]{Ishikawa2012PRL}%
  \BibitemOpen
  \bibfield  {author} {\bibinfo {author} {\bibfnamefont {K.~L.}\ \bibnamefont
  {Ishikawa}}\ and\ \bibinfo {author} {\bibfnamefont {K.}~\bibnamefont
  {Ueda}},\ }\href {\doibase 10.1103/PhysRevLett.108.033003} {\bibfield
  {journal} {\bibinfo  {journal} {Phys. Rev. Lett.}\ }\textbf {\bibinfo
  {volume} {108}},\ \bibinfo {pages} {033003} (\bibinfo {year}
  {2012})}\BibitemShut {NoStop}%
\bibitem [{\citenamefont {Sukiasyan}\ \emph {et~al.}(2012)\citenamefont
  {Sukiasyan}, \citenamefont {Ishikawa},\ and\ \citenamefont
  {Ivanov}}]{Suren2012PRA}%
  \BibitemOpen
  \bibfield  {author} {\bibinfo {author} {\bibfnamefont {S.}~\bibnamefont
  {Sukiasyan}}, \bibinfo {author} {\bibfnamefont {K.~L.}\ \bibnamefont
  {Ishikawa}}, \ and\ \bibinfo {author} {\bibfnamefont {M.}~\bibnamefont
  {Ivanov}},\ }\href {\doibase 10.1103/PhysRevA.86.033423} {\bibfield
  {journal} {\bibinfo  {journal} {Phys. Rev. A}\ }\textbf {\bibinfo {volume}
  {86}},\ \bibinfo {pages} {033423} (\bibinfo {year} {2012})}\BibitemShut
  {NoStop}%
\bibitem [{\citenamefont {Schultze}\ \emph {et~al.}(2010)\citenamefont
  {Schultze}, \citenamefont {Fie{\ss}}, \citenamefont {Karpowicz},
  \citenamefont {Gagnon}, \citenamefont {Korbman}, \citenamefont {Hofstetter},
  \citenamefont {Neppl}, \citenamefont {Cavalieri}, \citenamefont {Komninos},
  \citenamefont {Mercouris}, \citenamefont {Nicolaides}, \citenamefont
  {Pazourek}, \citenamefont {Nagele}, \citenamefont {Feist}, \citenamefont
  {Burgd\"orfer}, \citenamefont {Azzeer}, \citenamefont {Ernstorfer},
  \citenamefont {Kienberger}, \citenamefont {Kleineberg}, \citenamefont
  {Goulielmakis}, \citenamefont {Krausz},\ and\ \citenamefont
  {Yakovlev}}]{Schultze2010Science}%
  \BibitemOpen
  \bibfield  {author} {\bibinfo {author} {\bibfnamefont {M.}~\bibnamefont
  {Schultze}}, \bibinfo {author} {\bibfnamefont {M.}~\bibnamefont {Fie{\ss}}},
  \bibinfo {author} {\bibfnamefont {N.}~\bibnamefont {Karpowicz}}, \bibinfo
  {author} {\bibfnamefont {J.}~\bibnamefont {Gagnon}}, \bibinfo {author}
  {\bibfnamefont {M.}~\bibnamefont {Korbman}}, \bibinfo {author} {\bibfnamefont
  {M.}~\bibnamefont {Hofstetter}}, \bibinfo {author} {\bibfnamefont
  {S.}~\bibnamefont {Neppl}}, \bibinfo {author} {\bibfnamefont {A.~L.}\
  \bibnamefont {Cavalieri}}, \bibinfo {author} {\bibfnamefont {Y.}~\bibnamefont
  {Komninos}}, \bibinfo {author} {\bibfnamefont {T.}~\bibnamefont {Mercouris}},
  \bibinfo {author} {\bibfnamefont {C.~A.}\ \bibnamefont {Nicolaides}},
  \bibinfo {author} {\bibfnamefont {R.}~\bibnamefont {Pazourek}}, \bibinfo
  {author} {\bibfnamefont {S.}~\bibnamefont {Nagele}}, \bibinfo {author}
  {\bibfnamefont {J.}~\bibnamefont {Feist}}, \bibinfo {author} {\bibfnamefont
  {J.}~\bibnamefont {Burgd\"orfer}}, \bibinfo {author} {\bibfnamefont {A.~M.}\
  \bibnamefont {Azzeer}}, \bibinfo {author} {\bibfnamefont {R.}~\bibnamefont
  {Ernstorfer}}, \bibinfo {author} {\bibfnamefont {R.}~\bibnamefont
  {Kienberger}}, \bibinfo {author} {\bibfnamefont {U.}~\bibnamefont
  {Kleineberg}}, \bibinfo {author} {\bibfnamefont {E.}~\bibnamefont
  {Goulielmakis}}, \bibinfo {author} {\bibfnamefont {F.}~\bibnamefont
  {Krausz}}, \ and\ \bibinfo {author} {\bibfnamefont {V.~S.}\ \bibnamefont
  {Yakovlev}},\ }\href {\doibase 10.1126/science.1189401} {\bibfield  {journal}
  {\bibinfo  {journal} {Science}\ }\textbf {\bibinfo {volume} {328}},\ \bibinfo
  {pages} {1658} (\bibinfo {year} {2010})},\ \Eprint
  {http://arxiv.org/abs/http://www.sciencemag.org/content/328/5986/1658.full.pdf}
  {http://www.sciencemag.org/content/328/5986/1658.full.pdf} \BibitemShut
  {NoStop}%
\bibitem [{\citenamefont {Nagele}\ \emph {et~al.}(2012)\citenamefont {Nagele},
  \citenamefont {Pazourek}, \citenamefont {Feist},\ and\ \citenamefont
  {Burgd\"orfer}}]{Nagele2012PRA}%
  \BibitemOpen
  \bibfield  {author} {\bibinfo {author} {\bibfnamefont {S.}~\bibnamefont
  {Nagele}}, \bibinfo {author} {\bibfnamefont {R.}~\bibnamefont {Pazourek}},
  \bibinfo {author} {\bibfnamefont {J.}~\bibnamefont {Feist}}, \ and\ \bibinfo
  {author} {\bibfnamefont {J.}~\bibnamefont {Burgd\"orfer}},\ }\href {\doibase
  10.1103/PhysRevA.85.033401} {\bibfield  {journal} {\bibinfo  {journal} {Phys.
  Rev. A}\ }\textbf {\bibinfo {volume} {85}},\ \bibinfo {pages} {033401}
  (\bibinfo {year} {2012})}\BibitemShut {NoStop}%
\bibitem [{\citenamefont {Feist}\ \emph {et~al.}(2011)\citenamefont {Feist},
  \citenamefont {Nagele}, \citenamefont {Ticknor}, \citenamefont {Schneider},
  \citenamefont {Collins},\ and\ \citenamefont {Burgd\"orfer}}]{Feist2011PRL}%
  \BibitemOpen
  \bibfield  {author} {\bibinfo {author} {\bibfnamefont {J.}~\bibnamefont
  {Feist}}, \bibinfo {author} {\bibfnamefont {S.}~\bibnamefont {Nagele}},
  \bibinfo {author} {\bibfnamefont {C.}~\bibnamefont {Ticknor}}, \bibinfo
  {author} {\bibfnamefont {B.~I.}\ \bibnamefont {Schneider}}, \bibinfo {author}
  {\bibfnamefont {L.~A.}\ \bibnamefont {Collins}}, \ and\ \bibinfo {author}
  {\bibfnamefont {J.}~\bibnamefont {Burgd\"orfer}},\ }\href {\doibase
  10.1103/PhysRevLett.107.093005} {\bibfield  {journal} {\bibinfo  {journal}
  {Phys. Rev. Lett.}\ }\textbf {\bibinfo {volume} {107}},\ \bibinfo {pages}
  {093005} (\bibinfo {year} {2011})}\BibitemShut {NoStop}%
\bibitem [{\citenamefont {Ott}\ \emph {et~al.}(2014)\citenamefont {Ott},
  \citenamefont {Kaldun}, \citenamefont {Argenti}, \citenamefont {Raith},
  \citenamefont {Meyer}, \citenamefont {Laux}, \citenamefont {Zhang},
  \citenamefont {Bl{\"a}ttermann}, \citenamefont {Hagstotz}, \citenamefont
  {Ding}, \citenamefont {Heck}, \citenamefont {Madro{\~n}ero}, \citenamefont
  {Mart\'{\i}n},\ and\ \citenamefont {Pfeifer}}]{Ott2014Nature}%
  \BibitemOpen
  \bibfield  {author} {\bibinfo {author} {\bibfnamefont {C.}~\bibnamefont
  {Ott}}, \bibinfo {author} {\bibfnamefont {A.}~\bibnamefont {Kaldun}},
  \bibinfo {author} {\bibfnamefont {L.}~\bibnamefont {Argenti}}, \bibinfo
  {author} {\bibfnamefont {P.}~\bibnamefont {Raith}}, \bibinfo {author}
  {\bibfnamefont {K.}~\bibnamefont {Meyer}}, \bibinfo {author} {\bibfnamefont
  {M.}~\bibnamefont {Laux}}, \bibinfo {author} {\bibfnamefont {Y.}~\bibnamefont
  {Zhang}}, \bibinfo {author} {\bibfnamefont {A.}~\bibnamefont
  {Bl{\"a}ttermann}}, \bibinfo {author} {\bibfnamefont {S.}~\bibnamefont
  {Hagstotz}}, \bibinfo {author} {\bibfnamefont {T.}~\bibnamefont {Ding}},
  \bibinfo {author} {\bibfnamefont {R.}~\bibnamefont {Heck}}, \bibinfo {author}
  {\bibfnamefont {J.}~\bibnamefont {Madro{\~n}ero}}, \bibinfo {author}
  {\bibfnamefont {F.}~\bibnamefont {Mart\'{\i}n}}, \ and\ \bibinfo {author}
  {\bibfnamefont {T.}~\bibnamefont {Pfeifer}},\ }\href@noop {} {\bibfield
  {journal} {\bibinfo  {journal} {Nature}\ }\textbf {\bibinfo {volume} {516}},\
  \bibinfo {pages} {374} (\bibinfo {year} {2014})}\BibitemShut {NoStop}%
\bibitem [{\citenamefont {Parker}\ \emph {et~al.}(1996)\citenamefont {Parker},
  \citenamefont {Taylor}, \citenamefont {Clark},\ and\ \citenamefont
  {Blodgett-Ford}}]{Parker1996JPB}%
  \BibitemOpen
  \bibfield  {author} {\bibinfo {author} {\bibfnamefont {J.}~\bibnamefont
  {Parker}}, \bibinfo {author} {\bibfnamefont {K.~T.}\ \bibnamefont {Taylor}},
  \bibinfo {author} {\bibfnamefont {C.~W.}\ \bibnamefont {Clark}}, \ and\
  \bibinfo {author} {\bibfnamefont {S.}~\bibnamefont {Blodgett-Ford}},\ }\href
  {http://stacks.iop.org/0953-4075/29/i=2/a=002} {\bibfield  {journal}
  {\bibinfo  {journal} {J. Phys. B}\ }\textbf {\bibinfo {volume} {29}},\
  \bibinfo {pages} {L33} (\bibinfo {year} {1996})}\BibitemShut {NoStop}%
\bibitem [{\citenamefont {Parker}\ \emph {et~al.}(2000)\citenamefont {Parker},
  \citenamefont {Moore}, \citenamefont {Dundas},\ and\ \citenamefont
  {Taylor}}]{Parker2000JPB}%
  \BibitemOpen
  \bibfield  {author} {\bibinfo {author} {\bibfnamefont {J.~S.}\ \bibnamefont
  {Parker}}, \bibinfo {author} {\bibfnamefont {L.~R.}\ \bibnamefont {Moore}},
  \bibinfo {author} {\bibfnamefont {D.}~\bibnamefont {Dundas}}, \ and\ \bibinfo
  {author} {\bibfnamefont {K.~T.}\ \bibnamefont {Taylor}},\ }\href
  {http://stacks.iop.org/0953-4075/33/i=20/a=106} {\bibfield  {journal}
  {\bibinfo  {journal} {J. Phys. B}\ }\textbf {\bibinfo {volume} {33}},\
  \bibinfo {pages} {L691} (\bibinfo {year} {2000})}\BibitemShut {NoStop}%
\bibitem [{\citenamefont {Parker}\ \emph {et~al.}(2006)\citenamefont {Parker},
  \citenamefont {Doherty}, \citenamefont {Taylor}, \citenamefont {Schultz},
  \citenamefont {Blaga},\ and\ \citenamefont {DiMauro}}]{Parker2006PRL}%
  \BibitemOpen
  \bibfield  {author} {\bibinfo {author} {\bibfnamefont {J.~S.}\ \bibnamefont
  {Parker}}, \bibinfo {author} {\bibfnamefont {B.~J.~S.}\ \bibnamefont
  {Doherty}}, \bibinfo {author} {\bibfnamefont {K.~T.}\ \bibnamefont {Taylor}},
  \bibinfo {author} {\bibfnamefont {K.~D.}\ \bibnamefont {Schultz}}, \bibinfo
  {author} {\bibfnamefont {C.~I.}\ \bibnamefont {Blaga}}, \ and\ \bibinfo
  {author} {\bibfnamefont {L.~F.}\ \bibnamefont {DiMauro}},\ }\href {\doibase
  10.1103/PhysRevLett.96.133001} {\bibfield  {journal} {\bibinfo  {journal}
  {Phys. Rev. Lett.}\ }\textbf {\bibinfo {volume} {96}},\ \bibinfo {pages}
  {133001} (\bibinfo {year} {2006})}\BibitemShut {NoStop}%
\bibitem [{\citenamefont {Armstrong}\ \emph {et~al.}(2011)\citenamefont
  {Armstrong}, \citenamefont {Parker},\ and\ \citenamefont
  {Taylor}}]{Armstrong2011NJP}%
  \BibitemOpen
  \bibfield  {author} {\bibinfo {author} {\bibfnamefont {G.~S.~J.}\
  \bibnamefont {Armstrong}}, \bibinfo {author} {\bibfnamefont {J.~S.}\
  \bibnamefont {Parker}}, \ and\ \bibinfo {author} {\bibfnamefont {K.~T.}\
  \bibnamefont {Taylor}},\ }\href
  {http://stacks.iop.org/1367-2630/13/i=1/a=013024} {\bibfield  {journal}
  {\bibinfo  {journal} {New J. Phys.}\ }\textbf {\bibinfo {volume} {13}},\
  \bibinfo {pages} {013024} (\bibinfo {year} {2011})}\BibitemShut {NoStop}%
\bibitem [{\citenamefont {Ngoko~Djiokap}\ and\ \citenamefont
  {Starace}(2011)}]{NgokoDjiokap2011PRA}%
  \BibitemOpen
  \bibfield  {author} {\bibinfo {author} {\bibfnamefont {J.~M.}\ \bibnamefont
  {Ngoko~Djiokap}}\ and\ \bibinfo {author} {\bibfnamefont {A.~F.}\ \bibnamefont
  {Starace}},\ }\href {\doibase 10.1103/PhysRevA.84.013404} {\bibfield
  {journal} {\bibinfo  {journal} {Phys. Rev. A}\ }\textbf {\bibinfo {volume}
  {84}},\ \bibinfo {pages} {013404} (\bibinfo {year} {2011})}\BibitemShut
  {NoStop}%
\bibitem [{\citenamefont {Smyth}\ \emph {et~al.}(1998)\citenamefont {Smyth},
  \citenamefont {Parker},\ and\ \citenamefont {Taylor}}]{Smyth1998CPC}%
  \BibitemOpen
  \bibfield  {author} {\bibinfo {author} {\bibfnamefont {E.~S.}\ \bibnamefont
  {Smyth}}, \bibinfo {author} {\bibfnamefont {J.~S.}\ \bibnamefont {Parker}}, \
  and\ \bibinfo {author} {\bibfnamefont {K.}~\bibnamefont {Taylor}},\ }\href
  {\doibase http://dx.doi.org/10.1016/S0010-4655(98)00083-6} {\bibfield
  {journal} {\bibinfo  {journal} {Comput. Phys. Commun.}\ }\textbf {\bibinfo
  {volume} {114}},\ \bibinfo {pages} {1} (\bibinfo {year} {1998})}\BibitemShut
  {NoStop}%
\bibitem [{\citenamefont {Pindzola}\ and\ \citenamefont
  {Robicheaux}(1998{\natexlab{b}})}]{Pindzola1998PRA}%
  \BibitemOpen
  \bibfield  {author} {\bibinfo {author} {\bibfnamefont {M.~S.}\ \bibnamefont
  {Pindzola}}\ and\ \bibinfo {author} {\bibfnamefont {F.}~\bibnamefont
  {Robicheaux}},\ }\href {\doibase 10.1103/PhysRevA.57.318} {\bibfield
  {journal} {\bibinfo  {journal} {Phys. Rev. A}\ }\textbf {\bibinfo {volume}
  {57}},\ \bibinfo {pages} {318} (\bibinfo {year}
  {1998}{\natexlab{b}})}\BibitemShut {NoStop}%
\bibitem [{\citenamefont {Colgan}\ \emph
  {et~al.}(2004{\natexlab{a}})\citenamefont {Colgan}, \citenamefont
  {Pindzola},\ and\ \citenamefont {Robicheaux}}]{Colgan2004JPB}%
  \BibitemOpen
  \bibfield  {author} {\bibinfo {author} {\bibfnamefont {J.}~\bibnamefont
  {Colgan}}, \bibinfo {author} {\bibfnamefont {M.~S.}\ \bibnamefont
  {Pindzola}}, \ and\ \bibinfo {author} {\bibfnamefont {F.}~\bibnamefont
  {Robicheaux}},\ }\href {http://stacks.iop.org/0953-4075/37/i=23/a=L01}
  {\bibfield  {journal} {\bibinfo  {journal} {J. Phys. B}\ }\textbf {\bibinfo
  {volume} {37}},\ \bibinfo {pages} {L377} (\bibinfo {year}
  {2004}{\natexlab{a}})}\BibitemShut {NoStop}%
\bibitem [{\citenamefont {Pindzola}\ \emph {et~al.}(2007)\citenamefont
  {Pindzola}, \citenamefont {Robicheaux}, \citenamefont {Loch}, \citenamefont
  {Berengut}, \citenamefont {Topcu}, \citenamefont {Colgan}, \citenamefont
  {Foster}, \citenamefont {Griffin}, \citenamefont {Ballance}, \citenamefont
  {Schultz}, \citenamefont {Minami}, \citenamefont {Badnell}, \citenamefont
  {Witthoeft}, \citenamefont {Plante}, \citenamefont {Mitnik}, \citenamefont
  {Ludlow},\ and\ \citenamefont {Kleiman}}]{Pindzola2007JPB}%
  \BibitemOpen
  \bibfield  {author} {\bibinfo {author} {\bibfnamefont {M.~S.}\ \bibnamefont
  {Pindzola}}, \bibinfo {author} {\bibfnamefont {F.}~\bibnamefont
  {Robicheaux}}, \bibinfo {author} {\bibfnamefont {S.~D.}\ \bibnamefont
  {Loch}}, \bibinfo {author} {\bibfnamefont {J.~C.}\ \bibnamefont {Berengut}},
  \bibinfo {author} {\bibfnamefont {T.}~\bibnamefont {Topcu}}, \bibinfo
  {author} {\bibfnamefont {J.}~\bibnamefont {Colgan}}, \bibinfo {author}
  {\bibfnamefont {M.}~\bibnamefont {Foster}}, \bibinfo {author} {\bibfnamefont
  {D.~C.}\ \bibnamefont {Griffin}}, \bibinfo {author} {\bibfnamefont {C.~P.}\
  \bibnamefont {Ballance}}, \bibinfo {author} {\bibfnamefont {D.~R.}\
  \bibnamefont {Schultz}}, \bibinfo {author} {\bibfnamefont {T.}~\bibnamefont
  {Minami}}, \bibinfo {author} {\bibfnamefont {N.~R.}\ \bibnamefont {Badnell}},
  \bibinfo {author} {\bibfnamefont {M.~C.}\ \bibnamefont {Witthoeft}}, \bibinfo
  {author} {\bibfnamefont {D.~R.}\ \bibnamefont {Plante}}, \bibinfo {author}
  {\bibfnamefont {D.~M.}\ \bibnamefont {Mitnik}}, \bibinfo {author}
  {\bibfnamefont {J.~A.}\ \bibnamefont {Ludlow}}, \ and\ \bibinfo {author}
  {\bibfnamefont {U.}~\bibnamefont {Kleiman}},\ }\href
  {http://stacks.iop.org/0953-4075/40/i=7/a=R01} {\bibfield  {journal}
  {\bibinfo  {journal} {J. Phys. B}\ }\textbf {\bibinfo {volume} {40}},\
  \bibinfo {pages} {R39} (\bibinfo {year} {2007})}\BibitemShut {NoStop}%
\bibitem [{\citenamefont {Schneider}\ \emph {et~al.}(2011)\citenamefont
  {Schneider}, \citenamefont {Feist}, \citenamefont {Nagele}, \citenamefont
  {Pazourek}, \citenamefont {Hu}, \citenamefont {Collins},\ and\ \citenamefont
  {Burgd\"orfer}}]{Schneider2011QDI}%
  \BibitemOpen
  \bibfield  {author} {\bibinfo {author} {\bibfnamefont {B.~I.}\ \bibnamefont
  {Schneider}}, \bibinfo {author} {\bibfnamefont {J.}~\bibnamefont {Feist}},
  \bibinfo {author} {\bibfnamefont {S.}~\bibnamefont {Nagele}}, \bibinfo
  {author} {\bibfnamefont {R.}~\bibnamefont {Pazourek}}, \bibinfo {author}
  {\bibfnamefont {S.}~\bibnamefont {Hu}}, \bibinfo {author} {\bibfnamefont
  {L.~A.}\ \bibnamefont {Collins}}, \ and\ \bibinfo {author} {\bibfnamefont
  {J.}~\bibnamefont {Burgd\"orfer}},\ }\enquote {\bibinfo {title} {Recent
  advances in computational methods for the solution of the time-dependent
  schr\"odinger equation for the interaction of short, intense radiation with
  one and two electron systems: Application to he and ${\rm h_2^+}$},}\ in\
  \href@noop {} {\emph {\bibinfo {booktitle} {Quantum Dynamic Imaging}}},\
  \bibinfo {editor} {edited by\ \bibinfo {editor} {\bibfnamefont {A.~D.}\
  \bibnamefont {Bandrauk}}\ and\ \bibinfo {editor} {\bibfnamefont
  {M.}~\bibnamefont {Ivanov}}}\ (\bibinfo  {publisher} {Springer},\ \bibinfo
  {address} {New York},\ \bibinfo {year} {2011})\ Chap.~\bibinfo {chapter}
  {10}, pp.\ \bibinfo {pages} {149--208}\BibitemShut {NoStop}%
\bibitem [{\citenamefont {Blodgett-Ford}\ \emph {et~al.}(1993)\citenamefont
  {Blodgett-Ford}, \citenamefont {Parker},\ and\ \citenamefont
  {Clark}}]{Blodgett-Ford1993SILAP}%
  \BibitemOpen
  \bibfield  {author} {\bibinfo {author} {\bibfnamefont {S.}~\bibnamefont
  {Blodgett-Ford}}, \bibinfo {author} {\bibfnamefont {J.}~\bibnamefont
  {Parker}}, \ and\ \bibinfo {author} {\bibfnamefont {C.}~\bibnamefont
  {Clark}},\ }\enquote {\bibinfo {title} {Sequential vs. simultaneous
  ionization of two-electron atoms by intense laser radiation},}\ in\
  \href@noop {} {\emph {\bibinfo {booktitle} {Super-Intense Laser-Atom
  Physics}}},\ Vol.\ \bibinfo {volume} {316},\ \bibinfo {editor} {edited by\
  \bibinfo {editor} {\bibfnamefont {B.}~\bibnamefont {Piraux}}, \bibinfo
  {editor} {\bibfnamefont {A.}~\bibnamefont {L'Huillier}}, \ and\ \bibinfo
  {editor} {\bibfnamefont {K.}~\bibnamefont {Rz\c{a}\.{z}ewski}}}\ (\bibinfo
  {publisher} {Plenum},\ \bibinfo {address} {New York},\ \bibinfo {year}
  {1993})\ pp.\ \bibinfo {pages} {391--402}\BibitemShut {NoStop}%
\bibitem [{\citenamefont {Pindzola}\ and\ \citenamefont
  {Schultz}(1996)}]{Pindzola1996PRAa}%
  \BibitemOpen
  \bibfield  {author} {\bibinfo {author} {\bibfnamefont {M.~S.}\ \bibnamefont
  {Pindzola}}\ and\ \bibinfo {author} {\bibfnamefont {D.~R.}\ \bibnamefont
  {Schultz}},\ }\href {\doibase 10.1103/PhysRevA.53.1525} {\bibfield  {journal}
  {\bibinfo  {journal} {Phys. Rev. A}\ }\textbf {\bibinfo {volume} {53}},\
  \bibinfo {pages} {1525} (\bibinfo {year} {1996})}\BibitemShut {NoStop}%
\bibitem [{\citenamefont {Pindzola}\ and\ \citenamefont
  {Robicheaux}(1996)}]{Pindzola1996PRAb}%
  \BibitemOpen
  \bibfield  {author} {\bibinfo {author} {\bibfnamefont {M.~S.}\ \bibnamefont
  {Pindzola}}\ and\ \bibinfo {author} {\bibfnamefont {F.}~\bibnamefont
  {Robicheaux}},\ }\href {\doibase 10.1103/PhysRevA.54.2142} {\bibfield
  {journal} {\bibinfo  {journal} {Phys. Rev. A}\ }\textbf {\bibinfo {volume}
  {54}},\ \bibinfo {pages} {2142} (\bibinfo {year} {1996})}\BibitemShut
  {NoStop}%
\bibitem [{\citenamefont {Colgan}\ \emph
  {et~al.}(2004{\natexlab{b}})\citenamefont {Colgan}, \citenamefont
  {Pindzola},\ and\ \citenamefont {Robicheaux}}]{Colgan2004PRL}%
  \BibitemOpen
  \bibfield  {author} {\bibinfo {author} {\bibfnamefont {J.}~\bibnamefont
  {Colgan}}, \bibinfo {author} {\bibfnamefont {M.~S.}\ \bibnamefont
  {Pindzola}}, \ and\ \bibinfo {author} {\bibfnamefont {F.}~\bibnamefont
  {Robicheaux}},\ }\href {\doibase 10.1103/PhysRevLett.93.053201} {\bibfield
  {journal} {\bibinfo  {journal} {Phys. Rev. Lett.}\ }\textbf {\bibinfo
  {volume} {93}},\ \bibinfo {pages} {053201} (\bibinfo {year}
  {2004}{\natexlab{b}})}\BibitemShut {NoStop}%
\bibitem [{\citenamefont {Colgan}\ \emph {et~al.}(2005)\citenamefont {Colgan},
  \citenamefont {Pindzola},\ and\ \citenamefont {Robicheaux}}]{Colgan2005PRA}%
  \BibitemOpen
  \bibfield  {author} {\bibinfo {author} {\bibfnamefont {J.}~\bibnamefont
  {Colgan}}, \bibinfo {author} {\bibfnamefont {M.~S.}\ \bibnamefont
  {Pindzola}}, \ and\ \bibinfo {author} {\bibfnamefont {F.}~\bibnamefont
  {Robicheaux}},\ }\href {\doibase 10.1103/PhysRevA.72.022727} {\bibfield
  {journal} {\bibinfo  {journal} {Phys. Rev. A}\ }\textbf {\bibinfo {volume}
  {72}},\ \bibinfo {pages} {022727} (\bibinfo {year} {2005})}\BibitemShut
  {NoStop}%
\bibitem [{\citenamefont {Lagmago~Kamta}\ and\ \citenamefont
  {Starace}(2001)}]{Kamta2001PRL}%
  \BibitemOpen
  \bibfield  {author} {\bibinfo {author} {\bibfnamefont {G.}~\bibnamefont
  {Lagmago~Kamta}}\ and\ \bibinfo {author} {\bibfnamefont {A.~F.}\ \bibnamefont
  {Starace}},\ }\href {\doibase 10.1103/PhysRevLett.86.5687} {\bibfield
  {journal} {\bibinfo  {journal} {Phys. Rev. Lett.}\ }\textbf {\bibinfo
  {volume} {86}},\ \bibinfo {pages} {5687} (\bibinfo {year}
  {2001})}\BibitemShut {NoStop}%
\bibitem [{\citenamefont {Lagmago~Kamta}\ and\ \citenamefont
  {Starace}(2002)}]{Kamta2002PRA}%
  \BibitemOpen
  \bibfield  {author} {\bibinfo {author} {\bibfnamefont {G.}~\bibnamefont
  {Lagmago~Kamta}}\ and\ \bibinfo {author} {\bibfnamefont {A.~F.}\ \bibnamefont
  {Starace}},\ }\href {\doibase 10.1103/PhysRevA.65.053418} {\bibfield
  {journal} {\bibinfo  {journal} {Phys. Rev. A}\ }\textbf {\bibinfo {volume}
  {65}},\ \bibinfo {pages} {053418} (\bibinfo {year} {2002})}\BibitemShut
  {NoStop}%
\bibitem [{\citenamefont {Piraux}\ \emph {et~al.}(2003)\citenamefont {Piraux},
  \citenamefont {Bauer}, \citenamefont {Laulan},\ and\ \citenamefont
  {Bachau}}]{Piraux2003EPJD}%
  \BibitemOpen
  \bibfield  {author} {\bibinfo {author} {\bibfnamefont {B.}~\bibnamefont
  {Piraux}}, \bibinfo {author} {\bibfnamefont {J.}~\bibnamefont {Bauer}},
  \bibinfo {author} {\bibfnamefont {S.}~\bibnamefont {Laulan}}, \ and\ \bibinfo
  {author} {\bibfnamefont {H.}~\bibnamefont {Bachau}},\ }\href {\doibase
  10.1140/epjd/e2003-00063-3} {\bibfield  {journal} {\bibinfo  {journal} {Eur.
  Phys. J. D}\ }\textbf {\bibinfo {volume} {26}},\ \bibinfo {pages} {7}
  (\bibinfo {year} {2003})}\BibitemShut {NoStop}%
\bibitem [{\citenamefont {Laulan}\ and\ \citenamefont
  {Bachau}(2004)}]{Laulan2004PRA}%
  \BibitemOpen
  \bibfield  {author} {\bibinfo {author} {\bibfnamefont {S.}~\bibnamefont
  {Laulan}}\ and\ \bibinfo {author} {\bibfnamefont {H.}~\bibnamefont
  {Bachau}},\ }\href {\doibase 10.1103/PhysRevA.69.033408} {\bibfield
  {journal} {\bibinfo  {journal} {Phys. Rev. A}\ }\textbf {\bibinfo {volume}
  {69}},\ \bibinfo {pages} {033408} (\bibinfo {year} {2004})}\BibitemShut
  {NoStop}%
\bibitem [{\citenamefont {Boor}(2001)}]{Boor}%
  \BibitemOpen
  \bibfield  {author} {\bibinfo {author} {\bibfnamefont {C.~d.}\ \bibnamefont
  {Boor}},\ }\href@noop {} {\emph {\bibinfo {title} {A Practical Guide to
  Splines}}},\ \bibinfo {series} {Applied Mathematical Sciences}, Vol.~\bibinfo
  {volume} {27}\ (\bibinfo  {publisher} {Springer},\ \bibinfo {address} {New
  York},\ \bibinfo {year} {2001})\BibitemShut {NoStop}%
\bibitem [{\citenamefont {Press}\ \emph {et~al.}(2007)\citenamefont {Press},
  \citenamefont {Teukolsky}, \citenamefont {Vetterling},\ and\ \citenamefont
  {Flannery}}]{NumericalRecipes}%
  \BibitemOpen
  \bibfield  {author} {\bibinfo {author} {\bibfnamefont {W.~H.}\ \bibnamefont
  {Press}}, \bibinfo {author} {\bibfnamefont {S.~A.}\ \bibnamefont
  {Teukolsky}}, \bibinfo {author} {\bibfnamefont {W.~T.}\ \bibnamefont
  {Vetterling}}, \ and\ \bibinfo {author} {\bibfnamefont {B.~P.}\ \bibnamefont
  {Flannery}},\ }\href@noop {} {\emph {\bibinfo {title} {Numerical Recipes: The
  Art of Scientific Computing}}},\ \bibinfo {edition} {3rd}\ ed.\ (\bibinfo
  {publisher} {Cambridge University Press},\ \bibinfo {address} {New York},\
  \bibinfo {year} {2007})\BibitemShut {NoStop}%
\bibitem [{\citenamefont {Harumiya}\ \emph {et~al.}(2000)\citenamefont
  {Harumiya}, \citenamefont {Kawata}, \citenamefont {Kono},\ and\ \citenamefont
  {Fujimura}}]{Haruyama2000JCP}%
  \BibitemOpen
  \bibfield  {author} {\bibinfo {author} {\bibfnamefont {K.}~\bibnamefont
  {Harumiya}}, \bibinfo {author} {\bibfnamefont {I.}~\bibnamefont {Kawata}},
  \bibinfo {author} {\bibfnamefont {H.}~\bibnamefont {Kono}}, \ and\ \bibinfo
  {author} {\bibfnamefont {Y.}~\bibnamefont {Fujimura}},\ }\href {\doibase
  http://dx.doi.org/10.1063/1.1319348} {\bibfield  {journal} {\bibinfo
  {journal} {J. Chem. Phys.}\ }\textbf {\bibinfo {volume} {113}},\ \bibinfo
  {pages} {8953} (\bibinfo {year} {2000})}\BibitemShut {NoStop}%
\bibitem [{\citenamefont {Harumiya}\ \emph {et~al.}(2002)\citenamefont
  {Harumiya}, \citenamefont {Kono}, \citenamefont {Fujimura}, \citenamefont
  {Kawata},\ and\ \citenamefont {Bandrauk}}]{Haruyama2002PRA}%
  \BibitemOpen
  \bibfield  {author} {\bibinfo {author} {\bibfnamefont {K.}~\bibnamefont
  {Harumiya}}, \bibinfo {author} {\bibfnamefont {H.}~\bibnamefont {Kono}},
  \bibinfo {author} {\bibfnamefont {Y.}~\bibnamefont {Fujimura}}, \bibinfo
  {author} {\bibfnamefont {I.}~\bibnamefont {Kawata}}, \ and\ \bibinfo {author}
  {\bibfnamefont {A.~D.}\ \bibnamefont {Bandrauk}},\ }\href {\doibase
  10.1103/PhysRevA.66.043403} {\bibfield  {journal} {\bibinfo  {journal} {Phys.
  Rev. A}\ }\textbf {\bibinfo {volume} {66}},\ \bibinfo {pages} {043403}
  (\bibinfo {year} {2002})}\BibitemShut {NoStop}%
\bibitem [{\citenamefont {Kawata}\ and\ \citenamefont
  {Kono}(1999)}]{Kawata1999JCP}%
  \BibitemOpen
  \bibfield  {author} {\bibinfo {author} {\bibfnamefont {I.}~\bibnamefont
  {Kawata}}\ and\ \bibinfo {author} {\bibfnamefont {H.}~\bibnamefont {Kono}},\
  }\href {\doibase http://dx.doi.org/10.1063/1.480281} {\bibfield  {journal}
  {\bibinfo  {journal} {J. Chem. Phys.}\ }\textbf {\bibinfo {volume} {111}},\
  \bibinfo {pages} {9498} (\bibinfo {year} {1999})}\BibitemShut {NoStop}%
\bibitem [{\citenamefont {Ohtsuki}\ \emph {et~al.}(2001)\citenamefont
  {Ohtsuki}, \citenamefont {Sugawara}, \citenamefont {Kono},\ and\
  \citenamefont {Fujimura}}]{Ohtsuki2001BCSJ}%
  \BibitemOpen
  \bibfield  {author} {\bibinfo {author} {\bibfnamefont {Y.}~\bibnamefont
  {Ohtsuki}}, \bibinfo {author} {\bibfnamefont {M.}~\bibnamefont {Sugawara}},
  \bibinfo {author} {\bibfnamefont {H.}~\bibnamefont {Kono}}, \ and\ \bibinfo
  {author} {\bibfnamefont {Y.}~\bibnamefont {Fujimura}},\ }\href {\doibase
  10.1246/bcsj.74.1167} {\bibfield  {journal} {\bibinfo  {journal} {Bull. Chem.
  Soc. Jpn.}\ }\textbf {\bibinfo {volume} {74}},\ \bibinfo {pages} {1167}
  (\bibinfo {year} {2001})}\BibitemShut {NoStop}%
\bibitem [{\citenamefont {Colgan}\ \emph {et~al.}(2008)\citenamefont {Colgan},
  \citenamefont {Pindzola},\ and\ \citenamefont {Robicheaux}}]{Colgan2008JPB}%
  \BibitemOpen
  \bibfield  {author} {\bibinfo {author} {\bibfnamefont {J.}~\bibnamefont
  {Colgan}}, \bibinfo {author} {\bibfnamefont {M.~S.}\ \bibnamefont
  {Pindzola}}, \ and\ \bibinfo {author} {\bibfnamefont {F.}~\bibnamefont
  {Robicheaux}},\ }\href {http://stacks.iop.org/0953-4075/41/i=12/a=121002}
  {\bibfield  {journal} {\bibinfo  {journal} {J. Phys. B}\ }\textbf {\bibinfo
  {volume} {41}},\ \bibinfo {pages} {121002} (\bibinfo {year}
  {2008})}\BibitemShut {NoStop}%
\bibitem [{\citenamefont {Pindzola}\ \emph {et~al.}(2009)\citenamefont
  {Pindzola}, \citenamefont {Ludlow},\ and\ \citenamefont
  {Colgan}}]{Pindzola2009PRA}%
  \BibitemOpen
  \bibfield  {author} {\bibinfo {author} {\bibfnamefont {M.~S.}\ \bibnamefont
  {Pindzola}}, \bibinfo {author} {\bibfnamefont {J.~A.}\ \bibnamefont
  {Ludlow}}, \ and\ \bibinfo {author} {\bibfnamefont {J.}~\bibnamefont
  {Colgan}},\ }\href {\doibase 10.1103/PhysRevA.80.032707} {\bibfield
  {journal} {\bibinfo  {journal} {Phys. Rev. A}\ }\textbf {\bibinfo {volume}
  {80}},\ \bibinfo {pages} {032707} (\bibinfo {year} {2009})}\BibitemShut
  {NoStop}%
\bibitem [{\citenamefont {Lee}\ \emph {et~al.}(2010)\citenamefont {Lee},
  \citenamefont {Pindzola},\ and\ \citenamefont {Robicheaux}}]{Lee2010JPB}%
  \BibitemOpen
  \bibfield  {author} {\bibinfo {author} {\bibfnamefont {T.-G.}\ \bibnamefont
  {Lee}}, \bibinfo {author} {\bibfnamefont {M.~S.}\ \bibnamefont {Pindzola}}, \
  and\ \bibinfo {author} {\bibfnamefont {F.}~\bibnamefont {Robicheaux}},\
  }\href {http://stacks.iop.org/0953-4075/43/i=16/a=165601} {\bibfield
  {journal} {\bibinfo  {journal} {J. Phys. B}\ }\textbf {\bibinfo {volume}
  {43}},\ \bibinfo {pages} {165601} (\bibinfo {year} {2010})}\BibitemShut
  {NoStop}%
\bibitem [{\citenamefont {Simonsen}\ \emph {et~al.}(2012)\citenamefont
  {Simonsen}, \citenamefont {S\o{}rng\aa{}rd}, \citenamefont {Nepstad},\ and\
  \citenamefont {F\o{}rre}}]{Simonsen2012PRA}%
  \BibitemOpen
  \bibfield  {author} {\bibinfo {author} {\bibfnamefont {A.~S.}\ \bibnamefont
  {Simonsen}}, \bibinfo {author} {\bibfnamefont {S.~A.}\ \bibnamefont
  {S\o{}rng\aa{}rd}}, \bibinfo {author} {\bibfnamefont {R.}~\bibnamefont
  {Nepstad}}, \ and\ \bibinfo {author} {\bibfnamefont {M.}~\bibnamefont
  {F\o{}rre}},\ }\href {\doibase 10.1103/PhysRevA.85.063404} {\bibfield
  {journal} {\bibinfo  {journal} {Phys. Rev. A}\ }\textbf {\bibinfo {volume}
  {85}},\ \bibinfo {pages} {063404} (\bibinfo {year} {2012})}\BibitemShut
  {NoStop}%
\bibitem [{\citenamefont {Bachau}\ \emph {et~al.}(2001)\citenamefont {Bachau},
  \citenamefont {Cormier}, \citenamefont {Decleva}, \citenamefont {Hansen},\
  and\ \citenamefont {Mart\'{\i}n}}]{Bachau2001RPP}%
  \BibitemOpen
  \bibfield  {author} {\bibinfo {author} {\bibfnamefont {H.}~\bibnamefont
  {Bachau}}, \bibinfo {author} {\bibfnamefont {E.}~\bibnamefont {Cormier}},
  \bibinfo {author} {\bibfnamefont {P.}~\bibnamefont {Decleva}}, \bibinfo
  {author} {\bibfnamefont {J.~E.}\ \bibnamefont {Hansen}}, \ and\ \bibinfo
  {author} {\bibfnamefont {F.}~\bibnamefont {Mart\'{\i}n}},\ }\href
  {http://stacks.iop.org/0034-4885/64/i=12/a=205} {\bibfield  {journal}
  {\bibinfo  {journal} {Rep. Prog. Phys.}\ }\textbf {\bibinfo {volume} {64}},\
  \bibinfo {pages} {1815} (\bibinfo {year} {2001})}\BibitemShut {NoStop}%
\bibitem [{\citenamefont {Guan}\ \emph {et~al.}(2010)\citenamefont {Guan},
  \citenamefont {Bartschat},\ and\ \citenamefont {Schneider}}]{Guan2010PRA}%
  \BibitemOpen
  \bibfield  {author} {\bibinfo {author} {\bibfnamefont {X.}~\bibnamefont
  {Guan}}, \bibinfo {author} {\bibfnamefont {K.}~\bibnamefont {Bartschat}}, \
  and\ \bibinfo {author} {\bibfnamefont {B.~I.}\ \bibnamefont {Schneider}},\
  }\href {\doibase 10.1103/PhysRevA.82.041404} {\bibfield  {journal} {\bibinfo
  {journal} {Phys. Rev. A}\ }\textbf {\bibinfo {volume} {82}},\ \bibinfo
  {pages} {041404} (\bibinfo {year} {2010})}\BibitemShut {NoStop}%
\bibitem [{\citenamefont {Guan}\ \emph {et~al.}(2008)\citenamefont {Guan},
  \citenamefont {Bartschat},\ and\ \citenamefont {Schneider}}]{Guan2008PRA}%
  \BibitemOpen
  \bibfield  {author} {\bibinfo {author} {\bibfnamefont {X.}~\bibnamefont
  {Guan}}, \bibinfo {author} {\bibfnamefont {K.}~\bibnamefont {Bartschat}}, \
  and\ \bibinfo {author} {\bibfnamefont {B.~I.}\ \bibnamefont {Schneider}},\
  }\href {\doibase 10.1103/PhysRevA.77.043421} {\bibfield  {journal} {\bibinfo
  {journal} {Phys. Rev. A}\ }\textbf {\bibinfo {volume} {77}},\ \bibinfo
  {pages} {043421} (\bibinfo {year} {2008})}\BibitemShut {NoStop}%
\bibitem [{\citenamefont {Awasthi}\ \emph {et~al.}(2005)\citenamefont
  {Awasthi}, \citenamefont {Vanne},\ and\ \citenamefont
  {Saenz}}]{Awasthi2005JPB}%
  \BibitemOpen
  \bibfield  {author} {\bibinfo {author} {\bibfnamefont {M.}~\bibnamefont
  {Awasthi}}, \bibinfo {author} {\bibfnamefont {Y.~V.}\ \bibnamefont {Vanne}},
  \ and\ \bibinfo {author} {\bibfnamefont {A.}~\bibnamefont {Saenz}},\ }\href
  {http://stacks.iop.org/0953-4075/38/i=22/a=005} {\bibfield  {journal}
  {\bibinfo  {journal} {J. Phys. B}\ }\textbf {\bibinfo {volume} {38}},\
  \bibinfo {pages} {3973} (\bibinfo {year} {2005})}\BibitemShut {NoStop}%
\bibitem [{\citenamefont {Vanne}\ and\ \citenamefont
  {Saenz}(2004)}]{Vanne2004JPB}%
  \BibitemOpen
  \bibfield  {author} {\bibinfo {author} {\bibfnamefont {Y.~V.}\ \bibnamefont
  {Vanne}}\ and\ \bibinfo {author} {\bibfnamefont {A.}~\bibnamefont {Saenz}},\
  }\href {http://stacks.iop.org/0953-4075/37/i=20/a=005} {\bibfield  {journal}
  {\bibinfo  {journal} {J. Phys. B}\ }\textbf {\bibinfo {volume} {37}},\
  \bibinfo {pages} {4101} (\bibinfo {year} {2004})}\BibitemShut {NoStop}%
\bibitem [{\citenamefont {Awasthi}\ and\ \citenamefont
  {Saenz}(2010)}]{Awasthi2010PRA}%
  \BibitemOpen
  \bibfield  {author} {\bibinfo {author} {\bibfnamefont {M.}~\bibnamefont
  {Awasthi}}\ and\ \bibinfo {author} {\bibfnamefont {A.}~\bibnamefont
  {Saenz}},\ }\href {\doibase 10.1103/PhysRevA.81.063406} {\bibfield  {journal}
  {\bibinfo  {journal} {Phys. Rev. A}\ }\textbf {\bibinfo {volume} {81}},\
  \bibinfo {pages} {063406} (\bibinfo {year} {2010})}\BibitemShut {NoStop}%
\bibitem [{\citenamefont {Vanne}\ and\ \citenamefont
  {Saenz}(2008)}]{Vanne2008JMO}%
  \BibitemOpen
  \bibfield  {author} {\bibinfo {author} {\bibfnamefont {Y.~V.}\ \bibnamefont
  {Vanne}}\ and\ \bibinfo {author} {\bibfnamefont {A.}~\bibnamefont {Saenz}},\
  }\href {\doibase 10.1080/09500340802148979} {\bibfield  {journal} {\bibinfo
  {journal} {J. Mod. Opt.}\ }\textbf {\bibinfo {volume} {55}},\ \bibinfo
  {pages} {2665} (\bibinfo {year} {2008})},\ \Eprint
  {http://arxiv.org/abs/http://dx.doi.org/10.1080/09500340802148979}
  {http://dx.doi.org/10.1080/09500340802148979} \BibitemShut {NoStop}%
\bibitem [{\citenamefont {Vanne}\ and\ \citenamefont
  {Saenz}(2009)}]{Vanne2009PRA}%
  \BibitemOpen
  \bibfield  {author} {\bibinfo {author} {\bibfnamefont {Y.~V.}\ \bibnamefont
  {Vanne}}\ and\ \bibinfo {author} {\bibfnamefont {A.}~\bibnamefont {Saenz}},\
  }\href {\doibase 10.1103/PhysRevA.80.053422} {\bibfield  {journal} {\bibinfo
  {journal} {Phys. Rev. A}\ }\textbf {\bibinfo {volume} {80}},\ \bibinfo
  {pages} {053422} (\bibinfo {year} {2009})}\BibitemShut {NoStop}%
\bibitem [{\citenamefont {Vanne}\ and\ \citenamefont
  {Saenz}(2010)}]{Vanne2010PRA}%
  \BibitemOpen
  \bibfield  {author} {\bibinfo {author} {\bibfnamefont {Y.~V.}\ \bibnamefont
  {Vanne}}\ and\ \bibinfo {author} {\bibfnamefont {A.}~\bibnamefont {Saenz}},\
  }\href {\doibase 10.1103/PhysRevA.82.011403} {\bibfield  {journal} {\bibinfo
  {journal} {Phys. Rev. A}\ }\textbf {\bibinfo {volume} {82}},\ \bibinfo
  {pages} {011403} (\bibinfo {year} {2010})}\BibitemShut {NoStop}%
\bibitem [{\citenamefont {F\"orster}\ \emph {et~al.}(2014)\citenamefont
  {F\"orster}, \citenamefont {Vanne},\ and\ \citenamefont
  {Saenz}}]{Forster2014PRA}%
  \BibitemOpen
  \bibfield  {author} {\bibinfo {author} {\bibfnamefont {J.}~\bibnamefont
  {F\"orster}}, \bibinfo {author} {\bibfnamefont {Y.~V.}\ \bibnamefont
  {Vanne}}, \ and\ \bibinfo {author} {\bibfnamefont {A.}~\bibnamefont
  {Saenz}},\ }\href {\doibase 10.1103/PhysRevA.90.053424} {\bibfield  {journal}
  {\bibinfo  {journal} {Phys. Rev. A}\ }\textbf {\bibinfo {volume} {90}},\
  \bibinfo {pages} {053424} (\bibinfo {year} {2014})}\BibitemShut {NoStop}%
\bibitem [{\citenamefont {Dehghanian}\ \emph {et~al.}(2010)\citenamefont
  {Dehghanian}, \citenamefont {Bandrauk},\ and\ \citenamefont
  {Kamta}}]{Dehghanian2010PRA}%
  \BibitemOpen
  \bibfield  {author} {\bibinfo {author} {\bibfnamefont {E.}~\bibnamefont
  {Dehghanian}}, \bibinfo {author} {\bibfnamefont {A.~D.}\ \bibnamefont
  {Bandrauk}}, \ and\ \bibinfo {author} {\bibfnamefont {G.~L.}\ \bibnamefont
  {Kamta}},\ }\href {\doibase 10.1103/PhysRevA.81.061403} {\bibfield  {journal}
  {\bibinfo  {journal} {Phys. Rev. A}\ }\textbf {\bibinfo {volume} {81}},\
  \bibinfo {pages} {061403} (\bibinfo {year} {2010})}\BibitemShut {NoStop}%
\bibitem [{\citenamefont {Sanz-Vicario}\ \emph {et~al.}(2006)\citenamefont
  {Sanz-Vicario}, \citenamefont {Bachau},\ and\ \citenamefont
  {Mart\'{\i}n}}]{Sanz-Vicario2006PRA}%
  \BibitemOpen
  \bibfield  {author} {\bibinfo {author} {\bibfnamefont {J.~L.}\ \bibnamefont
  {Sanz-Vicario}}, \bibinfo {author} {\bibfnamefont {H.}~\bibnamefont
  {Bachau}}, \ and\ \bibinfo {author} {\bibfnamefont {F.}~\bibnamefont
  {Mart\'{\i}n}},\ }\href {\doibase 10.1103/PhysRevA.73.033410} {\bibfield
  {journal} {\bibinfo  {journal} {Phys. Rev. A}\ }\textbf {\bibinfo {volume}
  {73}},\ \bibinfo {pages} {033410} (\bibinfo {year} {2006})}\BibitemShut
  {NoStop}%
\bibitem [{\citenamefont {Palacios}\ \emph {et~al.}(2006)\citenamefont
  {Palacios}, \citenamefont {Bachau},\ and\ \citenamefont
  {Mart\'{\i}n}}]{Palacios2006PRL}%
  \BibitemOpen
  \bibfield  {author} {\bibinfo {author} {\bibfnamefont {A.}~\bibnamefont
  {Palacios}}, \bibinfo {author} {\bibfnamefont {H.}~\bibnamefont {Bachau}}, \
  and\ \bibinfo {author} {\bibfnamefont {F.}~\bibnamefont {Mart\'{\i}n}},\
  }\href {\doibase 10.1103/PhysRevLett.96.143001} {\bibfield  {journal}
  {\bibinfo  {journal} {Phys. Rev. Lett.}\ }\textbf {\bibinfo {volume} {96}},\
  \bibinfo {pages} {143001} (\bibinfo {year} {2006})}\BibitemShut {NoStop}%
\bibitem [{\citenamefont {Mart\'{\i}n}(1999)}]{Martin1999JPB}%
  \BibitemOpen
  \bibfield  {author} {\bibinfo {author} {\bibfnamefont {F.}~\bibnamefont
  {Mart\'{\i}n}},\ }\href {http://stacks.iop.org/0953-4075/32/i=16/a=201}
  {\bibfield  {journal} {\bibinfo  {journal} {J. Phys. B}\ }\textbf {\bibinfo
  {volume} {32}},\ \bibinfo {pages} {R197} (\bibinfo {year}
  {1999})}\BibitemShut {NoStop}%
\bibitem [{\citenamefont {P\'erez-Torres}\ \emph {et~al.}(2014)\citenamefont
  {P\'erez-Torres}, \citenamefont {Sanz-Vicario}, \citenamefont {Veyrinas},
  \citenamefont {Billaud}, \citenamefont {Picard}, \citenamefont {Elkharrat},
  \citenamefont {Poullain}, \citenamefont {Saquet}, \citenamefont {Lebech},
  \citenamefont {Houver}, \citenamefont {Mart\'{\i}n},\ and\ \citenamefont
  {Dowek}}]{Perez-Torres2014PRA}%
  \BibitemOpen
  \bibfield  {author} {\bibinfo {author} {\bibfnamefont {J.~F.}\ \bibnamefont
  {P\'erez-Torres}}, \bibinfo {author} {\bibfnamefont {J.~L.}\ \bibnamefont
  {Sanz-Vicario}}, \bibinfo {author} {\bibfnamefont {K.}~\bibnamefont
  {Veyrinas}}, \bibinfo {author} {\bibfnamefont {P.}~\bibnamefont {Billaud}},
  \bibinfo {author} {\bibfnamefont {Y.~J.}\ \bibnamefont {Picard}}, \bibinfo
  {author} {\bibfnamefont {C.}~\bibnamefont {Elkharrat}}, \bibinfo {author}
  {\bibfnamefont {S.~M.}\ \bibnamefont {Poullain}}, \bibinfo {author}
  {\bibfnamefont {N.}~\bibnamefont {Saquet}}, \bibinfo {author} {\bibfnamefont
  {M.}~\bibnamefont {Lebech}}, \bibinfo {author} {\bibfnamefont {J.~C.}\
  \bibnamefont {Houver}}, \bibinfo {author} {\bibfnamefont {F.}~\bibnamefont
  {Mart\'{\i}n}}, \ and\ \bibinfo {author} {\bibfnamefont {D.}~\bibnamefont
  {Dowek}},\ }\href {\doibase 10.1103/PhysRevA.90.043417} {\bibfield  {journal}
  {\bibinfo  {journal} {Phys. Rev. A}\ }\textbf {\bibinfo {volume} {90}},\
  \bibinfo {pages} {043417} (\bibinfo {year} {2014})}\BibitemShut {NoStop}%
\bibitem [{\citenamefont {Sansone}\ \emph {et~al.}(2010)\citenamefont
  {Sansone}, \citenamefont {Kelkensberg}, \citenamefont {Perez-Torres},
  \citenamefont {Morales}, \citenamefont {Kling}, \citenamefont {Siu},
  \citenamefont {Ghafur}, \citenamefont {Johnsson}, \citenamefont {Swoboda},
  \citenamefont {Benedetti}, \citenamefont {Ferrari}, \citenamefont {Lepine},
  \citenamefont {Sanz-Vicario}, \citenamefont {Zherebtsov}, \citenamefont
  {Znakovskaya}, \citenamefont {L'Huillier}, \citenamefont {Ivanov},
  \citenamefont {Nisoli}, \citenamefont {Mart\'{\i}n},\ and\ \citenamefont
  {Vrakking}}]{Sansone2010Nature}%
  \BibitemOpen
  \bibfield  {author} {\bibinfo {author} {\bibfnamefont {G.}~\bibnamefont
  {Sansone}}, \bibinfo {author} {\bibfnamefont {F.}~\bibnamefont
  {Kelkensberg}}, \bibinfo {author} {\bibfnamefont {J.~F.}\ \bibnamefont
  {Perez-Torres}}, \bibinfo {author} {\bibfnamefont {F.}~\bibnamefont
  {Morales}}, \bibinfo {author} {\bibfnamefont {M.~F.}\ \bibnamefont {Kling}},
  \bibinfo {author} {\bibfnamefont {W.}~\bibnamefont {Siu}}, \bibinfo {author}
  {\bibfnamefont {O.}~\bibnamefont {Ghafur}}, \bibinfo {author} {\bibfnamefont
  {P.}~\bibnamefont {Johnsson}}, \bibinfo {author} {\bibfnamefont
  {M.}~\bibnamefont {Swoboda}}, \bibinfo {author} {\bibfnamefont
  {E.}~\bibnamefont {Benedetti}}, \bibinfo {author} {\bibfnamefont
  {F.}~\bibnamefont {Ferrari}}, \bibinfo {author} {\bibfnamefont
  {F.}~\bibnamefont {Lepine}}, \bibinfo {author} {\bibfnamefont {J.~L.}\
  \bibnamefont {Sanz-Vicario}}, \bibinfo {author} {\bibfnamefont
  {S.}~\bibnamefont {Zherebtsov}}, \bibinfo {author} {\bibfnamefont
  {I.}~\bibnamefont {Znakovskaya}}, \bibinfo {author} {\bibfnamefont
  {A.}~\bibnamefont {L'Huillier}}, \bibinfo {author} {\bibfnamefont {M.~Y.}\
  \bibnamefont {Ivanov}}, \bibinfo {author} {\bibfnamefont {M.}~\bibnamefont
  {Nisoli}}, \bibinfo {author} {\bibfnamefont {F.}~\bibnamefont {Mart\'{\i}n}},
  \ and\ \bibinfo {author} {\bibfnamefont {M.~J.~J.}\ \bibnamefont
  {Vrakking}},\ }\href@noop {} {\bibfield  {journal} {\bibinfo  {journal}
  {Nature}\ }\textbf {\bibinfo {volume} {465}},\ \bibinfo {pages} {763}
  (\bibinfo {year} {2010})}\BibitemShut {NoStop}%
\bibitem [{\citenamefont {Nguyen-Dang}\ \emph {et~al.}(2009)\citenamefont
  {Nguyen-Dang}, \citenamefont {Peters}, \citenamefont {Wang},\ and\
  \citenamefont {Dion}}]{Nguyen-Dang2009CP}%
  \BibitemOpen
  \bibfield  {author} {\bibinfo {author} {\bibfnamefont {T.-T.}\ \bibnamefont
  {Nguyen-Dang}}, \bibinfo {author} {\bibfnamefont {M.}~\bibnamefont {Peters}},
  \bibinfo {author} {\bibfnamefont {S.-M.}\ \bibnamefont {Wang}}, \ and\
  \bibinfo {author} {\bibfnamefont {F.}~\bibnamefont {Dion}},\ }\href@noop {}
  {\bibfield  {journal} {\bibinfo  {journal} {Chem. Phys.}\ }\textbf {\bibinfo
  {volume} {366}},\ \bibinfo {pages} {71} (\bibinfo {year} {2009})}\BibitemShut
  {NoStop}%
\bibitem [{\citenamefont {Nguyen-Dang}\ and\ \citenamefont
  {Viau-Trudel}(2013)}]{Nguyen-Dang2013JCP}%
  \BibitemOpen
  \bibfield  {author} {\bibinfo {author} {\bibfnamefont {T.-T.}\ \bibnamefont
  {Nguyen-Dang}}\ and\ \bibinfo {author} {\bibfnamefont {J.}~\bibnamefont
  {Viau-Trudel}},\ }\href@noop {} {\bibfield  {journal} {\bibinfo  {journal}
  {J. Chem. Phys.}\ }\textbf {\bibinfo {volume} {139}},\ \bibinfo {pages}
  {244102} (\bibinfo {year} {2013})}\BibitemShut {NoStop}%
\bibitem [{\citenamefont {Miranda}\ \emph {et~al.}(2011)\citenamefont
  {Miranda}, \citenamefont {Fisher}, \citenamefont {Stella},\ and\
  \citenamefont {Horsfield}}]{Miranda2011JCPa}%
  \BibitemOpen
  \bibfield  {author} {\bibinfo {author} {\bibfnamefont {R.~P.}\ \bibnamefont
  {Miranda}}, \bibinfo {author} {\bibfnamefont {A.~J.}\ \bibnamefont {Fisher}},
  \bibinfo {author} {\bibfnamefont {L.}~\bibnamefont {Stella}}, \ and\ \bibinfo
  {author} {\bibfnamefont {A.~P.}\ \bibnamefont {Horsfield}},\ }\href {\doibase
  http://dx.doi.org/10.1063/1.3600397} {\bibfield  {journal} {\bibinfo
  {journal} {J. Chem. Phys.}\ }\textbf {\bibinfo {volume} {134}},\ \bibinfo
  {eid} {244101} (\bibinfo {year} {2011})}\BibitemShut {NoStop}%
\bibitem [{\citenamefont {Sato}\ and\ \citenamefont
  {Ishikawa}(2013)}]{Sato2013PRA}%
  \BibitemOpen
  \bibfield  {author} {\bibinfo {author} {\bibfnamefont {T.}~\bibnamefont
  {Sato}}\ and\ \bibinfo {author} {\bibfnamefont {K.~L.}\ \bibnamefont
  {Ishikawa}},\ }\href {\doibase 10.1103/PhysRevA.88.023402} {\bibfield
  {journal} {\bibinfo  {journal} {Phys. Rev. A}\ }\textbf {\bibinfo {volume}
  {88}},\ \bibinfo {pages} {023402} (\bibinfo {year} {2013})}\BibitemShut
  {NoStop}%
\bibitem [{\citenamefont {Sato}\ and\ \citenamefont
  {Ishikawa}(2015)}]{Sato2015PRA}%
  \BibitemOpen
  \bibfield  {author} {\bibinfo {author} {\bibfnamefont {T.}~\bibnamefont
  {Sato}}\ and\ \bibinfo {author} {\bibfnamefont {K.~L.}\ \bibnamefont
  {Ishikawa}},\ }\href {\doibase 10.1103/PhysRevA.91.023417} {\bibfield
  {journal} {\bibinfo  {journal} {Phys. Rev. A}\ }\textbf {\bibinfo {volume}
  {91}},\ \bibinfo {pages} {023417} (\bibinfo {year} {2015})}\BibitemShut
  {NoStop}%
\bibitem [{\citenamefont {Szabo}\ and\ \citenamefont
  {Ostlund}(1996)}]{Szabo1996}%
  \BibitemOpen
  \bibfield  {author} {\bibinfo {author} {\bibfnamefont {A.}~\bibnamefont
  {Szabo}}\ and\ \bibinfo {author} {\bibfnamefont {N.~S.}\ \bibnamefont
  {Ostlund}},\ }\href@noop {} {\emph {\bibinfo {title} {Modern Quantum
  Chemistry}}}\ (\bibinfo  {publisher} {Dover},\ \bibinfo {address} {Mineola,
  NY},\ \bibinfo {year} {1996})\BibitemShut {NoStop}%
\bibitem [{\citenamefont {Helgaker}\ \emph {et~al.}(2000)\citenamefont
  {Helgaker}, \citenamefont {J{\o}rgensen},\ and\ \citenamefont
  {Olsen}}]{Helgaker}%
  \BibitemOpen
  \bibfield  {author} {\bibinfo {author} {\bibfnamefont {T.}~\bibnamefont
  {Helgaker}}, \bibinfo {author} {\bibfnamefont {P.}~\bibnamefont
  {J{\o}rgensen}}, \ and\ \bibinfo {author} {\bibfnamefont {J.}~\bibnamefont
  {Olsen}},\ }\href@noop {} {\emph {\bibinfo {title} {Molecular
  Electronic-Structure Theory}}}\ (\bibinfo  {publisher} {Wiley},\ \bibinfo
  {address} {Chichester, UK},\ \bibinfo {year} {2000})\BibitemShut {NoStop}%
\bibitem [{\citenamefont {Frenkel}(1934)}]{Frenkel}%
  \BibitemOpen
  \bibfield  {author} {\bibinfo {author} {\bibfnamefont {J.}~\bibnamefont
  {Frenkel}},\ }\href@noop {} {\emph {\bibinfo {title} {Wave Mechanics:
  Advanced General Theory}}}\ (\bibinfo  {publisher} {Clarendon Press},\
  \bibinfo {address} {Oxford},\ \bibinfo {year} {1934})\BibitemShut {NoStop}%
\bibitem [{\citenamefont {L\"owdin}\ and\ \citenamefont
  {Mukherjee}(1972)}]{Loewdin1972CPL}%
  \BibitemOpen
  \bibfield  {author} {\bibinfo {author} {\bibfnamefont {P.-O.}\ \bibnamefont
  {L\"owdin}}\ and\ \bibinfo {author} {\bibfnamefont {P.}~\bibnamefont
  {Mukherjee}},\ }\href {\doibase
  http://dx.doi.org/10.1016/0009-2614(72)87127-6} {\bibfield  {journal}
  {\bibinfo  {journal} {Chem. Phys. Lett.}\ }\textbf {\bibinfo {volume} {14}},\
  \bibinfo {pages} {1 } (\bibinfo {year} {1972})}\BibitemShut {NoStop}%
\bibitem [{\citenamefont {Moccia}(1973)}]{Moccia1973IJQC}%
  \BibitemOpen
  \bibfield  {author} {\bibinfo {author} {\bibfnamefont {R.}~\bibnamefont
  {Moccia}},\ }\href {\doibase 10.1002/qua.560070414} {\bibfield  {journal}
  {\bibinfo  {journal} {Int. J. Quant. Chem.}\ }\textbf {\bibinfo {volume}
  {7}},\ \bibinfo {pages} {779} (\bibinfo {year} {1973})}\BibitemShut {NoStop}%
\bibitem [{\citenamefont {Caillat}\ \emph {et~al.}(2005)\citenamefont
  {Caillat}, \citenamefont {Zanghellini}, \citenamefont {Kitzler},
  \citenamefont {Koch}, \citenamefont {Kreuzer},\ and\ \citenamefont
  {Scrinzi}}]{Caillat2005PRA}%
  \BibitemOpen
  \bibfield  {author} {\bibinfo {author} {\bibfnamefont {J.}~\bibnamefont
  {Caillat}}, \bibinfo {author} {\bibfnamefont {J.}~\bibnamefont
  {Zanghellini}}, \bibinfo {author} {\bibfnamefont {M.}~\bibnamefont
  {Kitzler}}, \bibinfo {author} {\bibfnamefont {O.}~\bibnamefont {Koch}},
  \bibinfo {author} {\bibfnamefont {W.}~\bibnamefont {Kreuzer}}, \ and\
  \bibinfo {author} {\bibfnamefont {A.}~\bibnamefont {Scrinzi}},\ }\href
  {\doibase 10.1103/PhysRevA.71.012712} {\bibfield  {journal} {\bibinfo
  {journal} {Phys. Rev. A}\ }\textbf {\bibinfo {volume} {71}},\ \bibinfo
  {pages} {012712} (\bibinfo {year} {2005})}\BibitemShut {NoStop}%
\bibitem [{\citenamefont {Sato}\ and\ \citenamefont
  {Ishikawa}(2014)}]{Sato2014JPB}%
  \BibitemOpen
  \bibfield  {author} {\bibinfo {author} {\bibfnamefont {T.}~\bibnamefont
  {Sato}}\ and\ \bibinfo {author} {\bibfnamefont {K.~L.}\ \bibnamefont
  {Ishikawa}},\ }\href {http://stacks.iop.org/0953-4075/47/i=20/a=204031}
  {\bibfield  {journal} {\bibinfo  {journal} {J. Phys. B}\ }\textbf {\bibinfo
  {volume} {47}},\ \bibinfo {pages} {204031} (\bibinfo {year}
  {2014})}\BibitemShut {NoStop}%
\bibitem [{\citenamefont {Dahlen}\ and\ \citenamefont {van
  Leeuwen}(2001)}]{Dahlen2001PRA}%
  \BibitemOpen
  \bibfield  {author} {\bibinfo {author} {\bibfnamefont {N.~E.}\ \bibnamefont
  {Dahlen}}\ and\ \bibinfo {author} {\bibfnamefont {R.}~\bibnamefont {van
  Leeuwen}},\ }\href {\doibase 10.1103/PhysRevA.64.023405} {\bibfield
  {journal} {\bibinfo  {journal} {Phys. Rev. A}\ }\textbf {\bibinfo {volume}
  {64}},\ \bibinfo {pages} {023405} (\bibinfo {year} {2001})}\BibitemShut
  {NoStop}%
\bibitem [{\citenamefont {Nguyen}\ and\ \citenamefont
  {Bandrauk}(2006)}]{Nguyen2006PRA}%
  \BibitemOpen
  \bibfield  {author} {\bibinfo {author} {\bibfnamefont {N.~A.}\ \bibnamefont
  {Nguyen}}\ and\ \bibinfo {author} {\bibfnamefont {A.~D.}\ \bibnamefont
  {Bandrauk}},\ }\href {\doibase 10.1103/PhysRevA.73.032708} {\bibfield
  {journal} {\bibinfo  {journal} {Phys. Rev. A}\ }\textbf {\bibinfo {volume}
  {73}},\ \bibinfo {pages} {032708} (\bibinfo {year} {2006})}\BibitemShut
  {NoStop}%
\bibitem [{\citenamefont {Bunse-Gerstner}\ and\ \citenamefont
  {Gragg}(1988)}]{BunseGerstner1988JCAM}%
  \BibitemOpen
  \bibfield  {author} {\bibinfo {author} {\bibfnamefont {A.}~\bibnamefont
  {Bunse-Gerstner}}\ and\ \bibinfo {author} {\bibfnamefont {W.~B.}\
  \bibnamefont {Gragg}},\ }\href {\doibase
  http://dx.doi.org/10.1016/0377-0427(88)90386-X} {\bibfield  {journal}
  {\bibinfo  {journal} {J. Comput. Appl. Math.}\ }\textbf {\bibinfo {volume}
  {21}},\ \bibinfo {pages} {41 } (\bibinfo {year} {1988})}\BibitemShut
  {NoStop}%
\bibitem [{\citenamefont {Brics}\ and\ \citenamefont
  {Bauer}(2013)}]{Brics2013PRA}%
  \BibitemOpen
  \bibfield  {author} {\bibinfo {author} {\bibfnamefont {M.}~\bibnamefont
  {Brics}}\ and\ \bibinfo {author} {\bibfnamefont {D.}~\bibnamefont {Bauer}},\
  }\href@noop {} {\bibfield  {journal} {\bibinfo  {journal} {Phys. Rev. A}\
  }\textbf {\bibinfo {volume} {88}},\ \bibinfo {pages} {052514} (\bibinfo
  {year} {2013})}\BibitemShut {NoStop}%
\bibitem [{\citenamefont {Rapp}\ \emph
  {et~al.}(2014{\natexlab{a}})\citenamefont {Rapp}, \citenamefont {Brics},\
  and\ \citenamefont {Bauer}}]{Rapp2014PRAa}%
  \BibitemOpen
  \bibfield  {author} {\bibinfo {author} {\bibfnamefont {J.}~\bibnamefont
  {Rapp}}, \bibinfo {author} {\bibfnamefont {M.}~\bibnamefont {Brics}}, \ and\
  \bibinfo {author} {\bibfnamefont {D.}~\bibnamefont {Bauer}},\ }\href@noop {}
  {\bibfield  {journal} {\bibinfo  {journal} {Phys. Rev. A}\ }\textbf {\bibinfo
  {volume} {90}},\ \bibinfo {pages} {012518} (\bibinfo {year}
  {2014}{\natexlab{a}})}\BibitemShut {NoStop}%
\bibitem [{\citenamefont {Brics}\ \emph {et~al.}(2014)\citenamefont {Brics},
  \citenamefont {Rapp},\ and\ \citenamefont {Bauer}}]{Brics2014PRAb}%
  \BibitemOpen
  \bibfield  {author} {\bibinfo {author} {\bibfnamefont {M.}~\bibnamefont
  {Brics}}, \bibinfo {author} {\bibfnamefont {J.}~\bibnamefont {Rapp}}, \ and\
  \bibinfo {author} {\bibfnamefont {D.}~\bibnamefont {Bauer}},\ }\href@noop {}
  {\bibfield  {journal} {\bibinfo  {journal} {Phys. Rev. A}\ }\textbf {\bibinfo
  {volume} {90}},\ \bibinfo {pages} {053418} (\bibinfo {year}
  {2014})}\BibitemShut {NoStop}%
\bibitem [{\citenamefont {Rapp}\ \emph
  {et~al.}(2014{\natexlab{b}})\citenamefont {Rapp}, \citenamefont {Brics},\
  and\ \citenamefont {Bauer}}]{Rapp2014PRA}%
  \BibitemOpen
  \bibfield  {author} {\bibinfo {author} {\bibfnamefont {J.}~\bibnamefont
  {Rapp}}, \bibinfo {author} {\bibfnamefont {M.}~\bibnamefont {Brics}}, \ and\
  \bibinfo {author} {\bibfnamefont {D.}~\bibnamefont {Bauer}},\ }\href
  {\doibase 10.1103/PhysRevA.90.012518} {\bibfield  {journal} {\bibinfo
  {journal} {Phys. Rev. A}\ }\textbf {\bibinfo {volume} {90}},\ \bibinfo
  {pages} {012518} (\bibinfo {year} {2014}{\natexlab{b}})}\BibitemShut
  {NoStop}%
\bibitem [{\citenamefont {McWeeny}(1992)}]{McWeeny}%
  \BibitemOpen
  \bibfield  {author} {\bibinfo {author} {\bibfnamefont {R.}~\bibnamefont
  {McWeeny}},\ }\href@noop {} {\emph {\bibinfo {title} {Methods of Molecular
  Quantum Mechanics}}},\ \bibinfo {edition} {2nd}\ ed.\ (\bibinfo  {publisher}
  {Academic},\ \bibinfo {address} {San Diego, CA},\ \bibinfo {year}
  {1992})\BibitemShut {NoStop}%
\bibitem [{\citenamefont {Greenman}\ \emph {et~al.}(2010)\citenamefont
  {Greenman}, \citenamefont {Ho}, \citenamefont {Pabst}, \citenamefont
  {Kamarchik}, \citenamefont {Mazziotti},\ and\ \citenamefont
  {Santra}}]{Greenman2010PRA}%
  \BibitemOpen
  \bibfield  {author} {\bibinfo {author} {\bibfnamefont {L.}~\bibnamefont
  {Greenman}}, \bibinfo {author} {\bibfnamefont {P.~J.}\ \bibnamefont {Ho}},
  \bibinfo {author} {\bibfnamefont {S.}~\bibnamefont {Pabst}}, \bibinfo
  {author} {\bibfnamefont {E.}~\bibnamefont {Kamarchik}}, \bibinfo {author}
  {\bibfnamefont {D.~A.}\ \bibnamefont {Mazziotti}}, \ and\ \bibinfo {author}
  {\bibfnamefont {R.}~\bibnamefont {Santra}},\ }\href {\doibase
  10.1103/PhysRevA.82.023406} {\bibfield  {journal} {\bibinfo  {journal} {Phys.
  Rev. A}\ }\textbf {\bibinfo {volume} {82}},\ \bibinfo {pages} {023406}
  (\bibinfo {year} {2010})}\BibitemShut {NoStop}%
\bibitem [{\citenamefont {Rohringer}\ \emph {et~al.}(2006)\citenamefont
  {Rohringer}, \citenamefont {Gordon},\ and\ \citenamefont
  {Santra}}]{Rohringer2006PRA}%
  \BibitemOpen
  \bibfield  {author} {\bibinfo {author} {\bibfnamefont {N.}~\bibnamefont
  {Rohringer}}, \bibinfo {author} {\bibfnamefont {A.}~\bibnamefont {Gordon}}, \
  and\ \bibinfo {author} {\bibfnamefont {R.}~\bibnamefont {Santra}},\ }\href
  {\doibase 10.1103/PhysRevA.74.043420} {\bibfield  {journal} {\bibinfo
  {journal} {Phys. Rev. A}\ }\textbf {\bibinfo {volume} {74}},\ \bibinfo
  {pages} {043420} (\bibinfo {year} {2006})}\BibitemShut {NoStop}%
\bibitem [{\citenamefont {Pabst}(2013)}]{Pabst2013EPJST}%
  \BibitemOpen
  \bibfield  {author} {\bibinfo {author} {\bibfnamefont {S.}~\bibnamefont
  {Pabst}},\ }\href {\doibase 10.1140/epjst/e2013-01819-x} {\bibfield
  {journal} {\bibinfo  {journal} {Eur. Phys. J. Spec. Top.}\ }\textbf {\bibinfo
  {volume} {221}},\ \bibinfo {pages} {1} (\bibinfo {year} {2013})}\BibitemShut
  {NoStop}%
\bibitem [{\citenamefont {Pabst}\ \emph
  {et~al.}(2012{\natexlab{a}})\citenamefont {Pabst}, \citenamefont {Sytcheva},
  \citenamefont {Moulet}, \citenamefont {Wirth}, \citenamefont {Goulielmakis},\
  and\ \citenamefont {Santra}}]{Pabst2012PRAa}%
  \BibitemOpen
  \bibfield  {author} {\bibinfo {author} {\bibfnamefont {S.}~\bibnamefont
  {Pabst}}, \bibinfo {author} {\bibfnamefont {A.}~\bibnamefont {Sytcheva}},
  \bibinfo {author} {\bibfnamefont {A.}~\bibnamefont {Moulet}}, \bibinfo
  {author} {\bibfnamefont {A.}~\bibnamefont {Wirth}}, \bibinfo {author}
  {\bibfnamefont {E.}~\bibnamefont {Goulielmakis}}, \ and\ \bibinfo {author}
  {\bibfnamefont {R.}~\bibnamefont {Santra}},\ }\href {\doibase
  10.1103/PhysRevA.86.063411} {\bibfield  {journal} {\bibinfo  {journal} {Phys.
  Rev. A}\ }\textbf {\bibinfo {volume} {86}},\ \bibinfo {pages} {063411}
  (\bibinfo {year} {2012}{\natexlab{a}})}\BibitemShut {NoStop}%
\bibitem [{\citenamefont {Pabst}\ \emph {et~al.}(2011)\citenamefont {Pabst},
  \citenamefont {Greenman}, \citenamefont {Ho}, \citenamefont {Mazziotti},\
  and\ \citenamefont {Santra}}]{Pabst2011PRL}%
  \BibitemOpen
  \bibfield  {author} {\bibinfo {author} {\bibfnamefont {S.}~\bibnamefont
  {Pabst}}, \bibinfo {author} {\bibfnamefont {L.}~\bibnamefont {Greenman}},
  \bibinfo {author} {\bibfnamefont {P.~J.}\ \bibnamefont {Ho}}, \bibinfo
  {author} {\bibfnamefont {D.~A.}\ \bibnamefont {Mazziotti}}, \ and\ \bibinfo
  {author} {\bibfnamefont {R.}~\bibnamefont {Santra}},\ }\href {\doibase
  10.1103/PhysRevLett.106.053003} {\bibfield  {journal} {\bibinfo  {journal}
  {Phys. Rev. Lett.}\ }\textbf {\bibinfo {volume} {106}},\ \bibinfo {pages}
  {053003} (\bibinfo {year} {2011})}\BibitemShut {NoStop}%
\bibitem [{\citenamefont {Pabst}\ \emph
  {et~al.}(2012{\natexlab{b}})\citenamefont {Pabst}, \citenamefont {Greenman},
  \citenamefont {Mazziotti},\ and\ \citenamefont {Santra}}]{Pabst2012PRAb}%
  \BibitemOpen
  \bibfield  {author} {\bibinfo {author} {\bibfnamefont {S.}~\bibnamefont
  {Pabst}}, \bibinfo {author} {\bibfnamefont {L.}~\bibnamefont {Greenman}},
  \bibinfo {author} {\bibfnamefont {D.~A.}\ \bibnamefont {Mazziotti}}, \ and\
  \bibinfo {author} {\bibfnamefont {R.}~\bibnamefont {Santra}},\ }\href
  {\doibase 10.1103/PhysRevA.85.023411} {\bibfield  {journal} {\bibinfo
  {journal} {Phys. Rev. A}\ }\textbf {\bibinfo {volume} {85}},\ \bibinfo
  {pages} {023411} (\bibinfo {year} {2012}{\natexlab{b}})}\BibitemShut
  {NoStop}%
\bibitem [{\citenamefont {Pabst}\ and\ \citenamefont
  {Santra}(2013)}]{Pabst2013PRL}%
  \BibitemOpen
  \bibfield  {author} {\bibinfo {author} {\bibfnamefont {S.}~\bibnamefont
  {Pabst}}\ and\ \bibinfo {author} {\bibfnamefont {R.}~\bibnamefont {Santra}},\
  }\href {\doibase 10.1103/PhysRevLett.111.233005} {\bibfield  {journal}
  {\bibinfo  {journal} {Phys. Rev. Lett.}\ }\textbf {\bibinfo {volume} {111}},\
  \bibinfo {pages} {233005} (\bibinfo {year} {2013})}\BibitemShut {NoStop}%
\bibitem [{\citenamefont {Sytcheva}\ \emph {et~al.}(2012)\citenamefont
  {Sytcheva}, \citenamefont {Pabst}, \citenamefont {Son},\ and\ \citenamefont
  {Santra}}]{Sytcheva2012PRA}%
  \BibitemOpen
  \bibfield  {author} {\bibinfo {author} {\bibfnamefont {A.}~\bibnamefont
  {Sytcheva}}, \bibinfo {author} {\bibfnamefont {S.}~\bibnamefont {Pabst}},
  \bibinfo {author} {\bibfnamefont {S.-K.}\ \bibnamefont {Son}}, \ and\
  \bibinfo {author} {\bibfnamefont {R.}~\bibnamefont {Santra}},\ }\href
  {\doibase 10.1103/PhysRevA.85.023414} {\bibfield  {journal} {\bibinfo
  {journal} {Phys. Rev. A}\ }\textbf {\bibinfo {volume} {85}},\ \bibinfo
  {pages} {023414} (\bibinfo {year} {2012})}\BibitemShut {NoStop}%
\bibitem [{\citenamefont {Karamatskou}\ \emph {et~al.}(2013)\citenamefont
  {Karamatskou}, \citenamefont {Pabst},\ and\ \citenamefont
  {Santra}}]{Karamatskou2013PRA}%
  \BibitemOpen
  \bibfield  {author} {\bibinfo {author} {\bibfnamefont {A.}~\bibnamefont
  {Karamatskou}}, \bibinfo {author} {\bibfnamefont {S.}~\bibnamefont {Pabst}},
  \ and\ \bibinfo {author} {\bibfnamefont {R.}~\bibnamefont {Santra}},\ }\href
  {\doibase 10.1103/PhysRevA.87.043422} {\bibfield  {journal} {\bibinfo
  {journal} {Phys. Rev. A}\ }\textbf {\bibinfo {volume} {87}},\ \bibinfo
  {pages} {043422} (\bibinfo {year} {2013})}\BibitemShut {NoStop}%
\bibitem [{\citenamefont {Kato}\ and\ \citenamefont
  {Kono}(2004)}]{Kato2004CPL}%
  \BibitemOpen
  \bibfield  {author} {\bibinfo {author} {\bibfnamefont {T.}~\bibnamefont
  {Kato}}\ and\ \bibinfo {author} {\bibfnamefont {H.}~\bibnamefont {Kono}},\
  }\href {\doibase http://dx.doi.org/10.1016/j.cplett.2004.05.106} {\bibfield
  {journal} {\bibinfo  {journal} {Chem. Phys. Lett.}\ }\textbf {\bibinfo
  {volume} {392}},\ \bibinfo {pages} {533} (\bibinfo {year}
  {2004})}\BibitemShut {NoStop}%
\bibitem [{\citenamefont {Pindzola}\ \emph {et~al.}(1991)\citenamefont
  {Pindzola}, \citenamefont {Griffin},\ and\ \citenamefont
  {Bottcher}}]{Pindzola1991PRL}%
  \BibitemOpen
  \bibfield  {author} {\bibinfo {author} {\bibfnamefont {M.~S.}\ \bibnamefont
  {Pindzola}}, \bibinfo {author} {\bibfnamefont {D.~C.}\ \bibnamefont
  {Griffin}}, \ and\ \bibinfo {author} {\bibfnamefont {C.}~\bibnamefont
  {Bottcher}},\ }\href {\doibase 10.1103/PhysRevLett.66.2305} {\bibfield
  {journal} {\bibinfo  {journal} {Phys. Rev. Lett.}\ }\textbf {\bibinfo
  {volume} {66}},\ \bibinfo {pages} {2305} (\bibinfo {year}
  {1991})}\BibitemShut {NoStop}%
\bibitem [{\citenamefont {Ivanic}(2003)}]{Ivanic2003JCP-a}%
  \BibitemOpen
  \bibfield  {author} {\bibinfo {author} {\bibfnamefont {J.}~\bibnamefont
  {Ivanic}},\ }\href {\doibase http://dx.doi.org/10.1063/1.1615954} {\bibfield
  {journal} {\bibinfo  {journal} {J. Chem. Phys.}\ }\textbf {\bibinfo {volume}
  {119}},\ \bibinfo {pages} {9364} (\bibinfo {year} {2003})}\BibitemShut
  {NoStop}%
\bibitem [{\citenamefont {Miyagi}\ and\ \citenamefont
  {Madsen}(2013)}]{Miyagi2013PRA}%
  \BibitemOpen
  \bibfield  {author} {\bibinfo {author} {\bibfnamefont {H.}~\bibnamefont
  {Miyagi}}\ and\ \bibinfo {author} {\bibfnamefont {L.~B.}\ \bibnamefont
  {Madsen}},\ }\href {\doibase 10.1103/PhysRevA.87.062511} {\bibfield
  {journal} {\bibinfo  {journal} {Phys. Rev. A}\ }\textbf {\bibinfo {volume}
  {87}},\ \bibinfo {pages} {062511} (\bibinfo {year} {2013})}\BibitemShut
  {NoStop}%
\bibitem [{\citenamefont {Miyagi}\ and\ \citenamefont
  {Madsen}(2014)}]{Miyagi2014PRA}%
  \BibitemOpen
  \bibfield  {author} {\bibinfo {author} {\bibfnamefont {H.}~\bibnamefont
  {Miyagi}}\ and\ \bibinfo {author} {\bibfnamefont {L.~B.}\ \bibnamefont
  {Madsen}},\ }\href {\doibase 10.1103/PhysRevA.89.063416} {\bibfield
  {journal} {\bibinfo  {journal} {Phys. Rev. A}\ }\textbf {\bibinfo {volume}
  {89}},\ \bibinfo {pages} {063416} (\bibinfo {year} {2014})}\BibitemShut
  {NoStop}%
\bibitem [{\citenamefont {Bauch}\ \emph {et~al.}(2014)\citenamefont {Bauch},
  \citenamefont {S\o{}rensen},\ and\ \citenamefont {Madsen}}]{Bauch2014PRA}%
  \BibitemOpen
  \bibfield  {author} {\bibinfo {author} {\bibfnamefont {S.}~\bibnamefont
  {Bauch}}, \bibinfo {author} {\bibfnamefont {L.~K.}\ \bibnamefont
  {S\o{}rensen}}, \ and\ \bibinfo {author} {\bibfnamefont {L.~B.}\ \bibnamefont
  {Madsen}},\ }\href {\doibase 10.1103/PhysRevA.90.062508} {\bibfield
  {journal} {\bibinfo  {journal} {Phys. Rev. A}\ }\textbf {\bibinfo {volume}
  {90}},\ \bibinfo {pages} {062508} (\bibinfo {year} {2014})}\BibitemShut
  {NoStop}%
\bibitem [{\citenamefont {Olsen}\ \emph {et~al.}(1988)\citenamefont {Olsen},
  \citenamefont {Roos}, \citenamefont {J\o~rgensen},\ and\ \citenamefont
  {Jensen}}]{Olsen1988JCP}%
  \BibitemOpen
  \bibfield  {author} {\bibinfo {author} {\bibfnamefont {J.}~\bibnamefont
  {Olsen}}, \bibinfo {author} {\bibfnamefont {B.~O.}\ \bibnamefont {Roos}},
  \bibinfo {author} {\bibfnamefont {P.}~\bibnamefont {J\o~rgensen}}, \ and\
  \bibinfo {author} {\bibfnamefont {H.~J. r.~A.}\ \bibnamefont {Jensen}},\
  }\href {\doibase http://dx.doi.org/10.1063/1.455063} {\bibfield  {journal}
  {\bibinfo  {journal} {J. Chem. Phys.}\ }\textbf {\bibinfo {volume} {89}},\
  \bibinfo {pages} {2185} (\bibinfo {year} {1988})}\BibitemShut {NoStop}%
\bibitem [{\citenamefont {Haxton}\ and\ \citenamefont
  {McCurdy}(2015)}]{Haxton2015PRA}%
  \BibitemOpen
  \bibfield  {author} {\bibinfo {author} {\bibfnamefont {D.~J.}\ \bibnamefont
  {Haxton}}\ and\ \bibinfo {author} {\bibfnamefont {C.~W.}\ \bibnamefont
  {McCurdy}},\ }\href {\doibase 10.1103/PhysRevA.91.012509} {\bibfield
  {journal} {\bibinfo  {journal} {Phys. Rev. A}\ }\textbf {\bibinfo {volume}
  {91}},\ \bibinfo {pages} {012509} (\bibinfo {year} {2015})}\BibitemShut
  {NoStop}%
\bibitem [{\citenamefont {Bandrauk}\ \emph {et~al.}(2013)\citenamefont
  {Bandrauk}, \citenamefont {Fillion-Gourdeau},\ and\ \citenamefont
  {Lorin}}]{Bandrauk2013JPB}%
  \BibitemOpen
  \bibfield  {author} {\bibinfo {author} {\bibfnamefont {A.~D.}\ \bibnamefont
  {Bandrauk}}, \bibinfo {author} {\bibfnamefont {F.}~\bibnamefont
  {Fillion-Gourdeau}}, \ and\ \bibinfo {author} {\bibfnamefont
  {E.}~\bibnamefont {Lorin}},\ }\href
  {http://stacks.iop.org/0953-4075/46/i=15/a=153001} {\bibfield  {journal}
  {\bibinfo  {journal} {J. Phys. B}\ }\textbf {\bibinfo {volume} {46}},\
  \bibinfo {pages} {153001} (\bibinfo {year} {2013})}\BibitemShut {NoStop}%
\bibitem [{\citenamefont {Huber}\ and\ \citenamefont
  {Klamroth}(2011)}]{Huber2011JCP}%
  \BibitemOpen
  \bibfield  {author} {\bibinfo {author} {\bibfnamefont {C.}~\bibnamefont
  {Huber}}\ and\ \bibinfo {author} {\bibfnamefont {T.}~\bibnamefont
  {Klamroth}},\ }\href {\doibase http://dx.doi.org/10.1063/1.3530807}
  {\bibfield  {journal} {\bibinfo  {journal} {J. Chem. Phys.}\ }\textbf
  {\bibinfo {volume} {134}},\ \bibinfo {eid} {054113} (\bibinfo {year}
  {2011})}\BibitemShut {NoStop}%
\bibitem [{\citenamefont {Kvaal}(2012)}]{Kvaal2012JCP}%
  \BibitemOpen
  \bibfield  {author} {\bibinfo {author} {\bibfnamefont {S.}~\bibnamefont
  {Kvaal}},\ }\href {\doibase http://dx.doi.org/10.1063/1.4718427} {\bibfield
  {journal} {\bibinfo  {journal} {J. Chem. Phys.}\ }\textbf {\bibinfo {volume}
  {136}},\ \bibinfo {eid} {194109} (\bibinfo {year} {2012})}\BibitemShut
  {NoStop}%
\bibitem [{\citenamefont {Hutchinson}\ \emph {et~al.}(2010)\citenamefont
  {Hutchinson}, \citenamefont {Lysaght},\ and\ \citenamefont {van~der
  Hart}}]{Hutchinson2010JPB}%
  \BibitemOpen
  \bibfield  {author} {\bibinfo {author} {\bibfnamefont {S.}~\bibnamefont
  {Hutchinson}}, \bibinfo {author} {\bibfnamefont {M.~A.}\ \bibnamefont
  {Lysaght}}, \ and\ \bibinfo {author} {\bibfnamefont {H.~W.}\ \bibnamefont
  {van~der Hart}},\ }\href {http://stacks.iop.org/0953-4075/43/i=9/a=095603}
  {\bibfield  {journal} {\bibinfo  {journal} {J. Phys. B}\ }\textbf {\bibinfo
  {volume} {43}},\ \bibinfo {pages} {095603} (\bibinfo {year}
  {2010})}\BibitemShut {NoStop}%
\bibitem [{\citenamefont {Burke}\ and\ \citenamefont
  {Burke}(1997)}]{Burke1997JPB}%
  \BibitemOpen
  \bibfield  {author} {\bibinfo {author} {\bibfnamefont {P.~G.}\ \bibnamefont
  {Burke}}\ and\ \bibinfo {author} {\bibfnamefont {V.~M.}\ \bibnamefont
  {Burke}},\ }\href {http://stacks.iop.org/0953-4075/30/i=11/a=002} {\bibfield
  {journal} {\bibinfo  {journal} {J. Phys. B}\ }\textbf {\bibinfo {volume}
  {30}},\ \bibinfo {pages} {L383} (\bibinfo {year} {1997})}\BibitemShut
  {NoStop}%
\bibitem [{\citenamefont {Lysaght}\ \emph {et~al.}(2008)\citenamefont
  {Lysaght}, \citenamefont {Burke},\ and\ \citenamefont {van~der
  Hart}}]{Lysaght2008PRL}%
  \BibitemOpen
  \bibfield  {author} {\bibinfo {author} {\bibfnamefont {M.~A.}\ \bibnamefont
  {Lysaght}}, \bibinfo {author} {\bibfnamefont {P.~G.}\ \bibnamefont {Burke}},
  \ and\ \bibinfo {author} {\bibfnamefont {H.~W.}\ \bibnamefont {van~der
  Hart}},\ }\href {\doibase 10.1103/PhysRevLett.101.253001} {\bibfield
  {journal} {\bibinfo  {journal} {Phys. Rev. Lett.}\ }\textbf {\bibinfo
  {volume} {101}},\ \bibinfo {pages} {253001} (\bibinfo {year}
  {2008})}\BibitemShut {NoStop}%
\bibitem [{\citenamefont {Lysaght}\ \emph {et~al.}(2009)\citenamefont
  {Lysaght}, \citenamefont {van~der Hart},\ and\ \citenamefont
  {Burke}}]{Lysaght2009PRA}%
  \BibitemOpen
  \bibfield  {author} {\bibinfo {author} {\bibfnamefont {M.~A.}\ \bibnamefont
  {Lysaght}}, \bibinfo {author} {\bibfnamefont {H.~W.}\ \bibnamefont {van~der
  Hart}}, \ and\ \bibinfo {author} {\bibfnamefont {P.~G.}\ \bibnamefont
  {Burke}},\ }\href {\doibase 10.1103/PhysRevA.79.053411} {\bibfield  {journal}
  {\bibinfo  {journal} {Phys. Rev. A}\ }\textbf {\bibinfo {volume} {79}},\
  \bibinfo {pages} {053411} (\bibinfo {year} {2009})}\BibitemShut {NoStop}%
\bibitem [{\citenamefont {Lysaght}\ \emph {et~al.}(2011)\citenamefont
  {Lysaght}, \citenamefont {Moore}, \citenamefont {Nikolopoulos}, \citenamefont
  {Parker}, \citenamefont {van~der Hart},\ and\ \citenamefont
  {Taylor}}]{Lysaght-QDI}%
  \BibitemOpen
  \bibfield  {author} {\bibinfo {author} {\bibfnamefont {M.}~\bibnamefont
  {Lysaght}}, \bibinfo {author} {\bibfnamefont {L.}~\bibnamefont {Moore}},
  \bibinfo {author} {\bibfnamefont {L.}~\bibnamefont {Nikolopoulos}}, \bibinfo
  {author} {\bibfnamefont {J.}~\bibnamefont {Parker}}, \bibinfo {author}
  {\bibfnamefont {H.}~\bibnamefont {van~der Hart}}, \ and\ \bibinfo {author}
  {\bibfnamefont {K.}~\bibnamefont {Taylor}},\ }\enquote {\bibinfo {title}
  {{\rm Ab Initio} methods for few- and many-electron atomic systems in intense
  short-pulse laser light},}\ in\ \href@noop {} {\emph {\bibinfo {booktitle}
  {Quantum Dynamic Imaging}}},\ \bibinfo {editor} {edited by\ \bibinfo {editor}
  {\bibfnamefont {A.~D.}\ \bibnamefont {Bandrauk}}\ and\ \bibinfo {editor}
  {\bibfnamefont {M.}~\bibnamefont {Ivanov}}}\ (\bibinfo  {publisher}
  {Springer},\ \bibinfo {address} {New York},\ \bibinfo {year} {2011})\
  Chap.~\bibinfo {chapter} {8}, pp.\ \bibinfo {pages} {107--134}\BibitemShut
  {NoStop}%
\bibitem [{\citenamefont {Guan}\ \emph {et~al.}(2007)\citenamefont {Guan},
  \citenamefont {Zatsarinny}, \citenamefont {Bartschat}, \citenamefont
  {Schneider}, \citenamefont {Feist},\ and\ \citenamefont
  {Noble}}]{Guan2007PRA}%
  \BibitemOpen
  \bibfield  {author} {\bibinfo {author} {\bibfnamefont {X.}~\bibnamefont
  {Guan}}, \bibinfo {author} {\bibfnamefont {O.}~\bibnamefont {Zatsarinny}},
  \bibinfo {author} {\bibfnamefont {K.}~\bibnamefont {Bartschat}}, \bibinfo
  {author} {\bibfnamefont {B.~I.}\ \bibnamefont {Schneider}}, \bibinfo {author}
  {\bibfnamefont {J.}~\bibnamefont {Feist}}, \ and\ \bibinfo {author}
  {\bibfnamefont {C.~J.}\ \bibnamefont {Noble}},\ }\href {\doibase
  10.1103/PhysRevA.76.053411} {\bibfield  {journal} {\bibinfo  {journal} {Phys.
  Rev. A}\ }\textbf {\bibinfo {volume} {76}},\ \bibinfo {pages} {053411}
  (\bibinfo {year} {2007})}\BibitemShut {NoStop}%
\bibitem [{\citenamefont {Bartschat}\ \emph {et~al.}(2011)\citenamefont
  {Bartschat}, \citenamefont {Guan}, \citenamefont {Noble}, \citenamefont
  {Schneider},\ and\ \citenamefont {Zatsarinny}}]{Bartschat-QDI}%
  \BibitemOpen
  \bibfield  {author} {\bibinfo {author} {\bibfnamefont {K.}~\bibnamefont
  {Bartschat}}, \bibinfo {author} {\bibfnamefont {X.}~\bibnamefont {Guan}},
  \bibinfo {author} {\bibfnamefont {C.}~\bibnamefont {Noble}}, \bibinfo
  {author} {\bibfnamefont {B.}~\bibnamefont {Schneider}}, \ and\ \bibinfo
  {author} {\bibfnamefont {O.}~\bibnamefont {Zatsarinny}},\ }\enquote {\bibinfo
  {title} {Multi-photon single and double ionization of complex atoms by
  ultrashort intense laser pulses},}\ in\ \href@noop {} {\emph {\bibinfo
  {booktitle} {Quantum Dynamic Imaging}}},\ \bibinfo {editor} {edited by\
  \bibinfo {editor} {\bibfnamefont {A.~D.}\ \bibnamefont {Bandrauk}}\ and\
  \bibinfo {editor} {\bibfnamefont {M.}~\bibnamefont {Ivanov}}}\ (\bibinfo
  {publisher} {Springer},\ \bibinfo {address} {New York},\ \bibinfo {year}
  {2011})\ Chap.~\bibinfo {chapter} {2}, pp.\ \bibinfo {pages}
  {13--22}\BibitemShut {NoStop}%
\bibitem [{\citenamefont {Zatsarinny}\ and\ \citenamefont
  {Fischer}(2000)}]{Zatsarinny2000JPB}%
  \BibitemOpen
  \bibfield  {author} {\bibinfo {author} {\bibfnamefont {O.}~\bibnamefont
  {Zatsarinny}}\ and\ \bibinfo {author} {\bibfnamefont {C.~F.}\ \bibnamefont
  {Fischer}},\ }\href {http://stacks.iop.org/0953-4075/33/i=3/a=303} {\bibfield
   {journal} {\bibinfo  {journal} {J. Phys. B}\ }\textbf {\bibinfo {volume}
  {33}},\ \bibinfo {pages} {313} (\bibinfo {year} {2000})}\BibitemShut
  {NoStop}%
\bibitem [{\citenamefont {Zatsarinny}\ and\ \citenamefont
  {Bartschat}(2004)}]{Zatsarinny2004JPB}%
  \BibitemOpen
  \bibfield  {author} {\bibinfo {author} {\bibfnamefont {O.}~\bibnamefont
  {Zatsarinny}}\ and\ \bibinfo {author} {\bibfnamefont {K.}~\bibnamefont
  {Bartschat}},\ }\href {http://stacks.iop.org/0953-4075/37/i=10/a=013}
  {\bibfield  {journal} {\bibinfo  {journal} {J. Phys. B}\ }\textbf {\bibinfo
  {volume} {37}},\ \bibinfo {pages} {2173} (\bibinfo {year}
  {2004})}\BibitemShut {NoStop}%
\bibitem [{\citenamefont {Zatsarinny}(2006)}]{Zatsarinny2006CPC}%
  \BibitemOpen
  \bibfield  {author} {\bibinfo {author} {\bibfnamefont {O.}~\bibnamefont
  {Zatsarinny}},\ }\href {\doibase http://dx.doi.org/10.1016/j.cpc.2005.10.006}
  {\bibfield  {journal} {\bibinfo  {journal} {Comput. Phys. Commun.}\ }\textbf
  {\bibinfo {volume} {174}},\ \bibinfo {pages} {273 } (\bibinfo {year}
  {2006})}\BibitemShut {NoStop}%
\bibitem [{\citenamefont {Park}\ and\ \citenamefont
  {Light}(1986)}]{Park1986JCP}%
  \BibitemOpen
  \bibfield  {author} {\bibinfo {author} {\bibfnamefont {T.~J.}\ \bibnamefont
  {Park}}\ and\ \bibinfo {author} {\bibfnamefont {J.~C.}\ \bibnamefont
  {Light}},\ }\href {\doibase http://dx.doi.org/10.1063/1.451548} {\bibfield
  {journal} {\bibinfo  {journal} {J. Chem. Phys.}\ }\textbf {\bibinfo {volume}
  {85}},\ \bibinfo {pages} {5870} (\bibinfo {year} {1986})}\BibitemShut
  {NoStop}%
\bibitem [{\citenamefont {Schneider}\ and\ \citenamefont
  {Collins}(2005)}]{Schneider2005JNCS}%
  \BibitemOpen
  \bibfield  {author} {\bibinfo {author} {\bibfnamefont {B.~I.}\ \bibnamefont
  {Schneider}}\ and\ \bibinfo {author} {\bibfnamefont {L.~A.}\ \bibnamefont
  {Collins}},\ }\href {\doibase
  http://dx.doi.org/10.1016/j.jnoncrysol.2005.03.028} {\bibfield  {journal}
  {\bibinfo  {journal} {J. Non-Cryst. Solids}\ }\textbf {\bibinfo {volume}
  {351}},\ \bibinfo {pages} {1551 } (\bibinfo {year} {2005})},\ \bibinfo {note}
  {papers from the Michael Weinberg Symposium Michael Weinberg
  Symposium}\BibitemShut {NoStop}%
\bibitem [{\citenamefont {Saad}(2003)}]{Saad}%
  \BibitemOpen
  \bibfield  {author} {\bibinfo {author} {\bibfnamefont {Y.}~\bibnamefont
  {Saad}},\ }\href@noop {} {\emph {\bibinfo {title} {Iterative Methods for
  Sparse Linear Systems}}},\ \bibinfo {edition} {2nd}\ ed.\ (\bibinfo
  {publisher} {SIAM},\ \bibinfo {address} {Philadelphia},\ \bibinfo {year}
  {2003})\BibitemShut {NoStop}%
\bibitem [{\citenamefont {Kato}\ and\ \citenamefont
  {Yamanouchi}(2009)}]{Kato2009JCP}%
  \BibitemOpen
  \bibfield  {author} {\bibinfo {author} {\bibfnamefont {T.}~\bibnamefont
  {Kato}}\ and\ \bibinfo {author} {\bibfnamefont {K.}~\bibnamefont
  {Yamanouchi}},\ }\href {\doibase http://dx.doi.org/10.1063/1.3249967}
  {\bibfield  {journal} {\bibinfo  {journal} {The Journal of Chemical Physics}\
  }\textbf {\bibinfo {volume} {131}},\ \bibinfo {eid} {164118} (\bibinfo {year}
  {2009}),\ http://dx.doi.org/10.1063/1.3249967}\BibitemShut {NoStop}%
\bibitem [{\citenamefont {Mercouris}\ \emph {et~al.}(1994)\citenamefont
  {Mercouris}, \citenamefont {Komninos}, \citenamefont {Dionissopoulou},\ and\
  \citenamefont {Nicolaides}}]{Mercouris1994PRA}%
  \BibitemOpen
  \bibfield  {author} {\bibinfo {author} {\bibfnamefont {T.}~\bibnamefont
  {Mercouris}}, \bibinfo {author} {\bibfnamefont {Y.}~\bibnamefont {Komninos}},
  \bibinfo {author} {\bibfnamefont {S.}~\bibnamefont {Dionissopoulou}}, \ and\
  \bibinfo {author} {\bibfnamefont {C.~A.}\ \bibnamefont {Nicolaides}},\ }\href
  {\doibase 10.1103/PhysRevA.50.4109} {\bibfield  {journal} {\bibinfo
  {journal} {Phys. Rev. A}\ }\textbf {\bibinfo {volume} {50}},\ \bibinfo
  {pages} {4109} (\bibinfo {year} {1994})}\BibitemShut {NoStop}%
\bibitem [{\citenamefont {Mercouris}\ \emph {et~al.}(2010)\citenamefont
  {Mercouris}, \citenamefont {Komninos},\ and\ \citenamefont
  {Nicolaides}}]{Mercouris2010AQC}%
  \BibitemOpen
  \bibfield  {author} {\bibinfo {author} {\bibfnamefont {T.}~\bibnamefont
  {Mercouris}}, \bibinfo {author} {\bibfnamefont {Y.}~\bibnamefont {Komninos}},
  \ and\ \bibinfo {author} {\bibfnamefont {C.~A.}\ \bibnamefont {Nicolaides}},\
  }in\ \href {\doibase http://dx.doi.org/10.1016/S0065-3276(10)60006-8} {\emph
  {\bibinfo {booktitle} {Unstable States in the Continuous Spectra, Part I:
  Analysis, Concepts, Methods, and Results}}},\ \bibinfo {series} {Advances in
  Quantum Chemistry}, Vol.~\bibinfo {volume} {60},\ \bibinfo {editor} {edited
  by\ \bibinfo {editor} {\bibfnamefont {C.~A.}\ \bibnamefont {Nicolaides}}\
  and\ \bibinfo {editor} {\bibfnamefont {E.}~\bibnamefont {Brandas}}}\
  (\bibinfo  {publisher} {Academic Press},\ \bibinfo {year} {2010})\ pp.\
  \bibinfo {pages} {333 -- 405}\BibitemShut {NoStop}%
\end{thebibliography}%
\end{document}